\documentclass[12pt,letterpaper]{article}
\usepackage[utf8]{inputenc}
\usepackage[english]{babel}
\usepackage[vmargin=0.6in,hmargin={1in,1in},headsep=0.3in,headheight=1in]{geometry}
\usepackage{subfiles}
\usepackage{tikz}
\usepackage{fix-cm}
\usepackage{enumitem}
\usepackage{multicol}
\usepackage{amsmath}
\usepackage{amsthm,amssymb,amsfonts}
\usepackage{array}
\usepackage{relsize}
\usepackage{xcolor}
\usepackage{lastpage}
\usepackage{blindtext}
\usepackage[linktocpage=true]{hyperref}
\hypersetup{
	colorlinks,
	citecolor=blue,
	filecolor=black,
	linkcolor=blue,
	urlcolor=blue
}
\usepackage{graphicx}
\usepackage{subcaption}
\usepackage{wrapfig}
\usepackage{indentfirst}
\usepackage{hyperref}
\usepackage{soul}
\usepackage{empheq}
\usepackage{placeins}
\usepackage{mathabx}
\setlist{parsep=0pt,listparindent=\parindent}
\setlist[itemize]{noitemsep, topsep=0pt}
\setlist[enumerate]{noitemsep, topsep=0pt}
\setlist{parsep=0pt,listparindent=\parindent}

\stepcounter{footnote}
\graphicspath{{Images/}{../Images/}}
\catcode`\^=13\def^#1{\sp{#1}{}}
\catcode`\_=13\def_#1{\sb{#1}{}}

\title{\textbf{\Huge{Curvature Invariants for the Alcubierre and Nat\'ario Warp Drives}}}
\author{B.~Mattingly$^{1,2}$\footnote{mailto: \em\texttt{\href{Gerald_Cleaver@Baylor.edu}{Gerald\_Cleaver@Baylor.edu}}}, A.~Kar$^{1,2}$, M.~Gorban$^{1,2}$, W.~Julius$^{1,2}$, C.~K.~Watson$^{1,2}$, \\
M.~Ali$^{1,2}$, A.~Baas$^{1,2}$, C.~Elmore$^{1,2}$, J.~S.~Lee$^{1,2}$, B. Shakerin$^{1,2}$, \\
E.~W.~Davis$^{1,3}$ and G.~B.~Cleaver$^{1,2}$}
\date{}
\begin{document}
		
\maketitle
\vspace{-0.8cm}
\begin{center}
\begin{minipage}[c]{0.72\textwidth}
	$^1$\emph{Early Universe, Cosmology and Strings \textnormal{(EUCOS)} Group, Center for Astrophysics, Space Physics and Engineering Research \textnormal{(CASPER)}, Baylor University, Waco, TX 76798, USA}\\ \\
	$^2$\emph{Department of Physics, Baylor University, Waco, TX 76798, USA}\\ \\
	$^3$\emph{Institute for Advanced Studies at Austin, 11855 Research Blvd., Austin, TX 78759, USA}\\
\end{minipage}
\end{center}	

\begin{minipage}{0.92\textwidth}
	\textbf{\large{Abstract}:} A process for using curvature invariants is applied to evaluate the metrics for the Alcubierre and the Nat\'ario warp drives at a constant velocity. 
	Curvature invariants are independent of coordinate bases, so plotting these invariants will be free of coordinate mapping distortions. 
	As a consequence, they provide a novel perspective into complex spacetimes such as warp drives.
	Warp drives are the theoretical solutions to Einstein's field equations that allow the possibility for faster-than-light (FTL) travel.
	While their mathematics is well established, the visualisation of such spacetimes is unexplored.
	This paper uses the methods of computing and plotting the warp drive curvature invariants to reveal these spacetimes.
	The warp drive parameters of velocity, skin depth and radius are varied individually and then plotted to see each parameter's unique effect on the surrounding curvature.
	For each warp drive, this research shows a safe harbor and how the shape function forms the warp bubble.
	The curvature plots for the constant velocity Nat\'ario warp drive do not contain a wake or a constant curvature indicating that these are unique features of the accelerating Nat\'ario warp drive.
\end{minipage}
	
\section{Introduction}
    No particle may have a local velocity that exceeds the speed of light in vacuum, $c$, in Newtonian mechanics and special relativity. 
    However, general relativity~(GR) allows a particle's global velocity to exceed $c$ while its local velocity obeys the prior statement.
    Alcubierre noticed that spacetime itself may expand and contract at arbitrary rates \cite{Alcubierre:1994tu}.
    He proposed pairing a local contraction of spacetime in front of a spaceship with a local expansion of spacetime behind it.
    While the spaceship remains within its own light cone and its local velocity never exceeds $c$, the global velocity, which is defined as the proper spatial distance divided by proper time, may be much greater than $c$ due to the contraction and expansion of spacetime.
    Distant observers will perceive the ship to be moving at a global velocity greater than $c$, and the spaceship will be able to make a trip to a distant star in an arbitrarily short proper time. 
    He named the faster-than-light (FTL) propulsion mechanism based on this principle a ``warp drive.''  
    
    A spaceship using an FTL warp drive must obey eight prerequisites to carry a human to a distant star \cite{Davis}. 
    First, the rocket equation does not describe the portion of the flight undergoing FTL travel.
    Second, the trip duration to a distant star may be reduced to be less than one year, as seen by both observers inside of the warp, which are called passengers, and by stationary observers outside of the warp. 
    Third, the proper time measured by any passenger should not be dilated by relativistic effects. 
    Fourth, the magnitude of any tidal-gravity accelerations acting on the passengers will be less than $g_{\Earth}$, which is the acceleration of gravity near the Earth's surface. 
    Fifth, the local speed of any passengers should be less than $c$. 
    Sixth, the matter of the passengers must not couple with any material used to generate the FTL space warp. 
    Seventh, the FTL warp should not generate an event horizon. 
    Eighth, the passengers riding the FTL warp should not encounter a singularity inside or out of it.  
    
    The two most well known superluminal solutions to Einstein's equations that obey these eight requirements are traversable wormholes and warp drives ~\cite{Alcubierre:1994tu,Natario:2001tk,Krasnikov:1995ad,VanDenBroeck:1999sn,5,6,9,Loup,Loup2,11,Lobo2,Lobo3}. 
    While Einstein's equations allow for their possibility, each solution remains at the theoretical level.
    Furthermore, constructing either solution in a lab is not readily accessible due to engineering constraints.
    Instead, these solutions provide a rich environment to test the superluminal limits of GR.
    For example, superluminal spacetimes require exotic matter, which is defined as matter that violates a null energy condition (NEC) and appear to allow closed-timelike-curves (CTC). 
    While exotic matter may appear problematic, certain quantum fields, such as the fields giving rise to the Casimir effect and the cosmological fluid appear to violate several NECs \cite{Lobo2}.
    In this manner, these superluminal solutions may probe the boundaries of physics. 
    
    Research into FTL warp drives has advanced tremendously since Alcubierre's original proposal. 
    Krasnikov developed a non-tachyonic FTL warp bubble \cite{Krasnikov:1995ad}. 
    Van Den Broeck reduced the amount of energy required by Alcubierre's warp drive by positing a warp bubble with a microscopic surface area and a macroscopic volume inside \cite{VanDenBroeck:1999sn}. 
    His modification reduced the energy requirements to form the warp bubble to only a few solar masses and his geometry has more lenient violation of the NEC.
    Later, Nat\'ario presented the geodesic equations for the general warp drive spacetime, Equation~\eqref{12} and his warp drive spacetime, Equation~\eqref{17}, that required zero spacetime expansion to occur~\cite{Natario:2001tk}. 
    His warp drive ``slides" through the exterior spacetime at a constant global velocity by balancing a contraction of the distance in front of it with an expansion of the distance behind it.
    His proposal revealed the essential property of a warp drive to be the change in distances along the direction of motion, and not the expansion/contraction of spacetime.
    Recently, Loup expanded Nat\'ario's work to encompass a changing global velocity that would accelerate from rest to a multiple of $c$~\cite{Loup,Loup2}.
    Finally, recent research computed the complete Einstein tensor $G_{\mu\nu}$ for the Alcubierre warp drive and derived a constraint on the trace of the energy momentum tensor that satisfied the weak, strong, null and dominant energy conditions from a dust matter distribution as its source~\cite{Santos-Pereira:2020puq}.

    While much progress has been made developing the physics of a warp drive spacetime, visualizing the geometry of the spacetime is underdeveloped. 
    The outside of the warp bubble is causally disconnected from the interior~\cite{Alcubierre:1994tu,Natario:2001tk,Lobo2,Lobo3}.
    As a consequence, computer simulations of the spacetime surrounding the ship need to be developed to plot the flight and steer the warp bubble.
    To date, the only method to plot the surrounding spacetime is to compute the York time to map the surrounding volume expansion \cite{Alcubierre:1994tu,Lobo2}. 
    While the York time is appropriate when the 3-geometry 
    of the hypersurfaces is flat, it will not contain all information about the surrounding spacetime in non-flat 3-geometries. 
    Alternatively, curvature invariants allow a manifestly coordinate invariant characterization of certain geometrical properties of spacetime \cite{ZM}.
    By calculating and plotting a warp drive spacetime's curvature invariants, a more complete understanding of their geometrical properties may be obtained.

    Curvature invariants are scalar products of the Riemann, Ricci, Weyl tensors and/or their covariant derivatives.
    They are functions of the metric itself, the Riemann tensor, and its covariant derivatives, as proven by Christoffel \cite{Chris}. 
    They are important for studying several of  a spacetime's geometrical properties, such as curvature singularities, the Petrov type of the Weyl tensor and the Segre type of the trace-free Ricci tensor, and the equivalence problem \cite{ZM}.
    A spacetime requires a set of up to seventeen curvature invariants to completely describe its geometry once special non-degenerate cases are taken into account.
    The set of invariants proposed by Carminati and McLenaghan (CM) have several attractive properties, such as general independence, lowest possible degree, and a minimal independent set for any Petrov type and choice of Ricci Tensor \cite{CM}.
    For Class $B_1$ spacetimes, which include all hyperbolic spacetimes, such as the general warp drive line element, only four CM invariants, $(R,r_1, r_2,$ and $w_2)$, are necessary to form a complete set~\cite{Santosuosso:1998he}.

    Henry et al. recently studied the hidden interiors of the Kerr--Newman black hole by computing and plotting all seventeen of its curvature invariants~\cite{Henry}. 
    They exposed surprisingly complex structures inside the interior of the Kerr--Newman black hole, more so than what is normally suggested by textbook depictions using coordinate-dependent methods.
    In addition, curvature invariants have been calculated to study the event horizons of other black hole metrics \cite{13,14,16}.
    As another example, the curvature invariants were calculated to find a naked curvature singularity for a spacetime that is cylindrically symmetric, Petrov type D, and admits CTCs  \cite{Ahmed:2017pww}.
    This body of work prompted the present authors to calculate and plot the curvature invariants of several wormhole solutions and the accelerating Nat\'ario warp drive \cite{3,Mattingly:2020zzt}.
    The research in this paper continues the investigation of the curvature invariants of warp drive spacetimes by calculating and plotting them for the Alcubierre and Nat\'ario warp drives at a constant velocity.

\section{Method to Compute the Invariants}
	The calculation of the complete set of CM invariants requires a metric $g_{ij}$ and a null tetrad $(l_i,k_i,m_i,\bar{m}_i)$ as its inputs~\cite{Dinverno,Stephani:2003tm}. 
	The metric can be used to calculate the affine connection $\Gamma^i_{jk}$, the Riemann tensor $R^i_{jkl}$, the Ricci tensor $R_{ij}$, the Ricci scalar $R$, the trace free Ricci tensor $S_{ij}$ and the Weyl tensor $C_{ijkl}$. 
	The indices $\{i,j,...\}$ range from $\{0,3\}$ in $(3+1)$ dimensions.
    The Newman--Penrose (NP) curvature components specifically require  the null tetrad, the Ricci Tensor, and the Weyl Tensor. 
    The NP components are presented in~\cite{Stephani:2003tm}.
	The complete set of thirteen CM invariants are defined in \cite{CM}. 
	Only four of these invariants are required by the syzygies for Class B spacetimes: the Ricci Scalar, the first two Ricci invariants, and the real component of the Weyl Invariant J \cite{Santosuosso:1998he}. 
	In terms of the NP curvature coordinates, they are:
	\begin{align}
	R &= g_{ij} R^{ij}, \label{eq:R} \\
	\begin{split}
	r_1& = \frac{1}{4} S_i^j S_j^i \\ & = 2\Phi_{20}\Phi_{02}+2\Phi_{22}\Phi_{00}-4\Phi_{12}\Phi_{10}-4\Phi_{21}\Phi_{01}+4\Phi_{11}^2, \end{split} \label{eq:r1}
	\\
	\begin{split}
	r_2& = -\frac{1}{8} S_i^j S_k^i S_j^k \\ & = 6\Phi_{02}\Phi_{21}\Phi_{10}-6\Phi_{11}\Phi_{02}\Phi_{20}+6\Phi_{01}\Phi_{12}\Phi_{20}-6\Phi_{12}\Phi_{00}\Phi_{21} -6\Phi_{22}\Phi_{01}\Phi_{10}+6\Phi_{22}\Phi_{11}\Phi_{00}, \end{split} \label{eq:r2}
	\\
	w_2& = -\frac{1}{8} \bar{C}_{i j k l} \bar{C}^{i j m n} \bar{C}^{k l}{}_{m n} \nonumber \\ & = 6\Psi_4\Psi_0\Psi_2-6\Psi_2^3-6\Psi_1^2\Psi_4-6\Psi_3^2\Psi_0+12\Psi_2\Psi_1\Psi_3. \label{eq:w2}
	\end{align}
	The tetrad components of the traceless Ricci Tensor are $\Phi_{00}$ through $\Phi_{22}$~\cite{Stephani:2003tm}.
	The complex tetrad components $\Psi_0$ to $\Psi_5$ are the six complex coefficients of the Weyl Tensor, due to its tracelessness.

\section{Warp Drive Spacetimes}
    Alcubierre and Nat\'ario developed warp drive theory using $(3+1)$ ADM\linebreak formalism~\cite{Alcubierre:1994tu,Natario:2001tk}.
    It decomposes spacetime into space-like hyper-surfaces, parameterized by the value of an arbitrary time coordinate $dx^0$ \cite{ADM,Marqu}. 
    The proper time $d\tau=N(x^\alpha, x^0) \ dx^0$ separates two nearby hypersurfaces, $x^0$ and $x^0+dx^0$. 
    The $(3+1)$ ADM metric is 
    \begin{equation}
	g_{ij}=\begin{pmatrix}
	\ g_{0 0}&\ \ g_{0 \beta} \ \\
	\ g_{\alpha 0}&\ \ g_{\alpha\beta} \ \
	\end{pmatrix} =
	\begin{pmatrix}
	\ -N^2+N_\alpha N^\alpha &\ \ N_\beta \ \\
	\ N_\alpha&\ \ g_{\alpha\beta} \ \
	\end{pmatrix}.
	\label{11}
	\end{equation}
	
	$N$ is the lapse function between the hypersurfaces. 
    $N_\alpha$ is the shift vector for each of the hypersurface's internal 3-geometries denoted by the indices $\{\alpha,\beta,...\}$ that range from $\{1,3\}$. 
    $g_{\alpha\beta}$ is the $3D$ metric of the spatial geometry of each hypersurface.
    A spacetime that moves at an arbitrary but constant velocity corresponds to a choice  in the lapse function $N$, equaling the identity and the shift vector $N_\alpha$, being a time-dependent vector field.
	Specific choices of $N_\alpha$ may result in global velocities of the spacetime exceeding $c$. 
	This principle prompted the development of warp drive spacetimes.

	The definition of a warp drive spacetime moving at a constant velocity is a globally hyperbolic spacetime $(M, g)$, where $M=\mathbb{R}^4$ and $g$ are given by the line element
    \begin{equation}
        \text{d}s^2=-\text{d}t^2+\overset{3}{\sum_{\alpha=1}} (\text{d}x^\alpha-X^\alpha \text{d}t)^2. \label{12}
    \end{equation}
    for three unspecified bounded smooth functions $(X^\alpha)=(X,Y,Z)$ in Cartesian \linebreak coordinates~\cite{Alcubierre:1994tu,Natario:2001tk}.
    As it is a globally hyperbolic spacetime, it is classified as a $B_1$ \linebreak spacetime \cite{Santosuosso:1998he}.
    Therefore, the invariants in Equations~\eqref{eq:R}--\eqref{eq:w2} are the complete set needed to classify a warp drive spacetime.
    The shift vector is given by the equation 
    \begin{equation}
        \textbf{X}=X^\alpha\frac{\partial}{\partial x^\alpha}=X\frac{\partial}{\partial x}+Y\frac{\partial}{\partial y}+Z\frac{\partial}{\partial z}, \label{13}
    \end{equation}
    and it forms a time-dependent vector field in Euclidean 3-space~\cite{Natario:2001tk}.
    Each warp drive spacetime considered in this article corresponds to specific choices of \eqref{13}.
    The future pointing normal covector to its Cauchy surface is $n_i=-dt\Leftrightarrow n^i=\frac{\partial}{\partial t}+X^\alpha\frac{\partial}{\partial x^\alpha}=\frac{\partial}{\partial t}+\textbf{X}$.
    Any observer that travels along this covector is an Eulerian and a free-fall observer. 
    Finally, a warp bubble with a constant velocity of $v_s(t)$ along the positive $x$-axis results from a vector field with $X = 0$ in the interior and $X=-v_s(t)$ in the exterior regions. 
    The warp drive line elements considered in this paper fulfill these conditions.
    
\subsection{Alcubierre's Warp Drive with a Constant Velocity}
    The line element for the Alcubierre warp drive in parallel covariant $(3+1)$ ADM and in natural units ($G=c=1$) is \cite{Alcubierre:1994tu}
    \begin{equation}
        \text{d}s^2=-\text{d}t^2+(\text{d}x-v_s f(r_s)\text{d}t)^2+\text{d}y^2+\text{d}z^2. \label{14}
    \end{equation}
    
    This is one of the simplest choices for a warp drive, and it describes a warp bubble shaped by the function $f(r_s)$ and traveling along the x-axis of a Cartesian coordinate system at an arbitrary velocity $v_s$.
    The range of its coordinates is $(-\infty<(x,y,z)<\infty)$ with an origin at the beginning of the flight.
    The ship is assumed to move along the x-axis of a Cartesian coordinate system.
    As a consequence, the shift vector~Equation \eqref{13} will only have an x-component, i.e., $(X,Y,Z)=(-v_s(t)f(r_s(t)),0,0)$.
    The internal $3$-geometry of the hypersurfaces is flat, so $g_{\alpha\beta}=\delta_{\alpha\beta}$ in Equation~\eqref{11}.
    The velocity vector is given by $v_s(t)=\frac{dx_s(t)}{dt}$, where the subscript~``$s$'' denotes the position of the spaceship.
    $v_s(t)$ is the arbitrary speed with which the Eulerian observers inside the warp bubble move in relation to Eulerian observers outside the warp bubble. 
    The radial distance is $r_s(t)=\sqrt{(x-x_s(t))^2+y^2+z^2}$.
    This is the path an Eulerian observer takes starting inside the warp bubble and traveling to the outside of the bubble. 
    The Alcubierre warp drive continuous shape function, $f(r_s)$, defines the shape of the warp bubble. It is  
    \begin{equation}
        f(r_s)=\frac{\tanh{\sigma(r_s+\rho)}-\tanh{\sigma(r_s-\rho)}}{2\tanh{\sigma \rho}}. \label{15}
    \end{equation}
    where $\sigma$ is the skin depth of the warp bubble with units of inverse length and $\rho$ is the radius of the warp bubble.
    The parameters, $\sigma>0$ and $\rho>0$, are arbitrary apart from being positive.
    Appendix \ref{app:null} describes how to find the comoving null tetrads for both the Alcubierre and Nat\'ario line elements.
    The comoving null tetrad describes light rays traveling parallel with the warp bubble. 
    For Equation~\eqref{14}, it is 
    \begin{align} \label{16}
        l_i &=\frac{1}{\sqrt{2}}\begin{pmatrix}1+f(r_s)v_s\\-1\\0\\0\end{pmatrix}, &
        k_i&=\frac{1}{\sqrt{2}}\begin{pmatrix}1-f(r_s)v_s\\1\\0\\0\end{pmatrix},\nonumber \\ \\
        m_i&=\frac{1}{\sqrt{2}}\begin{pmatrix}0\\0\\1\\i\end{pmatrix}, & \bar{m}_i &=\frac{1}{\sqrt{2}}\begin{pmatrix}0\\0\\1\\-i \end{pmatrix}. \nonumber 
    \end{align}
    
    This choice for the null tetrad will position the origin of the invariants at the start of the flight.
    Any other appropriate choice will be related to what is derived in \linebreak Section~\ref{chp:5Nat} by polynomial functions.
    Applying Equations~\eqref{14} and \eqref{16} to the complete set of CM invariants, Equations~\eqref{eq:R}--\eqref{eq:w2}, gives the four CM invariants for the Alcubierre warp~drive.

\subsection{Nat\'ario Warp Drive at a Constant Velocity}
    Nat\'ario improved upon Alcubierre's work by constructing a warp drive spacetime such that no net expansion occurs \cite{Natario:2001tk}.
    His warp drive spacetime chose a shift vector rotated around the $x-$axis in spherical polar coordinates.
    In natural units, its line element is
    \begin{equation}
        \text{d}s^2=(1-X_{r_s}X^{r_s}-X_{\theta}X^{\theta})\text{d}t^2+2(X_{r_s}\text{d}r_s+X_{\theta}r_s \text{d}\theta)\text{d}t-\text{d}r_s^2-r_s^2\text{d}\theta^2-r_s^2\sin^2{\theta}\text{d}\phi^2. \label{17}
    \end{equation}
    
    The line element uses the standard spherical coordinates of $(0\leq r_s<\infty$;\ $0\leq\theta\leq\pi$;$\ 0\leq\varphi\leq 2\pi)$, and $(-\infty< t<\infty)$ \cite{Natario:2001tk}.
    
    The analysis in Section~\ref{chp:5Nat} relies on the following choices from Nat\'ario's original paper.
    The vector field in \eqref{13} is set to 
    \begin{equation}
        \textbf{X}\sim -v_s(t) \text{d}[n(r_s)r_s^2\sin^2{\theta}\text{d}\phi]\sim-2v_sn(r_s)\cos{\theta} \  \textbf{e}_{r_s}+v_s(2n(r_s)+r_sn'(r_s))\sin{\theta} \ \textbf{e}_{\theta}. \label{18}
    \end{equation}
    
    Nat\'ario applied exterior derivatives and the Hodge Star $\star$ product to the coordinates of Equation~\eqref{11} to transform them into spherical.
    A detailed outline of the derivation of Equation~\eqref{18} may be found in the appendices of \cite{Loup,Loup2}.

    The internal $3$-geometry is set to be flat, so $g_{\alpha\beta}=\delta_{\alpha\beta}$ in Equation~\eqref{11}.
   As in the previous section, $v_s(t)$ is the constant speed for the Eulerian observers and $n(r_s)$ is the shape function of the warp bubble. 
    The shape function is arbitrary, other than the conditions $n(r_s)=\frac{1}{2}$ for large $r$ and $n(r_s)=0$ for small $r$.
    Nat\'ario's chosen shape function is 
    \begin{equation}
        n(r_s)=\frac{1}{2}\Bigg[1-\frac{1}{2}\Big(1-\tanh[\sigma(r_s-\rho)]\Big)\Bigg], \label{19}
    \end{equation}
    where $\sigma$ is the skin depth of the bubble with units of inverse length and $\rho$ is the radius of the bubble \cite{Loup2}.
    The front of the warp bubble corresponds to $\cos{\theta}>0$, and a compression occurs there, centered at a distance of $\rho$ along the radial direction.
    The back of the warp bubble corresponds to $\cos{\theta}<0$, and an expansion occurs there, centered at a distance of $\rho$ along the perpendicular direction.
    The comoving null tetrad describes geodesics traveling parallel to the warp bubble.
    It is derived in 
    It is
    \begin{align} \label{20}
        l_i&=\frac{1}{\sqrt{2}}\begin{pmatrix}1+X_{r_s}\\-1\\0\\0\end{pmatrix}, &
        k_i&=\frac{1}{\sqrt{2}}\begin{pmatrix}1-X_{r_s}\\1\\0\\0\end{pmatrix}, \nonumber \\ \\
        m_i&=\frac{1}{\sqrt{2}}\begin{pmatrix}X_\theta\\0\\-r\\i r \sin{\theta}\end{pmatrix}, &
        \bar{m}_i&=\frac{1}{\sqrt{2}}\begin{pmatrix}X_\theta\\0\\-r\\-i r \sin{\theta}\end{pmatrix}. \nonumber
    \end{align}
    
    It is emphasized at this moment that this choice for the null tetrad will center the invariants on the harbor, which is defined as the flat portion of spacetime inside of the warp bubble, as it travels.
    Any other appropriate choice will be related to what is derived in Section \ref{chp:5Nat} by polynomial functions.
    Plugging Equations~\eqref{17} and \eqref{20} into the complete set of CM invariants, Equations~\eqref{eq:R}--\eqref{eq:w2}, the four CM invariants can be derived.

\section{Invariants for the Alcubierre Warp Drive} \label{chp:4Alc}
    The four curvature invariants in Eqs.~\eqref{eq:R} through \eqref{eq:w2} were computed and plotted in Mathematica\textsuperscript{\textregistered} for the Alcubierre line element Eq.~\eqref{14}. 
    Its invariant functions are
    \begin{align}
        R =& \frac{1}{2} \sigma ^2 \text{v}_s^2 \coth (\rho  \sigma ) \nonumber \\
            & \times \Bigg(4 \tanh \Big(\sigma \big(\rho +\sqrt{(x-t \text{v}_s)^2} \ \big)\Big) \text{sech}^2\Big(\sigma \big(\rho +\sqrt{(x-t \text{v}_s)^2} \ \big)\Big) \nonumber\\
            & \ \ \ -4 \tanh \Big(\sigma \big(\sqrt{(x-t \text{v}_s)^2}-\rho \big)\Big) \text{sech}^2\Big(\sigma  \big(\sqrt{(x-t \text{v}_s)^2}-\rho \big)\Big) \label{eq:4.8}\\
            & \ \ \ -2 \sinh (\rho  \sigma ) \cosh ^3(\rho  \sigma ) \text{sech}^4\Big(\sigma \big(\sqrt{(x-t \text{v}_s)^2}-\rho \big)\Big) \text{sech}^4\Big(\sigma \big(\rho +\sqrt{(x-t \text{v}_s)^2} \ \big)\Big) \nonumber\\
            & \ \ \ \times \bigg(\cosh \Big(2 \sigma \big(\sqrt{(x-t \text{v}_s)^2}-\rho \big)\Big) +\cosh \Big(2 \sigma \big(\rho +\sqrt{(x-t \text{v}_s)^2} \ \big)\Big) -2 \cosh \Big(4 \sigma \sqrt{(x-t \text{v}_s)^2} \ \Big)+4\bigg) \Bigg), \nonumber \\
        r_1 =& \frac{1}{16} \sigma^4 \text{v}_s^4 \Bigg(\cosh ^4(\rho  \sigma ) \text{sech}^4\Big(\sigma  \big(\sqrt{(x-t \text{v}_s)^2}-\rho \big)\Big) \text{sech}^4\Big(\sigma \big(\rho +\sqrt{(x-t \text{v}_s)^2} \ \big)\Big) \nonumber\\
            & \times \bigg(\cosh \Big(2 \sigma \big(\sqrt{(x-t \text{v}_s)^2}-\rho \big)\Big)+\cosh \Big(2 \sigma \big(\rho +\sqrt{(x-t \text{v}_s)^2} \ \big)\Big) -2 \cosh \Big(4 \sigma  \sqrt{(x-t \text{v}_s)^2} \ \Big)+4 \bigg)\nonumber\\
            & +2 \coth (\rho  \sigma ) \bigg(\tanh \Big(\sigma  \big(\sqrt{(x-t \text{v}_s)^2}-\rho \big)\Big) \text{sech}^2\Big(\sigma \big(\sqrt{(x-t \text{v}_s)^2}-\rho \big)\Big) \label{eq:4.9} \\
            & \ \ \ \ \ \ \ \ \ \ \ \ \ \ \ \ \ \ \ \ \ \ \ \ \ -\tanh \Big(\sigma \big(\rho +\sqrt{(x-t \text{v}_s)^2} \ \big)\Big) \text{sech}^2\Big(\sigma \big(\rho +\sqrt{(x-t \text{v}_s)^2} \ \big)\Big)\bigg)\Bigg)^2, \nonumber \\
        r_2 =& \ 0, \label{eq:4.10} \\
        w_2 =& -\frac{1}{288} \sigma ^6 \text{v}_s^6 \nonumber \\
            & \times \Bigg( 2 \coth (\rho  \sigma ) \bigg(\tanh \Big(\sigma  \big(\rho +\sqrt{(x-t \text{v}_s)^2} \ \big)\Big) \text{sech}^2\Big(\sigma  \big(\rho +\sqrt{(x-t \text{v}_s)^2}\ \big)\Big) \nonumber \\
            & \ \ \ \ \ \ \ \ \ \ \ \ \ \ \ \ \ \ \ \ \ \ \ \ \ -\tanh \Big(\sigma \big(\sqrt{(x-t \text{v}_s)^2}-\rho \big)\Big) \text{sech}^2\Big(\sigma  \big(\sqrt{(x-t \text{v}_s)^2}-\rho \big)\Big)\bigg) \label{eq:4.11}\\
            & \ \ \ \ -\cosh ^4(\rho  \sigma ) \text{sech}^4\Big(\sigma  \big(\sqrt{(x-t \text{v}_s)^2}-\rho \big)\Big) \text{sech}^4\Big(\sigma  \big(\rho +\sqrt{(x-t \text{v}_s)^2} \ \big)\Big) \nonumber \\
            & \ \ \ \  \times \bigg(\cosh \Big(2 \sigma  \big(\sqrt{(x-t \text{v}_s)^2}-\rho \big)\Big)+\cosh \Big(2 \sigma  \big(\rho +\sqrt{(x-t \text{v}_s)^2} \ \big)\Big) -2 \cosh \big(4 \sigma  \sqrt{(x-t \text{v}_s)^2}\ \big)+4\bigg) \Bigg)^3. \nonumber
    \end{align}
    
    While Equations~\eqref{eq:4.8}--\eqref{eq:4.11} are very complicated functions, several features are apparent from inspecting them directly.
    First, $r_2$ is zero.
    It will not be plotted in Appendix \ref{app:Alc} as its plots will all be of a single smooth disc and reveal no curvature.
    Next, each non-zero invariant depends only on the tetrad elements $t$ and $x$.
    The axes of all plots are chosen to be these tetrad components.
    In addition, each of the non-zero invariants is proportional to both the skin depth $\sigma^n$ and the velocity, $\text{v}_s^n$.
    It should be expected that the magnitude of the invariants will then increase as these two parameters increase, with $w_2$ increasing the most. 
    Finally, each invariant does not have any curvature singularities inside the spacetime manifold.
    In the next subsections, the non-zero CM invariants will be analyzed to see any individual effects of the parameters $v_s$, $\rho$ and $\sigma$.

    The invariant plots for the Alcubierre warp drive may be found in the following subsections and in Appendix \ref{app:Alc}.
    They were plotted in Mathematica\textsuperscript{\textregistered} using the \emph{Plot3D} function.
    The $x-$coordinate, $t-$coordinate, and the magnitude of the invariants are along each axis.
    Since natural units were selected, the plots were normalized, such that $c=1$ and a slope of $1$ in the $x$ vs. $t$ plane corresponds with the warp bubble traveling at light speed.
    The presence of spacetime curvature may be detected on the plots by locating where the invariant functions have a non-zero magnitude.
    When the invariant functions have a positive magnitude, the spacetime has a positive curvature and vice versa. 
    
    The plots show a small range over the possible values of the parameters to demonstrate many of the basic features of each invariant.
    First, the shape of each invariant resembles the ``top hat'' function along each time slice \cite{Alcubierre:1994tu}.
    However, there are some minor variations between each invariant.
    The Ricci scalar $R$ oscillates from a trough, to a peak, to a flat area, to a peak and back to a trough.
    The $r_1$ invariant simply has a peak with a flat area followed by another peak.
    The $w_2$ follows the reverse pattern, such as the $R$ wavering from a peak, into a trough, into a flat area, into a trough and returning back to a peak. 
    In each invariant, a ship could safely surf along the harbor. 
    The central harbor disappears in Figure~\ref{fig:4.2}d--f because the plots lack precision.
    By plotting more points and consequently taking longer computational time, the central features will be recovered.
    The harbor's width is less than the precision computed in these later plots.
    Inspecting the functions, the harbor remains, and choosing smaller time intervals allows it to reappear in the plots. 
    
\subsection{Invariant Plots of the Velocity for the Alcubierre Warp Drive} \label{chp4.1:vel}

    The plots for variable velocities are Figure~\ref{fig:4.2} in this subsection and Figures \ref{fig:4.3} and \ref{fig:4.4} in Appendix \ref{app:Alc}.
    Varying the velocity has several effects.
    First, it linearly increases the slope of the peaks and troughs of the warp bubble in the $x$ vs. $t$ plane.
    This increase corresponds with a constant Newtonian velocity for the warp bubble, and it may be observed that the warp bubble's velocity acts exactly like $v=x/t$ for an Eulerian observer.
    This observation is a consistency check that the program has encoded the invariants correctly.
    Second, the velocity causes the magnitude of the invariants to decrease exponentially.
    This observation is in contrast to what was predicted by inspecting the leading terms of the invariants.
    It can be concluded that the additional terms overpower the leading term.
    Next, the shape of the warp bubble remains constant throughout the flight and the only effect of time is to increase the slope in the $x$ vs. $t$ plane.
    Finally, the plots have no discontinuities, agreeing with the previous statement that no singularities exist in the invariant functions.
    The invariant plots reveal nothing that will affect a spaceship inside the harbor.
    \begin{figure}[hb]
		\begin{subfigure}{.48\linewidth}
    		\includegraphics[scale=0.25]{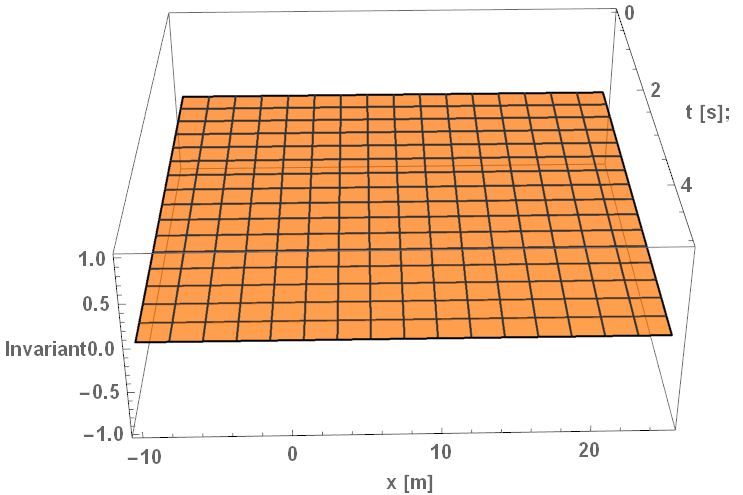}
    		\caption{Plot of Alcubierre $R$ with $v_s=0 \ c$}
    		\label{ARs8r1v0}
    	\end{subfigure}
    	~
    	\begin{subfigure}{.48\linewidth}
    		\includegraphics[scale=0.28]{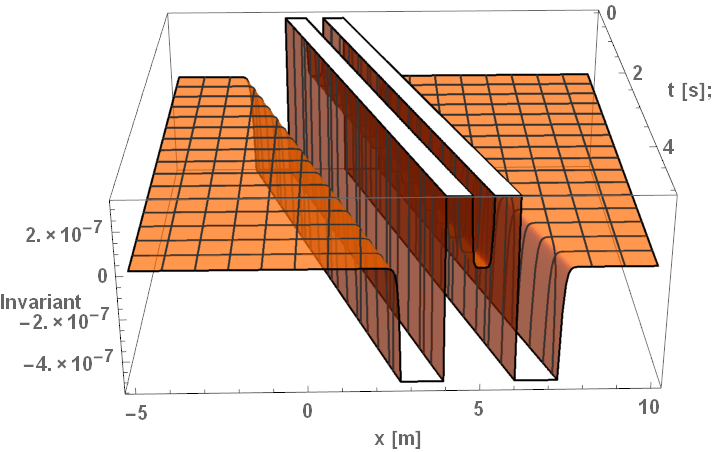}
    		\caption{Plot of Alcubierre $R$ with $v_s=1 \ c$}
    		\label{ARs8r1v1 a}
    	\end{subfigure}
    	\par \bigskip
    	\begin{subfigure}{.48\linewidth}
    		\includegraphics[scale=0.28]{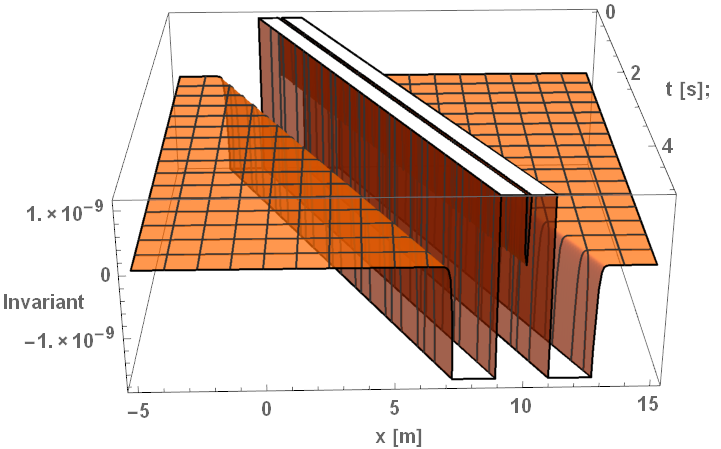}
    		\caption{Plot of Alcubierre $R$ with $v_s=2 \ c$}
    		\label{ARs8r1v2}
    	\end{subfigure}
    	~
    	\begin{subfigure}{.48\linewidth}
    		\includegraphics[scale=0.25]{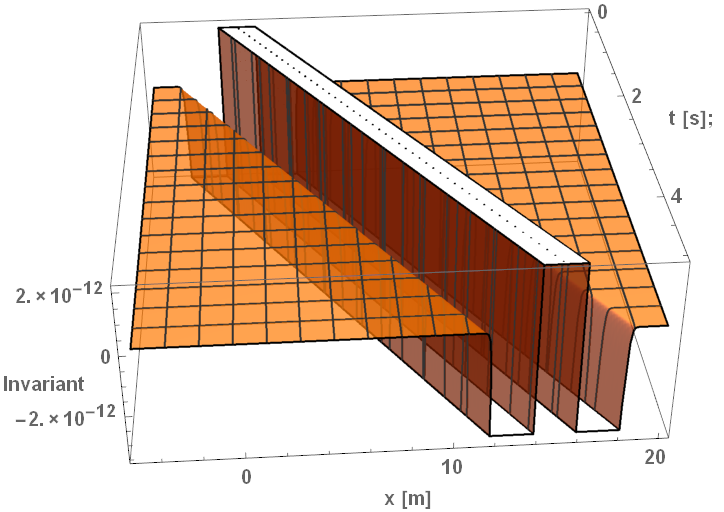}
    		\caption{Plot of Alcubierre $R$ with $v_s=3 \ c$}
    		\label{ARs8r1v3}
    	\end{subfigure}
    	\par \bigskip
    	\begin{subfigure}{.48\linewidth}
    		\includegraphics[scale=0.27]{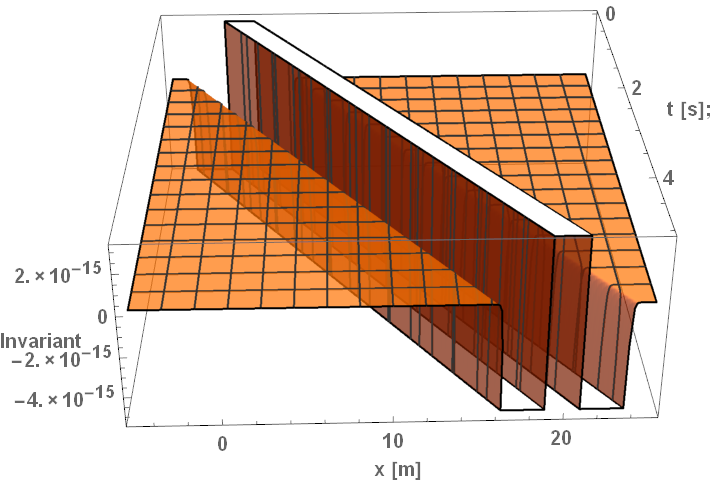}
    		\caption{Plot of Alcubierre $R$ with $v_s=4 \ c$}
    		\label{ARs8r1v4}
    	\end{subfigure}
    	~
    	\begin{subfigure}{.48\linewidth}
    		\includegraphics[scale=0.27]{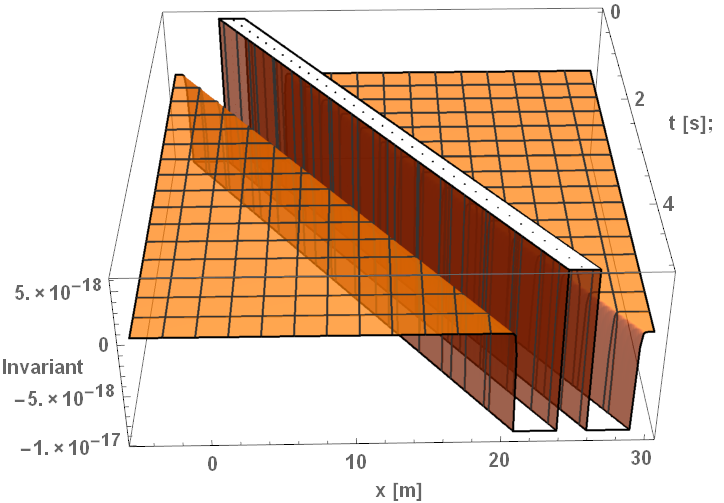}
    		\caption{Plot of Alcubierre $R$ with $v_s=5 \ c$}
    		\label{ARs8r1v5}
    	\end{subfigure}
    	\vspace{6pt}
	\caption{Plots of the $R$ invariants for the Alcubierre warp drive while varying its velocity.
	$\sigma$ = 8 m$^{-1}$ and $\rho=1$ m as Alcubierre originally suggested in his paper \cite{Alcubierre:1994tu}.
	Equation \eqref{14} is in natural units, so the speed of light was normalized out of the equation.
	The factor of $c$ was included in these captions to stress that the plots are of multiples of the speed of light.
	Their units are m s$^{-1}$.}\label{fig:4.2}
    \end{figure}
    
    The plots for variable velocities are Figs.~\ref{fig:4.2} in this subsection and \ref{fig:4.3} and \ref{fig:4.4} in Appendix \ref{app:Alc}.
    Varying the velocity has several effects.
    First, it linearly increases the slope of the peaks and troughs of the warp bubble in the $x$ vs. $t$ plane.
    This increase corresponds with a constant Newtonian velocity for the warp bubble.
    The warp bubble's velocity acts exactly like $v=d/t$ and is a good check that the program has been encoded correctly.
    The warp bubble will cover an increasing amount of distance over time as observed by an Eulerian observer.
    Second, the velocity causes the magnitude of the invariants to decrease exponentially.
    This observation is in contrast to what was predicted by inspecting the leading terms of the invariants.
    It can be concluded that the additional terms overpower the leading term.
    Next, the shape of the warp bubble remains constant throughout the flight and the only effect of time is to increase the slope in the $x$ vs. $t$ plane.
    Finally, the plots have no discontinuities, agreeing with the previous statement that no singularities exist in the invariant functions.
    The invariant plots reveal nothing that will affect a spaceship inside the harbor.
    \FloatBarrier

\subsection{Invariant plots of skin depth for the Alcubierre Warp Drive}
    The plots for varying skin depth are included in Figure~\ref{fig:4.5} in this subsection and Figures~\ref{fig:4.6} and \ref{fig:4.7} in Appendix \ref{app:Alc}.
    Repeating the method of Section \ref{chp4.1:vel}, the parameter $\sigma$ has been varied between values of $1$ m$^{-1}$ and $10$ m$^{-1}$ while maintaining the other parameters at the constant values of $\rho=1$  m  and $\sigma=8$  m$^{-1}$.
    Many of the features in these plots are the same as those discussed at the beginning of Section \ref{chp4.1:vel}; thus, the skin depth variation reveals two additional features.
    First, the plots advance towards the ``top hat'' function by slowly straightening out any dips.
    This feature is most notable in the plots of $r_1$ in \linebreak Figures~\ref{fig:4.5}a and \ref{fig:4.6}a.
    Multiple ripples occur in these two plots initially, but then  gradually smooth out as $\sigma$ increases.
    These unforeseen ripples could be the source of a rich internal structure inside the warp bubble similar to what was observed in the accelerating Nat\'ario warp drive \cite{Mattingly:2020zzt}.
    Second, the relative magnitude of the Ricci scalar and $r_1$ is several orders of magnitude greater than that of $w_2$.
    This can be seen as $\sigma\rightarrow 8$  m$^{-1}$, as the Ricci scalar goes to $10^{-9}$, $r_1$ goes to $10^{-11}$, and $w_2$ goes to the order of $10^{-28}$. 
    Consequently, the trace terms of the Riemann tensor that will have the greatest effect on the curvature are the Ricci scalar and $r_1$, which are members of the Ricci invariants in Equations~\eqref{eq:R} and \eqref{eq:w2}.
    The terms of the Weyl tensor will have negligible effects, since $w_2$ is a member of the Weyl tensor in Equation~\eqref{eq:w2}.
    The main effects of varying the skin depth is to decrease the magnitude of the warp bubble's curvature exponentially.
    This can be seen in each of the invariants as the magnitude decreases from being on the order of $10^{-3}$ to $10^{-28}$.
    The exponential decrease implies that thinner values for the warp bubble's skin depth $\sigma$ would propel itself at greater velocities due to the greater amount of curvature.
       \begin{figure}[ht]
    	\begin{subfigure}{.48\linewidth}
    		\includegraphics[scale=0.26]{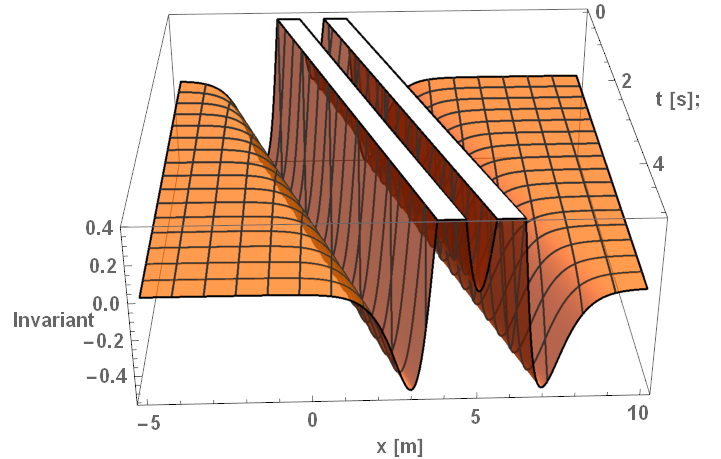}
    		\caption{Plot of Alcubierre $R$ with $\sigma=1$ m$^{-1}$}
    		\label{ARs1r1v1 b}
    	\end{subfigure}
    	~
    	\begin{subfigure}{.48\linewidth}
    		\includegraphics[scale=0.26]{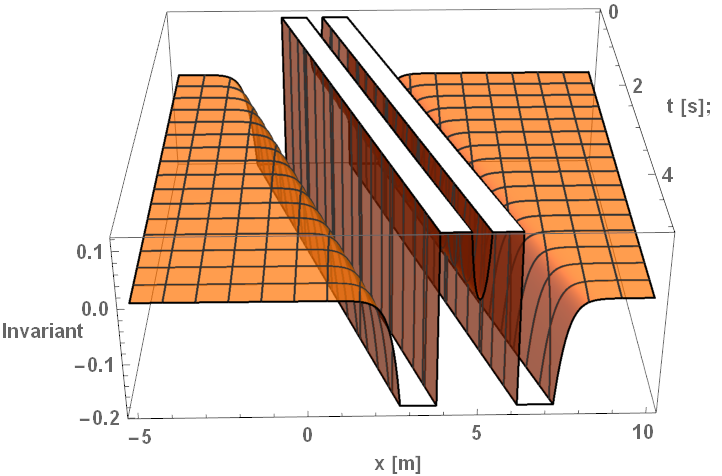}
    		\caption{Plot of Alcubierre $R$ with $\sigma=2$ m$^{-1}$}
    		\label{ARs2r1v1}
    	\end{subfigure}
    	\par \bigskip
    	\begin{subfigure}{.48\linewidth}
    		\includegraphics[scale=0.25]{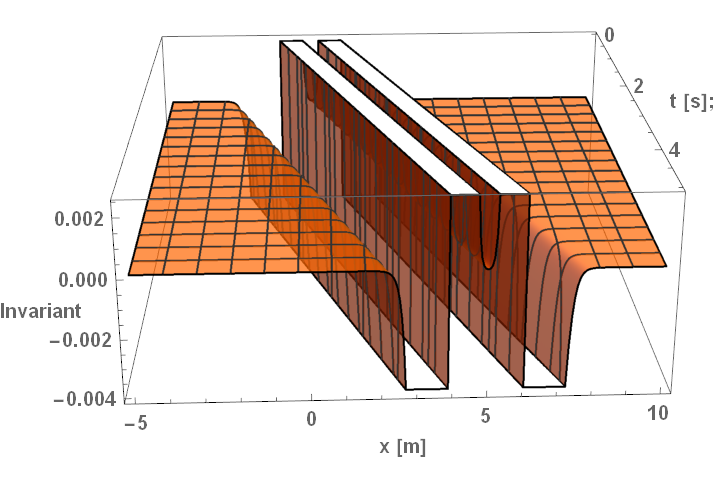}
    		\caption{Plot of Alcubierre $R$ with $\sigma=4$ m$^{-1}$}
    		\label{ARs4r1v1}
    	\end{subfigure}
    	~
    	\begin{subfigure}{.48\linewidth}
    		\includegraphics[scale=0.25]{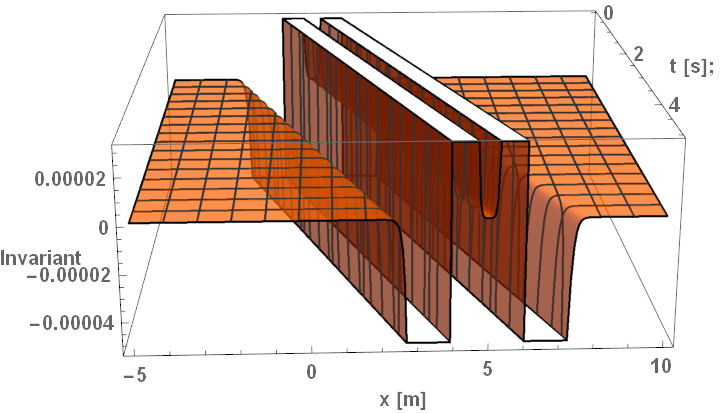}
    		\caption{Plot of Alcubierre $R$ with $\sigma=6$ m$^{-1}$}
    		\label{ARs6r1v1}
    	\end{subfigure}
    	\par \bigskip
    	\begin{subfigure}{.48\linewidth}
    		\includegraphics[scale=0.25]{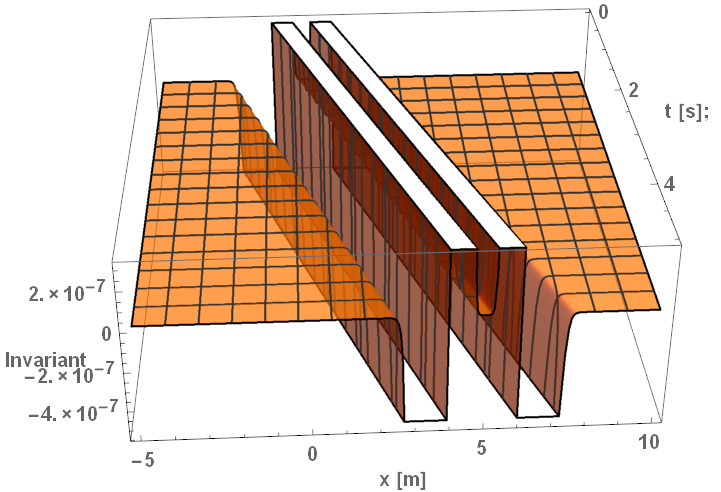}
    		\caption{Plot of Alcubierre $R$ with $\sigma=8$ m$^{-1}$}
    		\label{ARs8r1v1}
    	\end{subfigure}
    	~
    	\begin{subfigure}{.48\linewidth}
    		\includegraphics[scale=0.25]{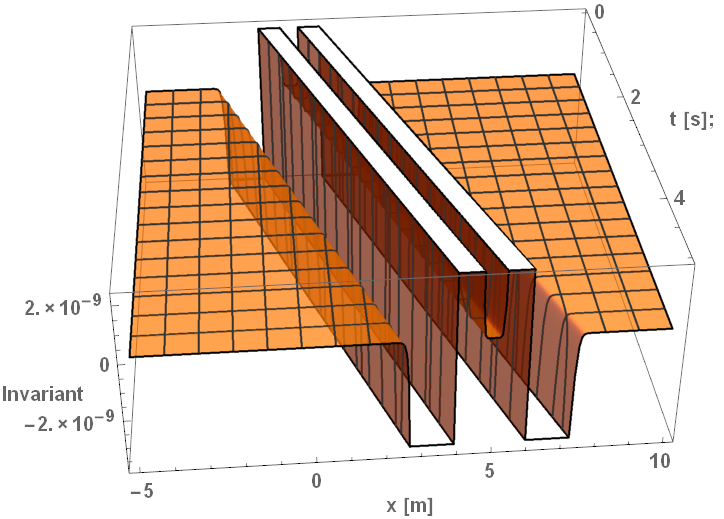}
    		\caption{Plot of Alcubierre $R$ with $\sigma=10$ m$^{-1}$}
    		\label{ARs10r1v1}
    	\end{subfigure}
    	\vspace{6pt}
	\caption{Plots of the $R$ invariant for the Alcubierre warp drive while varying skin depth.
	The parameters were chosen as $v_s=1 \ c$ and $\rho=1$  m  to match the parameters Alcubierre originally suggested in his paper \cite{Alcubierre:1994tu}.} \label{fig:4.5}
    \end{figure}
    \FloatBarrier

\subsection{Invariant plots of radius for the Alcubierre Warp Drive} \label{sec:AlcRad}
    The plots for varying the radius $\rho$ of the Alcubierre warp bubble are included in Figure~\ref{fig:4.8} in this subsection and Figures~\ref{fig:4.9}  and \ref{fig:4.10} in Appendix \ref{app:Alc}.
    Following the method in Section \ref{chp4.1:vel}, the parameter $\rho$ has been varied between values of $0.1$~m and $5$~m, while maintaining the other parameters at the constant values of $\sigma=8$  m$^{-1}$ and $v_s=1 c$.
    Many of the features in these plots are the same as those discussed at the beginning of \linebreak Section~\ref{chp4.1:vel}, but the variation of the radius does reveal an additional feature.
    The spatial size of the harbor inside the warp bubble is directly affected by the value of $\rho$.
    By inspecting the $x$-axis of each plot, the size of the harbor is of the same value as that of $\rho$.
    This behavior is as expected of the radius $\rho$, which confirms that the program is encoded correctly.
    Of greater interest, the magnitude of the invariants does not have a clear correlation with $\rho$. 
    As an example, consider the $r_1$ plots in Figure~\ref{fig:4.9}.
    When $\rho$ = 0~m, the $r_1$ invariant has its lowest magnitude of the order of $10^{-13}$.
    As the radius increases in the next four plots, the invariant increases to an order of $10^{-8}$.
    At the largest value $\rho$ = 5~m, the invariant decreases to an order of $10^{-9}$.
    Inspecting the invariant function itself in Equation~\eqref{eq:4.9}, $\rho$ does not seem to have a noticeable relationship that explains this behavior.
    In conclusion, the radius $\rho$ defines the size of the harbor and the warp bubble. 
    It must always be chosen large enough for the ship to be unaffected by the curvature of the warp bubble itself. 
  \begin{figure}[ht]
    	\begin{subfigure}{.45\linewidth}
    		\includegraphics[scale=0.25]{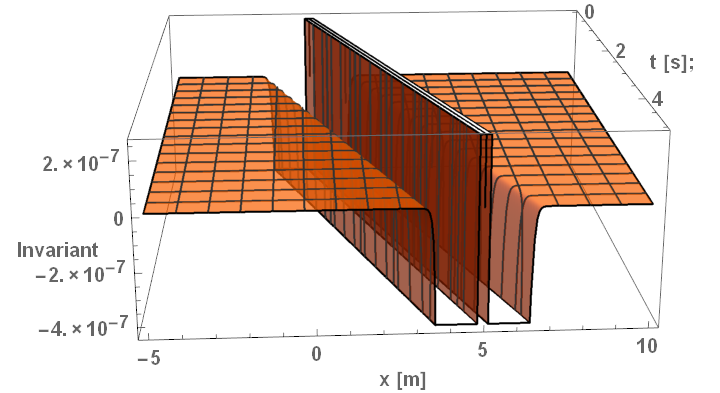}
    		\caption{Plot of Alcubierre $R$ with $\rho=0.1$  m }
    		\label{ARs8rp1v1}
    	\end{subfigure}
    	~
    	\begin{subfigure}{.45\linewidth}
    		\includegraphics[scale=0.25]{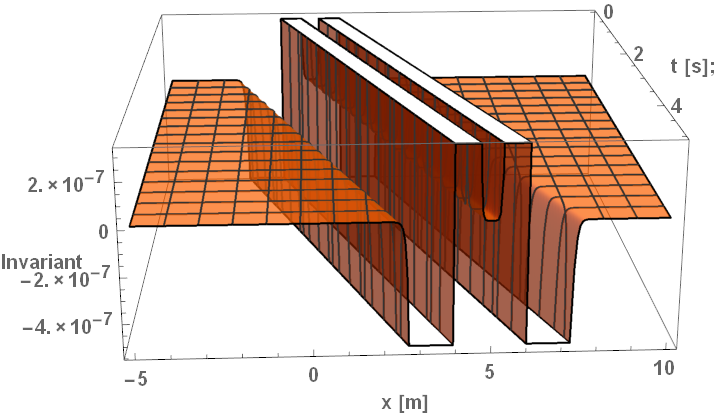}
    		\caption{Plot of Alcubierre $R$ with $\rho=1$  m }
    		\label{ARs8r10v1 c}
    	\end{subfigure}
    	\par \bigskip
    	\begin{subfigure}{.45\linewidth}
    		\includegraphics[scale=0.25]{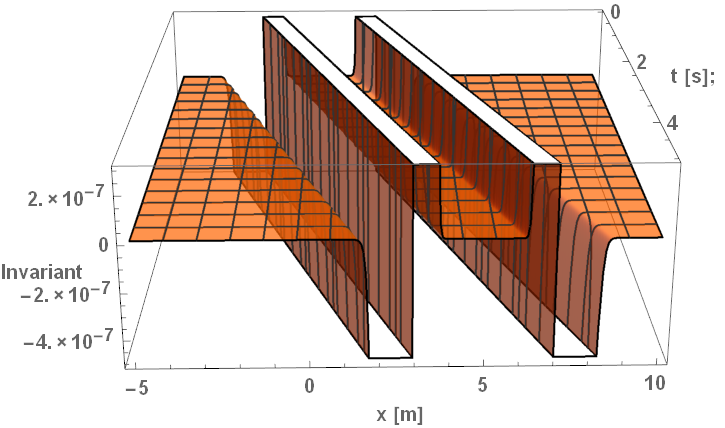}
    		\caption{Plot of Alcubierre $R$ with $\rho=2$  m }
    		\label{ARs8r2v1}
    	\end{subfigure}
    	~
    	\begin{subfigure}{.45\linewidth}
    		\includegraphics[scale=0.25]{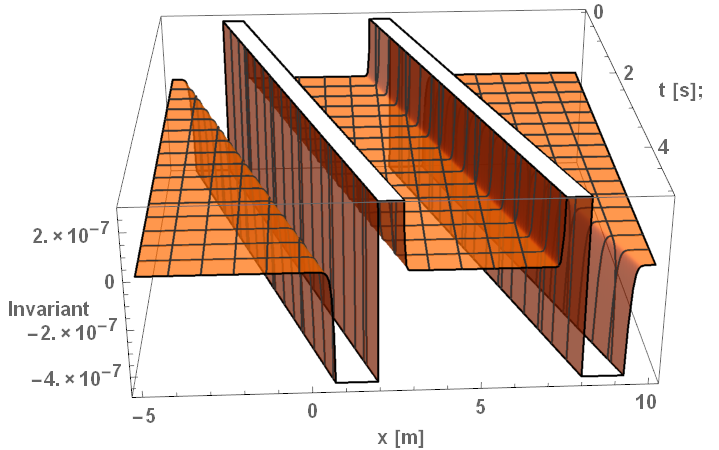}
    		\caption{Plot of Alcubierre $R$ with $\rho=3$  m }
    		\label{ARs8r3v1}
    	\end{subfigure}
    	\par \bigskip
    	\begin{subfigure}{.45\linewidth}
    		\includegraphics[scale=0.25]{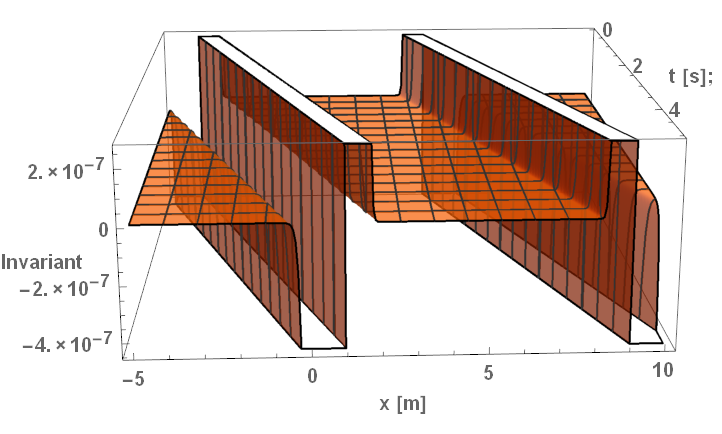}
    		\caption{Plot of Alcubierre $R$ with $\rho=4$  m }
    		\label{ARs8r4v1}
    	\end{subfigure}
    	~
    	\begin{subfigure}{.45\linewidth}
    		\includegraphics[scale=0.25]{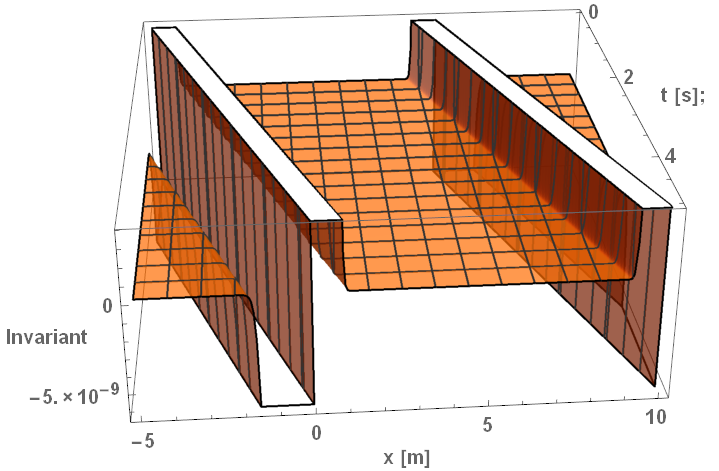}
    		\caption{Plot of Alcubierre $R$ with $\rho=5$  m }
    		\label{ARs8r5v1}
    	\end{subfigure}
    	\vspace{6pt}
	\caption{Plots of the $R$ invariant for the Alcubierre warp drive while varying radius. The other parameters were chosen as $\sigma=8$ m$^{-1}$ and $v_s=1 \ c$ to match the parameters Alcubierre originally suggested in his paper \cite{Alcubierre:1994tu}.} \label{fig:4.8}
    \end{figure}
    \FloatBarrier 
    
\section{Invariants for the Nat\'ario Warp Drive at a constant velocity} \label{chp:5Nat}
    The four curvature invariants in Equations~\eqref{eq:R}--\eqref{eq:w2} were computed and plotted in Mathematica\textsuperscript{\textregistered} for the Nat\'ario line element Equation~\eqref{17}.
    Its Ricci scalar is
    \begin{align} \label{eq:4.16}
        R =& -\frac{1}{8} \sigma ^2 v_s^2 \text{sech}^4\big(\sigma  (r-\rho )\big) \\
        & \times \Big(\cos (2 \theta )+r^2 \sigma ^2 \sin ^2(\theta ) \tanh ^2\big(\sigma  (r-\rho \big) -2 r \sigma  \sin ^2(\theta ) \tanh (\sigma  (r-\rho ))+2\Big). \nonumber
    \end{align}
    
    The Ricci scalar is included alone in this section as a demonstrative example due to its simplicity.
    The remaining three are significantly more complicated and are included in Appendix \ref{app:InvNat}.

   	As with the Alcubierre invariants, the Nat\'ario invariants are exceptionally complicated, but some features may be observed by inspecting their functions.
    The first significant difference between the two line elements is that the Nat\'ario invariants do not depend on time, but instead on the parameters $r$ and $\theta$.
    As a consequence, the coordinates for the plots are chosen to be $r$ and $\theta$.
    Since the warp bubble skims along the comoving null tetrad in Equation~\eqref{20}, they will show the shape of the bubble around the ship during flight.
    Similar to the Alcubierre invariants, each invariant is proportional to both $v_s^n$ and $\sigma^n$.
    The magnitude of the bubble's curvature will then increase exponentially with both velocity and skin depth.
    In addition, the Nat\'ario invariants are proportional to $\cos^n(\frac{\theta}{2})$ and $\text{sech}^n(\sigma(r-\rho))$.
    The warp bubble is shaped such that the curvature is at a maximum in front of the ship around $\frac{\theta}{2}=0$ and a minimum behind the ship around $\frac{\theta}{2}=\frac{\pi}{2}$.
    It is at a maximum for $r=\rho$ along the center of the warp bubble, since there $\text{sech}(0)=1$.
    Outside these values, the curvature should then fall off and go asymptotically to $0$.
    These features show that the Nat\'ario warp bubble also uses a ``top-hat'' function described in \cite{Alcubierre:1994tu}.
    Finally, there are no curvature singularities.
    The manifold is asymptotically flat and completely connected.
    The flight of such a warp bubble should be significantly less affected by any gravitational tidal forces, as compared to the Alcubierre metric in the previous section.
    The CM curvature invariants confirm that the Nat\'ario warp drive is a more realistic alternative to Alcubierre's.

    The invariant plots for the Nat\'ario warp drive may be found in the following subsections and in Appendix \ref{app:NatPlot}.
    They were plotted in Mathematica\textsuperscript{\textregistered} using the \emph{RevolutionPlot3D} function.
    The $r-$coordinate, $\theta-$coordinate, and the magnitude of the invariants are along each axis.
    The presence of spacetime curvature may be detected in a similar manner to the Alcubierre plots.
    When the invariant functions have a positive magnitude, the spacetime has a positive curvature and vice versa.
    For the Nat\'ario invariants, the spacetime curvature lies in the area around $r=\rho$.
    
    Despite the complexity of the invariants, the shape of the invariant plots is simple.
    It forms a very narrow and jagged ring, as in Figure~\ref{fig:4.11}.
    Precisely at the warp bubble's boundary located at $r=\rho$, the CM curvature invariants spike to non-zero magnitudes depending on the invariant.
    The Ricci scalar $R$ takes the form of a smooth disc outside the warp bubble.
    The shape of the $r_1$ invariant is that of a jagged disc at $r=\rho$.
    The disc has jagged edges in the negative direction, with sharp spikes at radial values $r=\rho$ and at polar angle values of $\theta=0$ and $\theta=\pi$.
    The shape of the $r_2$ invariant is that of a jagged disc at $r=\rho$.
    Its edges vary between positive and negative values depending on the polar angle $\theta$.
    Similarly, the shape of the $w_2$ invariant is that of a jagged disc at $r=\rho$.
    In front of the harbor ($\theta>0$), the invariant has rapidly changing negative values between $-1$ and $0$.
    Behind the harbor ($\theta<0$), the invariant has rapidly changing positive values between $0$ and $1$.
    The jagged edges of the plots must mean that the $r_1$, $r_2$ and $w_2$ invariants oscillate rapidly between values of $-1$ and $1$ along the circumference of the warp bubble.
    These oscillations must be occurring more rapidly than the program can plot.
    Outside of the warp bubble, the spacetime is more well behaved. 
    For values of $r\gg\rho$, the magnitude of each invariant is zero and the spacetime is asymptotically flat.
    For values of $r<\rho$ inside the warp bubble, the invariants' magnitude is also zero.
    As with the Alcubierre warp drive, this implies that there is a harbor unaffected by the curvature of the warp bubble.

    In contrast to the accelerating Nat\'ario warp drive, the constant velocity Nat\'ario warp drive does not feature either a wake or a constant non-zero curvature outside of the warp bubble for the Ricci Scalar \cite{Mattingly:2020zzt}.
    It can be concluded that these features are due to the acceleration of the warp drive.
    On the other hand, the invariant plots for the accelerating Nat\'ario line element contain features of the constant velocity plots. 
    Since the constant velocity plots are zero everywhere except at $r=\rho$, their impact is along the warp bubble's edge for the accelerating invariants.
    In the remainder of this section, the effect of each of the parameters: velocity $v_s$, skin depth $\sigma$, and radius $\rho$ is analyzed individually.

\subsection{Invariant Plots of the Velocity for the Nat\'ario Warp Drive}

    Figure~\ref{fig:4.11} in this subsection and Figures~\ref{fig:4.12} to \ref{fig:4.14} in Appendix \ref{app:NatPlot} plot the Nat\'ario invariants while varying the velocity.
    The plots reveal several new aspects of the invariant functions.
    First, the manifold is completely flat when $v_s= 0$  ms$^{-1}$ for each invariant, as expected.
    For the Ricci scalar, a non-zero velocity causes the invariant's magnitude to jump to a small negative value along the warp bubble's circumference at $r=\rho$.
    For $r_1$, an increase in velocity causes the magnitude of the invariant to swap from negative values to positive values as the velocity increases along the circumference of the warp bubble.
    For $r_2$, an increase in velocity causes the magnitude of the invariant to swap from positive values to negative values along the circumference of the warp bubble.
    For $w _2$, an increase in velocity changes the magnitude of the invariant between positive values to negative values as the velocity increases along the semicircle of the warp bubble behind the harbor.
    In front of the harbor, the $w_2$ invariant function remains negative regardless of the velocity along the warp bubble's circumference.

    The prediction of an exponential increase in the magnitude of the invariants due to the velocity is not consistent with the invariants' plots.
    A potential reason for this discrepancy is a dominant term inside each of the invariants that overcomes the exponential increase in the velocity.
    The dominant term in the invariant functions must either not depend on $v_s$ or the values for $\sigma$ are the dominant factor.
    The research in this paper may be extended to include either greater values of $v_s$ or lower values of the other parameters to further investigate this~discrepancy.

\subsection{Invariant Plots as a Function of the Skin Depth for the Nat\'ario Warp Drive}
    Figure~\ref{fig:4.16} plots the Nat\'ario invariants while changing the skin depth from $\sigma$ = 500, 000~m$^{-1}$ to $\sigma$ = 100, 000 m$^{-1}$.
    Notably, the shape of the invariants remains the same.
    Since $\text{sech}(r-\rho) \rightarrow 1$ as $(r-\rho) \rightarrow0$, the spike in the invariant functions match the limiting values of the $\text{sech}$ function and Equation~\eqref{19}.
    The dominant term(s) in the invariant function must then be proportional to $\text{sech}(r-\rho)$. 
    The $\sigma$ plots add further evidence that the shape of the CM invariants is a consequence of the ``top-hat'' function selected for the shape function in Equation~\eqref{19}.
    This conclusion indicates that the selection of the shape function will control the shape of the warp bubble.
    This analysis of the Nat\'ario skin depth reaches the same conclusion as discussed for the constant acceleration Nat\'ario invariants~\cite{Mattingly:2020zzt}. 
    
      \begin{figure}[ht]
    	\begin{subfigure}{.45\linewidth}
    	    
    		\includegraphics[scale=0.27]{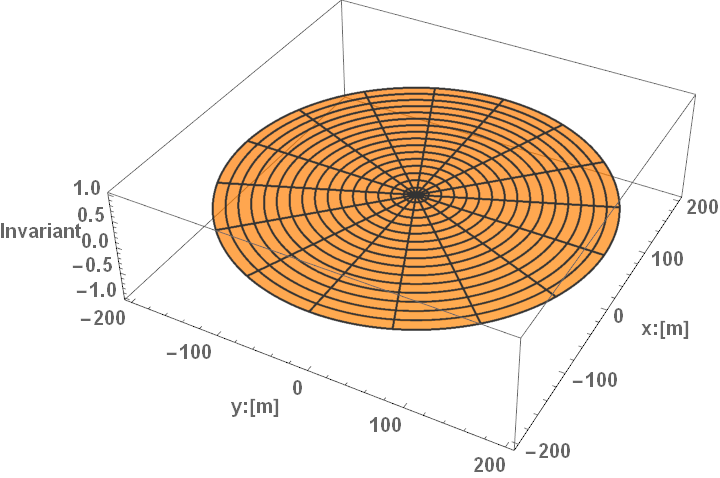}
    		\caption{Plot of Nat\'ario $R$ with $v=0.0 \ c$}
    		\label{fig:4.11a}
    	\end{subfigure}
    	~
    	\begin{subfigure}{.55\linewidth}
    	    \centering
    		\includegraphics[scale=0.28]{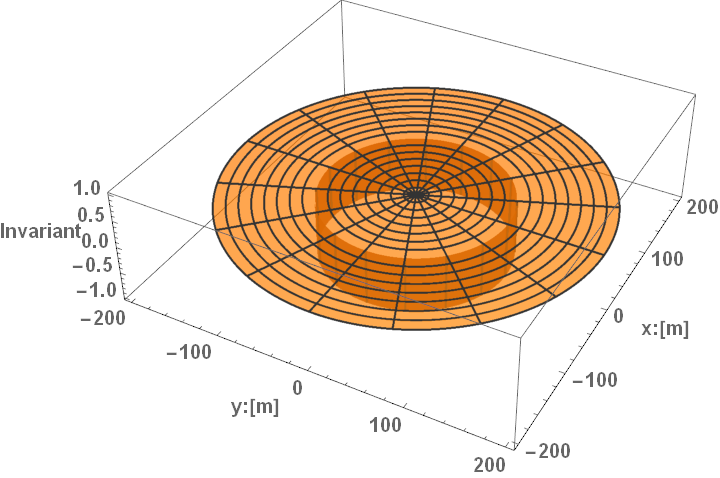}
    		\caption{Plot of Nat\'ario $R$ with $v=0.01 \ c$}
    		\label{fig:4.11b}
    	\end{subfigure}
    	\par \bigskip
    	\begin{subfigure}{.45\linewidth}
    	    \centering
    		\includegraphics[scale=0.28]{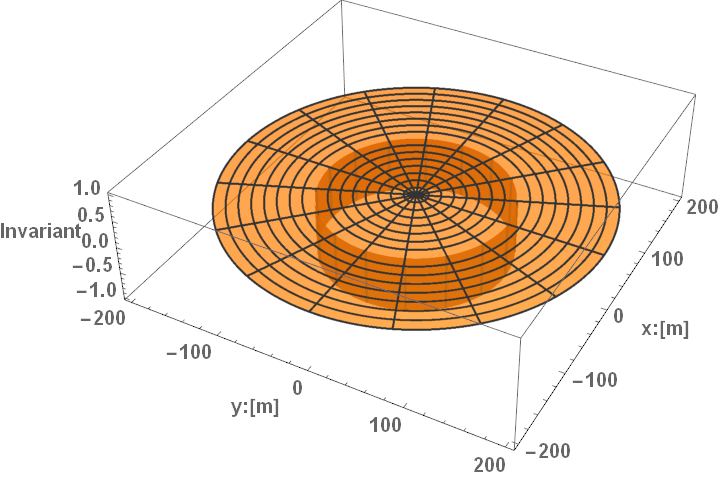}
    		\caption{Plot of Nat\'ario $R$ with $v=0.1 \ c$}
    		\label{fig:4.11c}
    	\end{subfigure}
    	~
    	\begin{subfigure}{.55\linewidth}
    	    \centering
    		\includegraphics[scale=0.28]{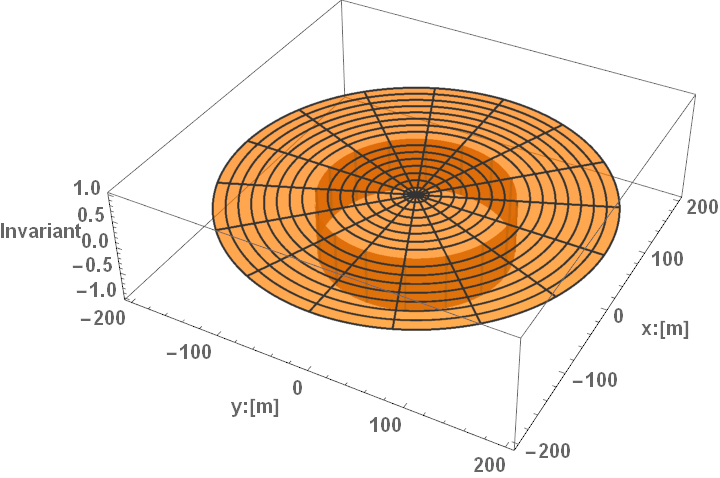}
    		\caption{Plot of Nat\'ario $R$ with $v=1 \ c$}
    		\label{fig:4.11d}
    	\end{subfigure}
    	\par \bigskip
    	\begin{subfigure}{.45\linewidth}
    	    \centering
    		\includegraphics[scale=0.28]{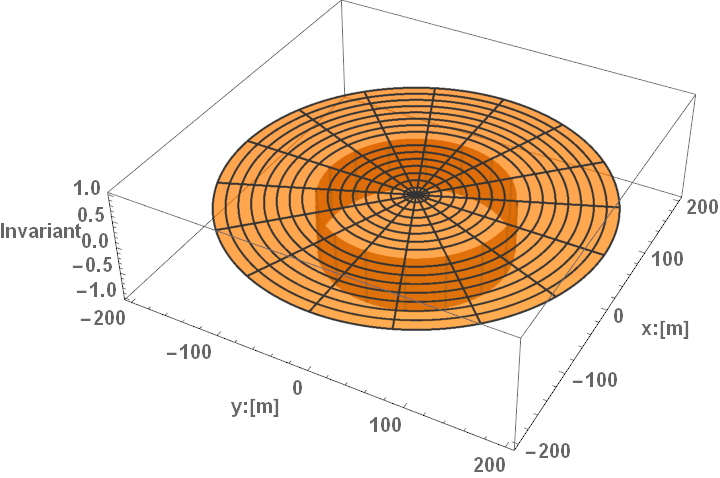}
    		\caption{Plot of Nat\'ario $R$ with $v=10 \ c$}
    		\label{fig:4.11e}
    	\end{subfigure}
    	~
    	\begin{subfigure}{.55\linewidth}
    	    \centering
    		\includegraphics[scale=0.28]{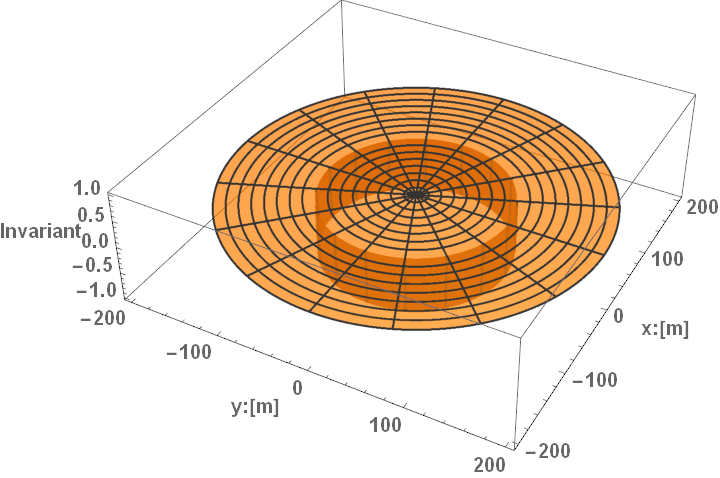}
    		\caption{Plot of Nat\'ario $R$ with $v=100 \ c$}
    		\label{fig:4.11f}
    	\end{subfigure}
    	\vspace{6pt}
	\caption{Velocity evolution of R, the Ricci scalar for the Nat\'ario warp drive at a constant velocity.
	The other parameters are set to $\sigma$ = 50, 000~$\frac{1}{\mathrm{m}}$ and $\rho$ = 100~m.
	Equation \eqref{17} is in natural units, so the speed of light was normalized out of the equation.
	The factor of $c$ was included in these captions to stress that the plots are of multiples of the speed of light.
	Their units are ms$^{-1}$.} \label{fig:4.11}
    \end{figure}
    \FloatBarrier
\subsection{Invariant Plots as a Function of the Radius for the Nat\'ario Warp Drive} \label{chp4.1:radius}
    As with the Alcubierre plots in Section \ref{sec:AlcRad}, the main effect of changing $\rho$ is that it increases the size of both the warp bubble and the Nat\'ario safe harbor.
    As $\rho$ increases from $\rho=50$~m  to $\rho=100$  m  in Figure~\ref{fig:4.18}, the sizes of the safe harbor and the bubble double.
    The spacing between the fringes in $r_1$, $r_2$, and $w_2$ is not affected by changing $\rho$, nor is the shape of the bubble.
    The plots confirm that the parameter $\rho$ moderates the size of the Nat\'ario warp bubble in the same fashion as Section \ref{sec:AlcRad} and the constant acceleration Nat\'ario warp drive~\cite{Mattingly:2020zzt}.
        
\section{Conclusion}
    This paper demonstrates how computing and plotting the curvature invariants for various parameters of the Alcubierre and Nat\'ario warp drive spacetimes can reveal geometrical features of the underlying curvature.
    The invariant functions contain no curvature singularities in either warp drive spacetime manifold.
    As a consequence of the warp drive spacetimes being able to being fully characterized by their invariants, they are $\mathcal{I}$-non-degenerate, and a horizon would be detected by a curvature singularity \cite{13,16}.
    Since no curvature singularities exist, the inside and outside of the warp drive must be connected.
    A potential coordinate transformation may be discovered that will remove the known causal discontinuity from them \cite{Lobo2,Lobo3}.

    While the individual invariant functions are complex and require significant computational time, their plots can be quickly scanned and understood.
    The plots give the magnitude of curvature at each point around the ship. 
    Where the plots' magnitudes are large, space is greatly curved and vice versa.
    For example, the curvature invariants' plots reveal a safe harbor for a ship to travel inside the warp bubble and an asymptotically flat space outside the bubble in all cases.
    Additionally, by observing the changes in slopes of the plots, the rate at which spacetime is being folded may be approximated.
    This information will help in mapping the spacetime around the ship and aid potential navigation.   

    Next, the effect of the free parameters of the warp drive spacetimes were analyzed for the curvature invariant functions and plots.
    The free parameters were varied individually to see each invariant. 
    At the radial position $r=\rho$ of the warp bubble(s), the curvature invariants have local maxima or minima identifying that the location of the warp bubble is at a radial distance of $\rho$ from the harbor.
    The sharp peaks around the radial position are due to the shape function converging to the ``top-hat'' function described by Alcubierre.
    For the Alcubierre warp drive, the warp bubbles resemble two troughs with simple internal structures.
    For the constant velocity Nat\'ario warp drive, the warp bubbles peak around $r=\rho$ while displaying rich internal structures.
    The internal structures of these warp bubbles are novel and require more diligent research to discover their effects on a warp drive's flight. 
    With this knowledge, a computer will be able to calculate the needed values of each free parameter to control the size, shape, and flight of a warp bubble.

    Of particular interest is how the invariants of the constant velocity Nat\'ario warp drive compare to the invariants of the Nat\'ario warp drive at a constant acceleration \cite{Mattingly:2020zzt}.
    The main difference between the two is the lack of a wake for each invariant and a constant non-zero curvature outside of the warp bubble for the Ricci scalar. 
    The other parameters of skin depth $\sigma$ and radius $\rho$ behave similarly for both warp drives.
    It can be concluded that the wake and constant non-zero curvature will appear as the warp drive undergoes an acceleration to reach a velocity greater than $c$.
    Potentially, the wake produced by the constant acceleration of spacetime may produce high frequency gravitational waves.
    As a physical consequence, these waves would only be detectable during the accelerating phase of a warp drive's flight, not during the constant velocity phase.
    More research is needed to explore these features and their impact on a warp drive, its flight and its surrounding spacetime.

    Computing and plotting the invariant functions has significant advantages for the inspection and potential navigation of warp drives.
    However, further investigation remains for warp drive curvature invariants.
    Those presented in this paper may be used to determine each warp drive spacetime's Petrov type of its Weyl tensor and Segre type of its Ricci tensor.
    As mentioned previously, plotting the invariants has the advantages that they are free from coordinate mapping distortions, divergences, discontinuities or other artifacts of the chosen coordinates.  
    Once the invariant plots reveal the location of any artifacts, their position can be related mathematically to the standard tensors, and their effect(s) on an object's motion can be analyzed. 
    The invariant plots properly illustrate the entire underlying spacetime, independent of a chosen coordinate system.
    A second advantage is the relative ease with which the invariants can be plotted. 
    Software packages exist or can be developed to calculate the standard tensors. 
    The aforementioned tensors lead to a chosen basis of invariants. 
    While the CM invariants were chosen in this paper, other sets of invariants exist, such as the Cartan invariants and the Witten and Petrov invariants \cite{16,8}.
    It is an open challenge to inspect the curvature of the warp drive spacetimes in these invariant sets.
    For example, the Cartan invariants may be computed for the Alcubierre and Nat\'ario warp drives.
    Then, their invariants may be compared with the invariants of other spacetimes, such as wormholes for equivalence.
    It is expected that the main features identified in this paper will also hold in these different bases.

    In addition to inspecting different invariant bases, further work can be done in mapping warp drive spacetimes, such as Alcubierre's at a constant acceleration, Krasnikov's at either constant velocity or constant acceleration, or Van Den Broeck's at either constant velocity or constant acceleration \cite{Krasnikov:1995ad,VanDenBroeck:1999sn,Loup2}.
    In addition, the lapse functions for the Krasnikov and Van Den Broeck's warp drives would need to be identified and then their accelerating line elements could be derived.
    After plotting their line elements for the invariants, the equivalence problem may be considered for all warp drive spacetimes.
    After their calculation, each proposed warp drive's curvature invariants may be compared and contrasted to their corresponding invariants at a constant velocity, as discussed in this paper.
    In this manner, any equivalent spacetime may be identified and any additional subclasses of warp drive spacetimes will be revealed.

    Another potential application of this research is in solving the geodesic equation for the warp drive spacetimes \cite{Natario:2001tk}.
    The affine connection, Riemann tensor, Ricci tensor, Weyl tensor, and Ricci scalar were all calculated in the process of finding the curvature invariants for the Alcubierre and Nat\'ario warp drives.
    Simple modifications of the Mathematica\textsuperscript{\textregistered} program would allow the geodesic structure and deviation to be found for these warp drives. 
    In many ways, the work considered in this paper is just the beginning to using curvature invariants to analyze warp drive spacetimes.
    
    \newpage

\section{Acknowledgements}
E.~W.~Davis would like to thank the Institute for Advanced Studies at Austin for supporting this work. B.~Mattingly would like to thank D.~D.~McNutt for beneficial discussions. 

\appendix
\section{Null Vectors of the Alcubierre and Nat\'ario Line Elements}\label{app:null}
    A null tetrad contains two real null vectors, $\bf{k}$ and $\bf{l}$, and two complex conjugate null vectors, $\bf{m}$ and $\bf{\bar{m}}$ that satisfy the following algebraic relationships 
    \cite{Stephani:2003tm}:
    \begin{align}
        \bf{e}_i &=(\bf{m},\bf{\bar{m}},\bf{l}, \bf{k}) \label{21}, \\
        g_{ij} &=2m_{(i}\bar{m}_{j)}-2k_{(i}l_{j)}=\begin{pmatrix}0 & 1 & 0 & 0 \\ 1 & 0 & 0 & 0 \\ 0 & 0 & 0 & -1 \\ 0 & 0 & -1 & 0 \end{pmatrix}. \label{22}
    \end{align}
    If an orthonormal tetrad, $\bf{E}_a$, exists for a given metric, it can be related to a complex null tetrad \eqref{21} by:
    \begin{align} \label{23}
        l_i &=\frac{1}{\sqrt{2}}(E_1 + E_2), & 
        k_i &=\frac{1}{\sqrt{2}}(E_1 - E_2), \nonumber \\ \\
        m_i &=\frac{1}{\sqrt{2}}(E_1 + i E_2), &
        \bar{m}_i &=\frac{1}{\sqrt{2}}(E_1 - i E_2). \nonumber
    \end{align}
    The orthonormal tetrad for the Alcubierre line element in \eqref{14} is:
    \begin{align} \label{24}
        E_1&=\begin{pmatrix}1 \ 0 \ 0 \ 0\end{pmatrix}, &
        E_2&=\begin{pmatrix}v_s f(r_s) \ -1 \ 0 \ 0\end{pmatrix}, \nonumber \\ \\
        E_3&=\begin{pmatrix}0 \ 0 \ 1 \ 0\end{pmatrix}, & 
        E_4&=\begin{pmatrix}0 \ 0 \ 0 \ 1 \end{pmatrix}. \nonumber
    \end{align}
    The Nat\'ario line element in \eqref{17} has an orthonormal tetrad:
    \begin{align} \label{25}
        E_1&=\begin{pmatrix}1 \ 0 \ 0 \ 0\end{pmatrix}, &
        E_2&=\begin{pmatrix}X_{r_s} \ -1 \ 0 \ 0\end{pmatrix}, \nonumber \\ \\
        E_3&=\begin{pmatrix}X_{\theta} \ 0 \ -r \ 0\end{pmatrix}, & 
        E_4&=\begin{pmatrix}0 \ 0 \ 0 \ r \sin{\theta}\end{pmatrix}. \nonumber 
    \end{align}
    Using Mathematica\textsuperscript{\textregistered}, it can be verified that 
    \begin{equation}
        g_{ij}=E_i\cdot E_j, \label{26}
    \end{equation}
    and by applying the equations \eqref{23} to \eqref{25} and \eqref{26} the null vectors in \eqref{16} and \eqref{20} result respectively. 
\newpage 

\section{Invariants for the Nat\'ario Line Element at Constant Velocity} \label{app:InvNat}
    The four curvature invariants in Equations~\eqref{eq:R}--\eqref{eq:w2} were computed and plotted in Mathematica\textsuperscript{\textregistered} for the Nat\'ario line element Equation~\eqref{17}.
    The Ricci scalar is included alone in the main body of text in Equation \eqref{eq:4.16}.
    The remaining three are significantly more complicated and are included in herein.
    They share the features of the Ricci Scalar identified in Section \ref{chp:5Nat}.
    \begin{align*} 
        r_1 &= \frac{1}{1024 r^2}v_s^2 \sigma ^2 \cos ^8(\frac{\theta }{2}) \text{sech}^4((r-\rho ) \sigma ) \Bigg((32 r^4 v_s^2 \sigma ^4 (\sin (\frac{\theta }{2}) \\
        &
        -\sin (\frac{3 \theta }{2}))^2 \tanh ^6((r-\rho ) \sigma ) +64 r^3 v_s^2 \sigma ^3 (r \sigma -1) (\sin (\frac{\theta }{2}) \\
        & -\sin (\frac{3 \theta }{2}))^2 \tanh ^5((r-\rho ) \sigma ) \\
        &-4 r^2 \sigma ^2 (-r^2 \sigma ^2 v_s^2+4 r \sigma  v_s^2+(r^2 \sigma ^2 -4 r \sigma -4) \cos (4 \theta ) v_s^2-12 v_s^2+16 r^2 \sigma ^2 \\
        & -8 (v_s^2+2 r^2 \sigma ^2) \cos (2 \theta )) \sec ^2(\frac{\theta }{2}) \tanh ^4((r-\rho ) \sigma )+2 r \sigma  (-4 r^2 \sigma ^2 v_s^2+48 r \sigma  v_s^2 \\
        & +(4 r^2 \sigma ^2+16 r \sigma +3) \cos (4 \theta ) v_s^2-7 v_s^2+192 r^2 \sigma ^2 \\
        & +4 (v_s^2 (8 r \sigma +1)-48 r^2 \sigma ^2) \cos (2 \theta )) \sec ^2(\frac{\theta }{2}) \tanh ^3((r-\rho ) \sigma ) \\
        & +(48 r^2 \sigma ^2 v_s^2-28 r \sigma  v_s^2+(16 r^2 \sigma ^2+12 r \sigma -3) \cos (4 \theta ) v_s^2+7 v_s^2-768 r^2 \sigma ^2 \\
        & +4 ((8 r^2 \sigma ^2+4 r \sigma -1) v_s^2+160 r^2 \sigma ^2) \cos (2 \theta )) \sec ^2(\frac{\theta }{2}) \tanh ^2((r-\rho ) \sigma ) \\
        & +2 (-7 r \sigma  v_s^2+3 (r \sigma -1) \cos (4 \theta ) v_s^2+7 v_s^2+320 r \sigma \\
        & + 4 (v_s^2 (r \sigma -1)-16 r \sigma ) \cos (2 \theta )) \sec ^2(\frac{\theta }{2}) \tanh ((r-\rho ) \sigma ) \\
        & -2 (3 \cos (2 \theta )+5) (\cos (2 \theta ) v_s^2-v_s^2+32) \sec ^2(\frac{\theta }{2})) \sec ^6(\frac{\theta }{2}) \\
        & -16 r \sigma  \text{sech}^2((r-\rho ) \sigma ) \tan ^2(\frac{\theta }{2}) (2 r^3 v_s^2 \sigma ^3 (3 \cos (2 \theta )-1) \tanh ^4((r-\rho ) \sigma ) \\
        & +r^2 v_s^2 \sigma ^2 (2 r \sigma +(10 r \sigma -9) \cos (2 \theta )+9) \tanh ^3((r-\rho ) \sigma ) \\
        & +r \sigma  ((4 r^2 \sigma ^2+5 r \sigma -9) v_s^2 \\
        & +(4 r^2 \sigma ^2-13 r \sigma +9) \cos (2 \theta ) v_s^2-32 r^2 \sigma ^2) \tanh ^2((r-\rho ) \sigma ) \\
        & +((-4 r^2 \sigma ^2-5 r \sigma +9) v_s^2-(4 r^2 \sigma ^2-13 r \sigma -7) \cos (2 \theta ) v_s^2+96 r^2 \sigma ^2) \tanh ((r-\rho ) \sigma ) \\
        & +9 v_s^2+4 r v_s^2 \sigma -32 r \sigma +7 v_s^2 \cos (2 \theta ) +4 r v_s^2 \sigma  \cos (2 \theta )) \\
        &\sec ^4(\frac{\theta }{2})+\frac{1}{4} r^2 \sigma ^2 \text{sech}^4((r-\rho ) \sigma ) ((4 r^2 \sigma ^2 v_s^2+16 r \sigma  v_s^2 -(4 r^2 \sigma ^2+16 r \sigma -3) \cos (4 \theta ) v_s^2\\
        &+109 v_s^2-64 r^2 \sigma ^2-4 (v_s^2-16 r^2 \sigma ^2) \cos (2 \theta )) \sec ^8(\frac{\theta }{2}) \\
        & -416 r^2 v_s^2 \sigma ^2 (\cos (2 \theta )-2) \tan ^2(\frac{\theta }{2}) \tanh ^2((r-\rho ) \sigma ) \sec ^4(\frac{\theta }{2}) \\
        & -64 r v_s^2 \sigma  (2 r \sigma +2 r \cos (2 \theta ) \sigma +17) \tan ^2(\frac{\theta }{2}) \tanh ((r-\rho ) \sigma ) \sec ^4(\frac{\theta }{2}) \\
        & +704 r^4 v_s^2 \sigma ^4 \tan ^4(\frac{\theta }{2}) \tanh ^4((r-\rho ) \sigma )-2816 r^3 v_s^2 \sigma ^3 \tan ^4(\frac{\theta }{2}) \tanh ^3((r-\rho ) \sigma ))\Bigg), 
        \tag{B.1} \label{eq:B.1}
    \end{align*}
    ~
    \newpage
    ~
    \begin{align*}  
        r_2 & = -\frac{1}{32768 r^3}3 v_s^4 \sigma ^3 \cos ^{12}(\frac{\theta }{2}) \text{sech}^6((r-\rho ) \sigma ) (\frac{1}{2} (\tanh ((r-\rho ) \sigma )+1) \\
        & \times (512 r^5 v_s^2 \sigma ^5 \cos ^4(\theta ) \tan ^2(\frac{\theta }{2}) \tanh ^7((r-\rho ) \sigma )+32 r^4 v_s^2 \sigma ^4 (4 r \sigma + \\
        & \times (4 r \sigma -3) \cos (2 \theta )-5) \sec ^2(\frac{\theta }{2}) (\sin (\frac{\theta }{2})-\sin (\frac{3 \theta }{2}))^2 \tanh ^6((r-\rho ) \sigma ) \\
        & +4 r^3 \sigma ^3 (2 r^2 \sigma ^2 v_s^2-10 r \sigma  v_s^2-2 (r^2 \sigma ^2-5 r \sigma -4) \cos (4 \theta ) v_s^2-r^2 \sigma ^2 \cos (6 \theta ) v_s^2 \\
        & +3 r \sigma  \cos (6 \theta ) v_s^2+4 \cos (6 \theta ) v_s^2+24 v_s^2-64 r^2 \sigma ^2 \\ 
        & +((r^2 \sigma ^2-3 r \sigma +28) v_s^2+64 r^2 \sigma ^2) \cos (2 \theta )) \sec ^4(\frac{\theta }{2}) \tanh ^5((r-\rho ) \sigma ) \\
        & +2 r^2 \sigma ^2 (-10 r^2 \sigma ^2 v_s^2+96 r \sigma  v_s^2+3 r^2 \sigma ^2 \cos (6 \theta ) v_s^2+16 r \sigma  \cos (6 \theta ) v_s^2 \\
        & +4 \cos (6 \theta ) v_s^2-20 v_s^2+816 r^2 \sigma ^2+(v_s^2 (-3 r^2 \sigma ^2+112 r \sigma +12)-832 r^2 \sigma ^2) \cos (2 \theta ) \\
        & +2 ((5 r^2 \sigma ^2+16 r \sigma +2) v_s^2+8 r^2 \sigma ^2) \cos (4 \theta )) \sec ^4(\frac{\theta }{2}) \tanh ^4((r-\rho ) \sigma ) \\
        & + r \sigma  (96 r^2 \sigma ^2 v_s^2-80 r \sigma  v_s^2+16 r^2 \sigma ^2 \cos (6 \theta ) v_s^2+16 r \sigma  \cos (6 \theta ) v_s^2-3 \cos (6 \theta ) v_s^2\\
        & +22 v_s^2 -3968 r^2 \sigma ^2+((112 r^2 \sigma ^2+48 r \sigma -13) v_s^2+3072 r^2 \sigma ^2) \cos (2 \theta )\\
        & +2 (v_s^2 (16 r^2 \sigma ^2+8 r \sigma -3) -64 r^2 \sigma ^2) \cos (4 \theta )) \sec ^4(\frac{\theta }{2}) \tanh ^3((r-\rho ) \sigma ) \\
        & +2 (-20 r^2 \sigma ^2 v_s^2+22 r \sigma  v_s^2+4 r^2 \sigma ^2 \cos (6 \theta ) v_s^2
         -3 r \sigma  \cos (6 \theta ) v_s^2-\cos (6 \theta ) v_s^2 \\
        & -2 v_s^2+2448 r^2 \sigma ^2+(v_s^2 (12 r^2 \sigma ^2-13 r \sigma +1)-448 r^2 \sigma ^2) \cos (2 \theta ) \\
        & +2 ((2 r^2 \sigma ^2-3 r \sigma +1) v_s^2+24 r^2 \sigma ^2) \cos (4 \theta )) \sec ^4(\frac{\theta }{2}) \tanh ^2((r-\rho ) \sigma ) \\
        & -(-22 r \sigma  v_s^2+3 r \sigma  \cos (6 \theta ) v_s^2+4 \cos (6 \theta ) v_s^2+8 v_s^2+3136 r \sigma +((13 r \sigma -4) v_s^2 \\
        & +1024 r \sigma ) \cos (2 \theta ) +(v_s^2 (6 r \sigma -8)-64 r \sigma ) \cos (4 \theta )) \sec ^4(\frac{\theta }{2}) \tanh ((r-\rho ) \sigma ) \\
        & +32 (\cos (4 \theta ) v_s^2-v_s^2+112 \cos (2 \theta )+144) \tan ^2(\frac{\theta }{2})) \sec ^8(\frac{\theta }{2}) \\
        & +2 r^2 \sigma ^2 \text{sech}^4((r-\rho ) \sigma ) \tan ^2(\frac{\theta }{2}) (-32 r^5 v_s^2 \sigma ^5 (\cos (2 \theta )-3) \tan ^2(\frac{\theta }{2}) \tanh ^6((r-\rho ) \sigma ) \\
        & -16 r^4 v_s^2 \sigma ^4 (-2 r \sigma +(6 r \sigma -7) \cos (2 \theta )+31) \tan ^2(\frac{\theta }{2}) \tanh ^5((r-\rho ) \sigma ) \\
        & +r^3 \sigma ^3 (-4 r^2 \sigma ^2 v_s^2-57 r \sigma  v_s^2+(4 r^2 \sigma ^2-19 r \sigma -9) \cos (4 \theta ) v_s^2+141 v_s^2+64 r^2 \sigma ^2 \\
        & +4 (v_s^2 (19 r \sigma -37)-16 r^2 \sigma ^2) \cos (2 \theta )) \sec ^4(\frac{\theta }{2}) \tanh ^4((r-\rho ) \sigma ) \\
        & -r^2 \sigma ^2 (-12 r^2 \sigma ^2 v_s^2-69 r \sigma  v_s^2+(12 r^2 \sigma ^2+17 r \sigma -15) \cos (4 \theta ) v_s^2+215 v_s^2+320 r^2 \sigma ^2 \\
        & +4 (3 v_s^2 (19 r \sigma -6)-80 r^2 \sigma ^2) \cos (2 \theta )) \sec ^4(\frac{\theta }{2}) \tanh ^3((r-\rho ) \sigma )\\
        & -r \sigma  (60 r^2 \sigma ^2 v_s^2+119 r \sigma  v_s^2 +(4 r^2 \sigma ^2-31 r \sigma +29) \cos (4 \theta ) v_s^2-157 v_s^2-512 r^2 \sigma ^2 \\
        & +8 r \sigma  ((8 r \sigma -23) v_s^2+64 r \sigma ) \cos (2 \theta )) \sec ^4(\frac{\theta }{2}) \tanh ^2((r-\rho ) \sigma ) \\
        & +(-128 r^3 \sigma ^3+12 r^3 v_s^2 \sigma ^3-416 r^2 \sigma ^2+94 r^2 v_s^2 \sigma ^2+113 r v_s^2 \sigma -98 v_s^2 \\
        & + 4 ((4 r^3 \sigma ^3+28 r^2 \sigma ^2-12 r \sigma -19) v_s^2+136 r^2 \sigma ^2) \cos (2 \theta ) \\
        & +v_s^2 (4 r^3 \sigma ^3+18 r^2 \sigma ^2-33 r \sigma -18) \cos (4 \theta )) \sec ^4(\frac{\theta }{2}) \tanh ((r-\rho ) \sigma ) \\
        & +2 (-r^2 \sigma ^2 v_s^2-22 r \sigma  v_s^2+(r^2 \sigma ^2-2 r \sigma -9) \cos (4 \theta ) v_s^2-49 v_s^2+144 r^2 \sigma ^2+224 r \sigma \\
        & +(16 r \sigma  (7 r \sigma +6)-2 v_s^2 (12 r \sigma +19)) \cos (2 \theta )) \sec ^4(\frac{\theta }{2})) \sec ^4(\frac{\theta }{2}) \\
        & +\frac{1}{8} r \sigma  \text{sech}^2((r-\rho ) \sigma ) (r^2 \sigma ^2 (72 r^2 \sigma ^2 v_s^2-664 r \sigma  v_s^2+44 r^2 \sigma ^2 \cos (6 \theta ) v_s^2\\
        & +44 r \sigma  \cos (6 \theta ) v_s^2 +69 \cos (6 \theta ) v_s^2+750 v_s^2-3904 r^2 \sigma ^2 \\
        &+((-108 r^2 \sigma ^2+532 r \sigma +139) v_s^2+5376 r^2 \sigma ^2) \cos (2 \theta ) \\
        & -2 ((4 r^2 \sigma ^2-44 r \sigma -97) v_s^2 +736 r^2 \sigma ^2) \cos (4 \theta )) \tanh ^4((r-\rho ) \sigma ) \sec ^8(\frac{\theta }{2}) \\
        & +2 r \sigma  (512 r^3 \sigma ^3-16 r^3 v_s^2 \sigma ^3+8 r^3 v_s^2 \cos (6 \theta ) \sigma ^3+2240 r^2 \sigma ^2-130 r^2 v_s^2 \sigma ^2 \\
        & +5 r^2 v_s^2 \cos (6 \theta ) \sigma ^2 +726 r v_s^2 \sigma +73 r v_s^2 \cos (6 \theta ) \sigma -186 v_s^2 \\
        & +(v_s^2 (-8 r^3 \sigma ^3+139 r^2 \sigma ^2+135 r \sigma +137) -256 r^2 \sigma ^2 (2 r \sigma +13)) \cos (2 \theta ) \\
        & +2 ((8 r^3 \sigma ^3-7 r^2 \sigma ^2+109 r \sigma +13) v_s^2+544 r^2 \sigma ^2) \cos (4 \theta ) \\
        & +23 v_s^2 \cos (6 \theta )) \tanh ^3((r-\rho ) \sigma ) \sec ^8(\frac{\theta }{2}) \\
        & +(-3968 r^3 \sigma ^3+48 r^3 v_s^2 \sigma ^3-8 r^3 v_s^2 \cos (6 \theta ) \sigma ^3 -3328 r^2 \sigma ^2+654 r^2 v_s^2 \sigma ^2 \\
        & +85 r^2 v_s^2 \cos (6 \theta ) \sigma ^2-728 r v_s^2 \sigma +100 r v_s^2 \cos (6 \theta ) \sigma +112 v_s^2 \\
        & +((8 r^3 \sigma ^3+123 r^2 \sigma ^2+540 r \sigma -82) v_s^2+3584 r^2 \sigma ^2 (r \sigma +2)) \cos (2 \theta ) \\
        & +((-48 r^3 \sigma ^3+290 r^2 \sigma ^2+88 r \sigma -32) v_s^2+384 r^2 \sigma ^2 (r \sigma -6)) \cos (4 \theta ) \\
        & +2 v_s^2 \cos (6 \theta )) \tanh ^2((r-\rho ) \sigma ) \sec ^8(\frac{\theta }{2})+2 (8 r \sigma  v_s^2+4 r \sigma  \cos (6 \theta ) v_s^2 \\
        & +\cos (6 \theta ) v_s^2+56 v_s^2 -704 r \sigma +(256 (r \sigma +7)-v_s^2 (4 r \sigma +41)) \cos (2 \theta ) \\
        & -8 (v_s^2 (r \sigma +2) -4 (14 r \sigma +3)) \cos (4 \theta )+1184) \sec ^8(\frac{\theta }{2}) \\
        & +2 (-24 r^2 \sigma ^2 v_s^2-170 r \sigma  v_s^2+4 r^2 \sigma ^2 \cos (6 \theta ) v_s^2 \\
        & +31 r \sigma  \cos (6 \theta ) v_s^2+2 \cos (6 \theta ) v_s^2+112 v_s^2+2112 r^2 \sigma ^2-192 r \sigma \\
        & -((4 r^2 \sigma ^2-129 r \sigma +82) v_s^2+256 r \sigma  (5 r \sigma +12)) \cos (2 \theta )+2 ((12 r^2 \sigma ^2+5 r \sigma -16) v_s^2 \\
        & +32 r \sigma  (3-13 r \sigma )) \cos (4 \theta )) \tanh ((r-\rho ) \sigma ) \sec ^8(\frac{\theta }{2}) \\
        & -64 r^4 \sigma ^4 (-r^2 \sigma ^2 v_s^2+8 r \sigma  v_s^2+(r^2 \sigma ^2-8 r \sigma +3) \cos (4 \theta ) v_s^2-23 v_s^2+16 r^2 \sigma ^2 \\
        & -4 (3 v_s^2+4 r^2 \sigma ^2) \cos (2 \theta )) \tan ^2(\frac{\theta }{2}) \tanh ^6((r-\rho ) \sigma ) \sec ^4(\frac{\theta }{2}) \\
        & +32 r^3 \sigma ^3 (-8 r^2 \sigma ^2 v_s^2+68 r \sigma  v_s^2+(8 r^2 \sigma ^2-20 r \sigma -13) \cos (4 \theta ) v_s^2-119 v_s^2+256 r^2 \sigma ^2 \\
        & + 4 (v_s^2 (4 r \sigma -15)-64 r^2 \sigma ^2) \cos (2 \theta )) \tan ^2(\frac{\theta }{2}) \tanh ^5((r-\rho ) \sigma ) \sec ^4(\frac{\theta }{2}) \\
        & +2048 r^6 v_s^2 \sigma ^6 \cos ^2(\theta ) \tan ^4(\frac{\theta }{2}) \tanh ^8((r-\rho ) \sigma ) \\
        & +4096 r^5 v_s^2 \sigma ^5 (r \sigma -2) \cos ^2(\theta ) \tan ^4(\frac{\theta }{2}) \tanh ^7((r-\rho ) \sigma )) \sec ^4(\frac{\theta }{2}) \\
        & +r^3 \sigma ^3 \text{sech}^6((r-\rho ) \sigma ) (-\frac{1}{16} (-(3 (2 r^2 \sigma ^2+8 r \sigma -69) v_s^2+256 r^2 \sigma ^2) \cos (2 \theta ) \\
        & +4 (3 (r^2 \sigma ^2+4 r \sigma +11) v_s^2+40 r^2 \sigma ^2) \cos (4 \theta )+3 (v_s^2 (2 r^2 \sigma ^2+8 r \sigma +3) \cos (6 \theta ) \\
        & -4 (v_s^2 (r^2 \sigma ^2+4 r \sigma -7)-8 r^2 \sigma ^2))) \sec ^{12}(\frac{\theta }{2})-\frac{1}{2} r^2 \sigma ^2 (-4 r^2 \sigma ^2 v_s^2-32 r \sigma  v_s^2 \\
        & +(4 r^2 \sigma ^2+32 r \sigma +45) \cos (4 \theta ) v_s^2-201 v_s^2+64 r^2 \sigma ^2 \\
        & +4 (69 v_s^2-16 r^2 \sigma ^2) \cos (2 \theta )) \tan ^2(\frac{\theta }{2}) \tanh ^2((r-\rho ) \sigma ) \sec ^8(\frac{\theta }{2}) \\
        & +r \sigma  (-4 r^2 \sigma ^2 v_s^2-52 r \sigma  v_s^2+(4 r^2 \sigma ^2+4 r \sigma +33) \cos (4 \theta ) v_s^2-61 v_s^2+64 r^2 \sigma ^2 \\
        & + (v_s^2 (52-48 r \sigma )-64 r^2 \sigma ^2) \cos (2 \theta )) \tan ^2(\frac{\theta }{2}) \tanh ((r-\rho ) \sigma ) \sec ^8(\frac{\theta }{2}) \\
        & -16 r^4 v_s^2 \sigma ^4 (20 \cos (2 \theta )-27) \tan ^4(\frac{\theta }{2}) \tanh ^4((r-\rho ) \sigma ) \sec ^4(\frac{\theta }{2}) \\
        & -64 r^3 v_s^2 \sigma ^3 (r \sigma +(r \sigma -4) \cos (2 \theta )+13) \tan ^4(\frac{\theta }{2}) \tanh ^3((r-\rho ) \sigma ) \sec ^4(\frac{\theta }{2}) \\
        & +192 r^6 v_s^2 \sigma ^6 \tan ^6(\frac{\theta }{2}) \tanh ^6((r-\rho ) \sigma )-1152 r^5 v_s^2 \sigma ^5 \tan ^6(\frac{\theta }{2}) \tanh ^5((r-\rho ) \sigma ))),
        \tag{B.2} \label{eq:B.2}
    \end{align*}
    ~
    \newpage
    ~
    \begin{align*} 
        w_2 &= \frac{1}{2415919104}v_s^4 \sigma ^3 \text{sech}^6((r-\rho ) \sigma ) (-32 v_s^2 \sigma ^3 (r v_s \sigma  \sin (\theta ) \text{sech}^2((r-\rho ) \sigma ) \\
        & +2 (v_s \cos (\theta )+v_s \sin (\theta )+v_s (\cos (\theta )+\sin (\theta )) \tanh ((r-\rho ) \sigma )+4))^3 \\
        & \times (\text{sech}^2((r-\rho ) \sigma ) (6 \cos ^2(\theta ) +7 \sin ^2(\theta )+4 r^2 \sigma ^2 \sin ^2(\theta ) \tanh ^2((r-\rho ) \sigma ) \\
        & -8 r \sigma  \sin ^2(\theta ) \tanh ((r-\rho ) \sigma )) \\
        & -3 (3 \cos (2 \theta )+1) \tanh ((r-\rho ) \sigma ) (\tanh ((r-\rho ) \sigma )+1))^3 \\
        & +\frac{27 i}{r^3} \text{sech}^8((r-\rho ) \sigma ) \sin ^4(\theta ) (\cosh (2 (r-\rho ) \sigma ) (4+4 i)+(4+4 i)+(1+i) v_s \cos (\theta ) \\
        & +i v_s \cos (\theta +2 i (r-\rho ) \sigma )+v_s \cos (\theta -2 i (r-\rho ) \sigma )+(1+i) v_s \sin (\theta ) \\
        & +(1+i) r v_s \sigma  \sin (\theta ) +v_s \sin (\theta +2 i (r-\rho ) \sigma )\\
        &+i v_s \sin (\theta -2 i (r-\rho ) \sigma ))^2 (-r^2 \sigma ^2 v_s^2-2 r \sigma  v_s^2+r^2 \sigma ^2 \cos (2 \theta ) v_s^2 \\
        & +2 r \sigma  \cos (2 \theta ) v_s^2-2 i \cos (2 \theta +2 i (r-\rho ) \sigma ) v_s^2+(1-i) r \sigma  \cos (2 \theta +2 i (r-\rho ) \sigma ) v_s^2 \\
        & -i \cos (2 \theta +4 i (r-\rho ) \sigma ) v_s^2+2 i \cos (2 \theta -2 i (r-\rho ) \sigma ) v_s^2\\
        & +(1+i) r \sigma  \cos (2 \theta -2 i (r-\rho ) \sigma ) v_s^2 +i \cos (2 \theta -4 i (r-\rho ) \sigma ) v_s^2-2 r \sigma  \sin (2 \theta ) v_s^2\\
        &-2 \sin (2 \theta ) v_s^2-(1+i) r \sigma  \sin (2 \theta +2 i (r-\rho ) \sigma ) v_s^2 -2 \sin (2 \theta +2 i (r-\rho ) \sigma ) v_s^2\\
        &-\sin (2 \theta +4 i (r-\rho ) \sigma ) v_s^2-(1-i) r \sigma  \sin (2 \theta -2 i (r-\rho ) \sigma ) v_s^2 \\
        & -2 \sin (2 \theta -2 i (r-\rho ) \sigma ) v_s^2-\sin (2 \theta -4 i (r-\rho ) \sigma ) v_s^2-2 r \sigma  \sinh (2 (r-\rho ) \sigma ) v_s^2 \\
        & -4 \sinh (2 (r-\rho ) \sigma ) v_s^2-2 \sinh (4 (r-\rho ) \sigma ) v_s^2-2 v_s^2-12 \cos (\theta ) v_s\\
        &-(8+4 i) \cos (\theta +2 i (r-\rho ) \sigma ) v_s-(2+2 i) \cos (\theta +4 i (r-\rho ) \sigma ) v_s \\
        & -(8-4 i) \cos (\theta -2 i (r-\rho ) \sigma ) v_s-(2-2 i) \cos (\theta -4 i (r-\rho ) \sigma ) v_s -8 r \sigma  \sin (\theta ) v_s \\
        &-12 \sin (\theta ) v_s -(8-4 i) \sin (\theta +2 i (r-\rho ) \sigma ) v_s-4 r \sigma  \sin (\theta +2 i (r-\rho ) \sigma ) v_s\\
        &-(2-2 i) \sin (\theta +4 i (r-\rho ) \sigma ) v_s -(8+4 i) \sin (\theta -2 i (r-\rho ) \sigma ) v_s\\
        &-4 r \sigma  \sin (\theta -2 i (r-\rho ) \sigma ) v_s-(2+2 i) \sin (\theta -4 i (r-\rho ) \sigma ) v_s \\
        & -2 ((r \sigma +2) v_s^2+16) \cosh (2 (r-\rho ) \sigma )\\
        &-2 (v_s^2+4) \cosh (4 (r-\rho ) \sigma )-24) (\tanh ((r-\rho ) \sigma )+1) \\
        & \times (r \sigma  \tanh ((r-\rho ) \sigma )-1) ((4 r \sigma -v_s (r^2 \sigma ^2-r \sigma +1) \cos (\theta )) \tanh ^3((r-\rho ) \sigma ) \\
        & -3 (2 r^2 \sigma ^2+v_s (r^2 \sigma ^2-r \sigma +1) \cos (\theta )-2) \tanh ^2((r-\rho ) \sigma ) \\
        & -3 (v_s (r^2 \sigma ^2-r \sigma +1) \cos (\theta )-4 r \sigma ) \tanh ((r-\rho ) \sigma )-2 r^2 \sigma ^2-r^2 v_s \sigma ^2 \cos (\theta )\\
        & -v_s \cos (\theta ) +r v_s \sigma  \cos (\theta )+\text{sech}^2((r-\rho ) \sigma ) (6 r^2 \sigma ^2+v_s (3 r^2 \sigma ^2+r \sigma -3) \cos (\theta ) \\
        & + (4 r \sigma +v_s (3 r^2 \sigma ^2+r \sigma -1) \cos (\theta )) \tanh ((r-\rho ) \sigma )+6)+2)^2\\
        & + \frac{2304}{r^2}v_s^2(\sigma  \sin ^4(\theta ) (\tanh ((r-\rho ) \sigma )+1)^2 (r \sigma  \tanh ((r-\rho ) \sigma )-1)^2\\
        & \times (r v_s \sigma  \sin (\theta ) \text{sech}^2((r-\rho ) \sigma ) \\
        & +2 (v_s \cos (\theta )+v_s \sin (\theta )+v_s (\cos (\theta )+\sin (\theta )) \tanh ((r-\rho ) \sigma )+4)) \\
        &\times (\text{sech}^2((r-\rho ) \sigma ) (6 \cos ^2(\theta ) +7 \sin ^2(\theta )+4 r^2 \sigma ^2 \sin ^2(\theta ) \tanh ^2((r-\rho ) \sigma )\\
        &-8 r \sigma  \sin ^2(\theta ) \tanh ((r-\rho ) \sigma ))\\
        & -3 (3 \cos (2 \theta )+1) \tanh ((r-\rho ) \sigma ) (\tanh ((r-\rho ) \sigma )+1))\\
        &\times (r^2 v_s^2 \sigma ^2 \sin ^2(\theta ) \text{sech}^4((r-\rho ) \sigma ) \\
        & +2 r v_s \sigma  (2 \sin (\theta ) (v_s \cos (\theta )+v_s \sin (\theta )+2)+v_s (2 \sin ^2(\theta )+\sin (2 \theta ))\\
        &\times \tanh ((r-\rho ) \sigma )) \text{sech}^2((r-\rho ) \sigma ) \\
        & +2 (-\text{sech}^2((r-\rho ) \sigma ) (\cos (\theta )+\sin (\theta ))^2 v_s^2+(\cos (\theta )+\sin (\theta ))^2 \tanh ^2((r-\rho ) \sigma ) v_s^2 \\
        & +6 \cos (\theta ) \sin (\theta ) v_s^2+3 v_s^2+8 \cos (\theta ) v_s+8 \sin (\theta ) v_s \\
        & +4 (\cos (\theta )+\sin (\theta )) (v_s \cos (\theta )+v_s \sin (\theta )+2) \tanh ((r-\rho ) \sigma ) v_s+16))) \\
        & -\frac{144 \sigma  \sin ^2(\theta )}{r^2} (r v_s \sigma  \sin (\theta ) \text{sech}^2((r-\rho ) \sigma )+2 (v_s \cos (\theta ) +v_s \sin (\theta ) \\
        & +v_s (\cos (\theta )+\sin (\theta )) \tanh ((r-\rho ) \sigma )+4))^3 (\text{sech}^2((r-\rho ) \sigma ) (6 \cos ^2(\theta )+7 \sin ^2(\theta ) \\
        & +4 r^2 \sigma ^2 \sin ^2(\theta ) \tanh ^2((r-\rho ) \sigma )-8 r \sigma  \sin ^2(\theta ) \tanh ((r-\rho ) \sigma )) \\ 
        & -3 (3 \cos (2 \theta )+1) \tanh ((r-\rho ) \sigma ) (\tanh ((r-\rho ) \sigma )+1)) \\
        & \times (r \sigma  (2 r \sigma +v_s (r \sigma +1) \cos (\theta )) \text{sech}^2((r-\rho ) \sigma ) \\
        & -2 (r^2 v_s \sigma ^2 \cos (\theta ) \tanh ^3((r-\rho ) \sigma )+r \sigma  (2 r \sigma +v_s (r \sigma -1) \cos (\theta )) \tanh ^2((r-\rho ) \sigma ) \\
        & +((v_s-r v_s \sigma ) \cos (\theta )-4 r \sigma ) \tanh ((r-\rho ) \sigma )+v_s \cos (\theta )-2)) \\
        & \times (\frac{1}{2} r^2 v_s^2 \sigma ^2 (r \sigma +1) \sin (2 \theta ) \text{sech}^4((r-\rho ) \sigma )+r \sigma  (-r^2 v_s^2 \sigma ^2 \sin (2 \theta ) \tanh ^3((r-\rho ) \sigma ) \\
        & -r v_s^2 \sigma  (r \sigma -1) \sin (2 \theta ) \tanh ^2((r-\rho ) \sigma )+2 v_s^2 \cos (\theta ) (r \sigma  \cos (\theta )+\cos (\theta ) \\
        & + 2 r \sigma  \sin (\theta )) \tanh ((r-\rho ) \sigma )+2 v_s^2 (r \sigma +1) \cos ^2(\theta )+4 v_s (r \sigma +1) \cos (\theta ) \\
        & + r \sigma  (v_s^2 \sin (2 \theta )-8)) \text{sech}^2((r-\rho ) \sigma )\\
        &-2 (r^2 v_s^2 \sigma ^2 (\cos (2 \theta )+\sin (2 \theta )+1) \tanh ^4((r-\rho ) \sigma ) \\
        & +2 r v_s \sigma  \cos (\theta ) (2 r \sigma +v_s (2 r \sigma -1) \cos (\theta )+v_s (2 r \sigma -1) \sin (\theta )) \tanh ^3((r-\rho ) \sigma ) \\
        & + (-8 r^2 \sigma ^2+4 r v_s (r \sigma -1) \cos (\theta ) \sigma +2 v_s^2 (r \sigma -1)^2 \cos ^2(\theta ) \\
        & + v_s^2 (r \sigma -1)^2 \sin (2 \theta )) \tanh ^2((r-\rho ) \sigma )+(-2 (r \sigma -2) \cos ^2(\theta ) v_s^2 \\
        & -(r \sigma -2) \sin (2 \theta ) v_s^2-4 (r \sigma -1) \cos (\theta ) v_s+16 r \sigma ) \tanh ((r-\rho ) \sigma )+2 v_s^2 \cos ^2(\theta ) \\
        & +4 v_s \cos (\theta )+v_s^2 \sin (2 \theta )+8))\\
        & +\frac{864}{r^3} \sin ^4(\theta ) (\tanh ((r-\rho ) \sigma )+1) (r \sigma  \tanh ((r-\rho ) \sigma )-1) (r v_s \sigma  \sin (\theta ) \text{sech}^2((r-\rho ) \sigma )\\
        &+2 (v_s \cos (\theta )+v_s \sin (\theta )+v_s (\cos (\theta )+\sin (\theta )) \tanh ((r-\rho ) \sigma )+4))^2 \\
        & \times (\frac{1}{2} r^2 v_s^2 \sigma ^2 (r \sigma +1) \sin (2 \theta ) \text{sech}^4((r-\rho ) \sigma )+r \sigma  (-r^2 v_s^2 \sigma ^2 \sin (2 \theta ) \tanh ^3((r-\rho ) \sigma ) \\
        & -r v_s^2 \sigma  (r \sigma -1) \sin (2 \theta ) \tanh ^2((r-\rho ) \sigma )+2 v_s^2 \cos (\theta ) (r \sigma  \cos (\theta )+\cos (\theta )+2 r \sigma  \sin (\theta )) \\
        & \times \tanh ((r-\rho ) \sigma )+2 v_s^2 (r \sigma +1) \cos ^2(\theta )+4 v_s (r \sigma +1) \cos (\theta )\\
        & +r \sigma  (v_s^2 \sin (2 \theta )-8)) \text{sech}^2((r-\rho ) \sigma ) \\
        & -2 (r^2 v_s^2 \sigma ^2 (\cos (2 \theta )+\sin (2 \theta )+1) \tanh ^4((r-\rho ) \sigma )\\
        &+2 r v_s \sigma  \cos (\theta ) (2 r \sigma +v_s (2 r \sigma -1) \cos (\theta ) +v_s (2 r \sigma -1) \sin (\theta )) \tanh ^3((r-\rho ) \sigma )\\
        &+(-8 r^2 \sigma ^2+4 r v_s (r \sigma -1) \cos (\theta ) \sigma +2 v_s^2 (r \sigma -1)^2 \cos ^2(\theta ) \\
        & +v_s^2 (r \sigma -1)^2 \sin (2 \theta )) \tanh ^2((r-\rho ) \sigma )\\
        & +(-2 (r \sigma -2) \cos ^2(\theta ) v_s^2-(r \sigma -2) \sin (2 \theta ) v_s^2 -4 (r \sigma -1) \cos (\theta ) v_s\\
        & +16 r \sigma ) \tanh ((r-\rho ) \sigma )+2 v_s^2 \cos ^2(\theta )+4 v_s \cos (\theta )+v_s^2 \sin (2 \theta )+8))^2).
        \tag{B.3} \label{eq:B.3}
    \end{align*}

\newpage
\FloatBarrier
\section{Invariant Plots for the Alcubierre Warp Drive} \label{app:Alc}

    \begin{figure}[ht]
	\begin{subfigure}{.48\linewidth}
		\includegraphics[scale=0.25]{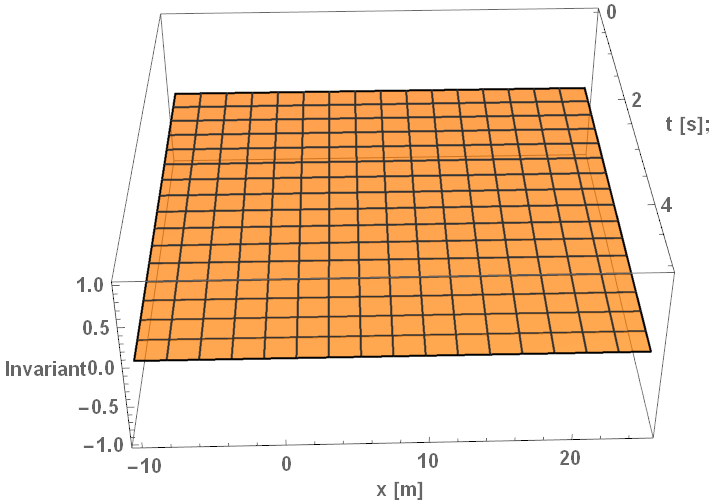}
		\caption{Plot of Alcubierre $r_1$ with and $v_s=0 \ c$}
		\label{Ar1s8r1v0}
	\end{subfigure}
	~
	\begin{subfigure}{.48\linewidth}
		\includegraphics[scale=0.23]{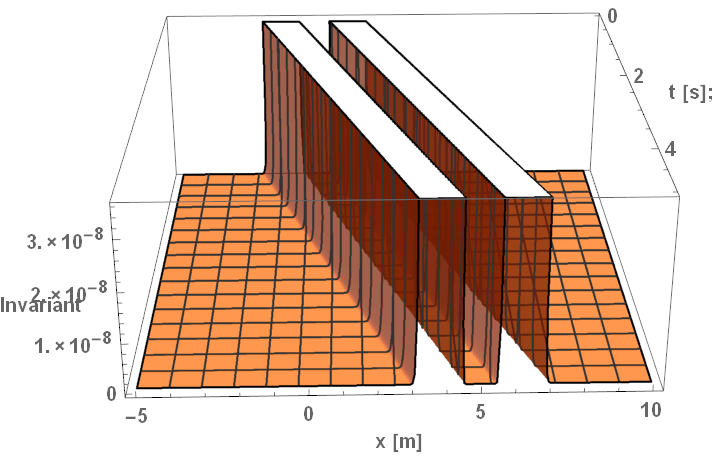}
		\caption{Plot of Alcubierre $r_1$ with $v_s=1 \ c$}
		\label{Ar1s8r1v1 a}
	\end{subfigure}
	\par \bigskip
	\begin{subfigure}{.48\linewidth}
		\includegraphics[scale=0.23]{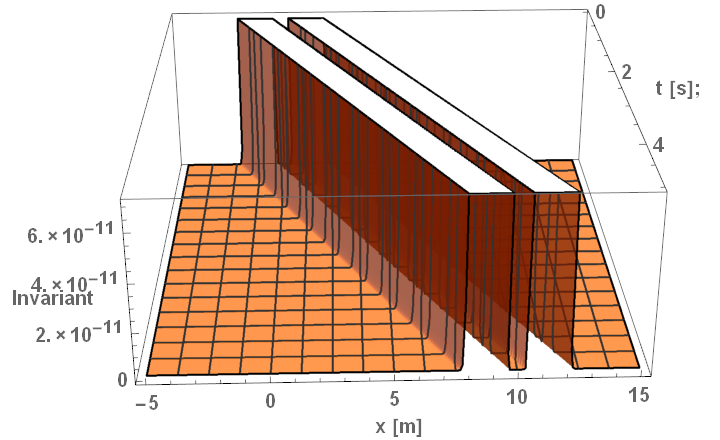}
		\caption{Plot of Alcubierre $r_1$ with $v_s=2 \ c$}
		\label{Ar1s8r1v2}
	\end{subfigure}
	~
	\begin{subfigure}{.48\linewidth}
		\includegraphics[scale=0.23]{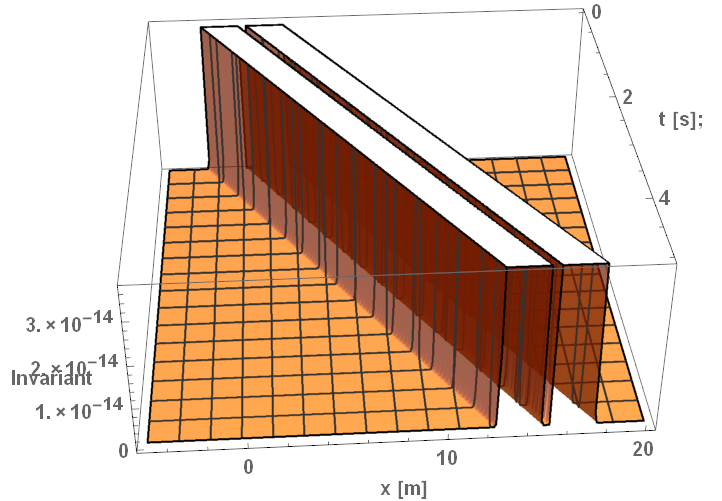}
		\caption{Plot of Alcubierre $r_1$ with $v_s=3 \ c$}
		\label{Ar1s8r1v3}
	\end{subfigure}
	\par \bigskip
	\begin{subfigure}{.48\linewidth}
		\includegraphics[scale=0.23]{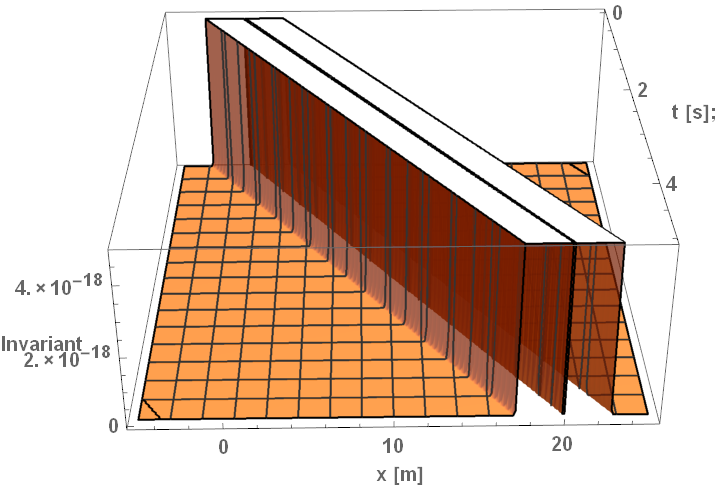}
		\caption{Plot of Alcubierre $r_1$ with $v_s=4 \ c$}
		\label{Ar1s8r1v4}
	\end{subfigure}
	~
	\begin{subfigure}{.48\linewidth}
		\includegraphics[scale=0.23]{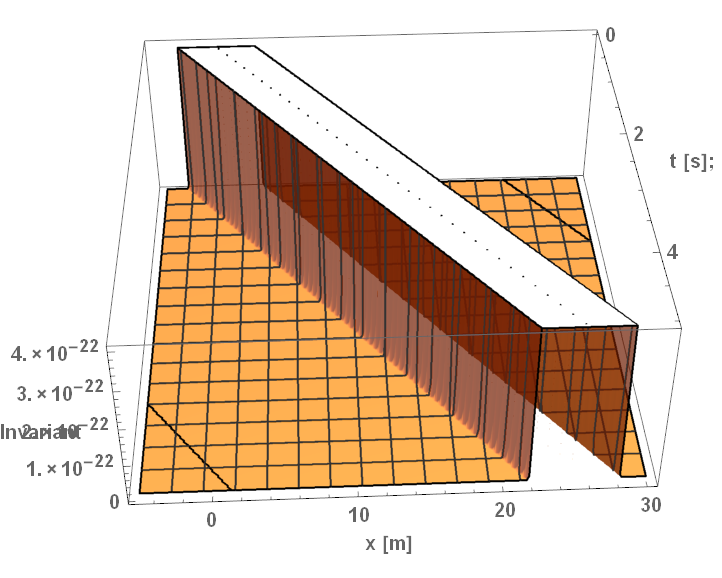}
		\caption{Plot of Alcubierre $r_1$ with $v_s=5 \ c$}
		\label{Ar1s8r1v5}
	\end{subfigure}
	\vspace{6pt}
	\caption{Plots of the $r_1$ invariants for the Alcubierre warp drive while varying velocity. 
	 The other parameters were chosen as $\sigma=8 \ m^{-1}$ and $\rho=1$  m  to match the parameters Alcubierre originally suggested in his paper \cite{Alcubierre:1994tu}.
	Equation \eqref{14} is in natural units, so the speed of light was normalized out of the equation.
	The factor of $c$ was included in these captions to stress that the plots are of multiples of the speed of light.
	Their units are ms$^{-1}$.} \label{fig:4.3}
    \end{figure}
    
    \newpage
    
    \begin{figure}[ht]
	\begin{subfigure}{.48\linewidth}
		\includegraphics[scale=0.24]{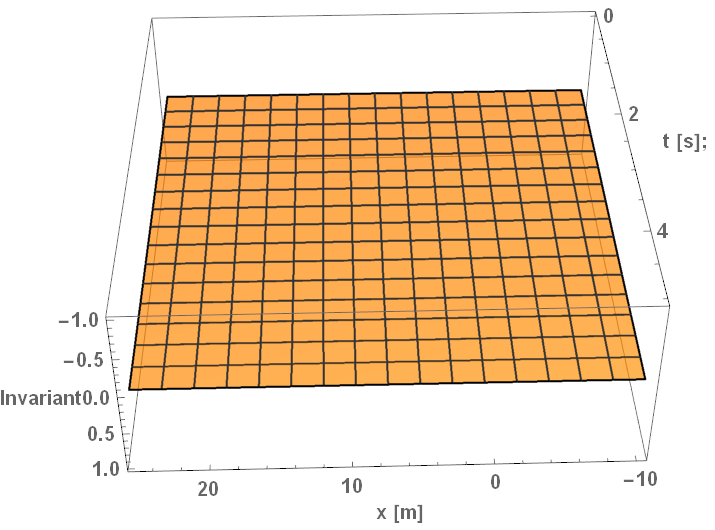}
		\caption{Plot of Alcubierre $w_2$ with $v_s=0 \ c$}
		\label{Aw2s8r1v0}
	\end{subfigure}
	~
	\begin{subfigure}{.48\linewidth}
		\includegraphics[scale=0.28]{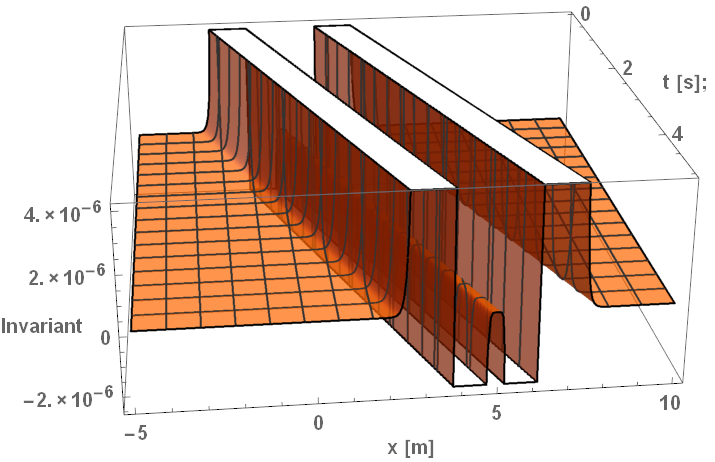}
		\caption{Plot of Alcubierre $w_2$ with $v_s=1 \ c$}
		\label{Aw2s8r1v1 a}
	\end{subfigure}
	\par \bigskip
	\begin{subfigure}{.48\linewidth}
		\includegraphics[scale=0.28]{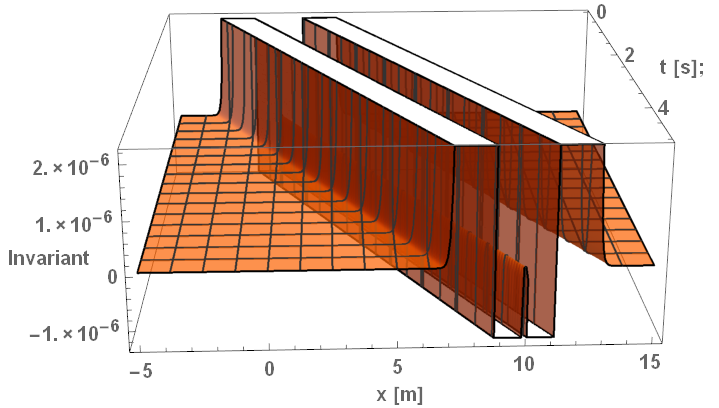}
		\caption{Plot of Alcubierre $w_2$ with $v_s=2 \ c$}
		\label{Aw2s8r1v2}
	\end{subfigure}
	~
	\begin{subfigure}{.48\linewidth}
		\includegraphics[scale=0.27]{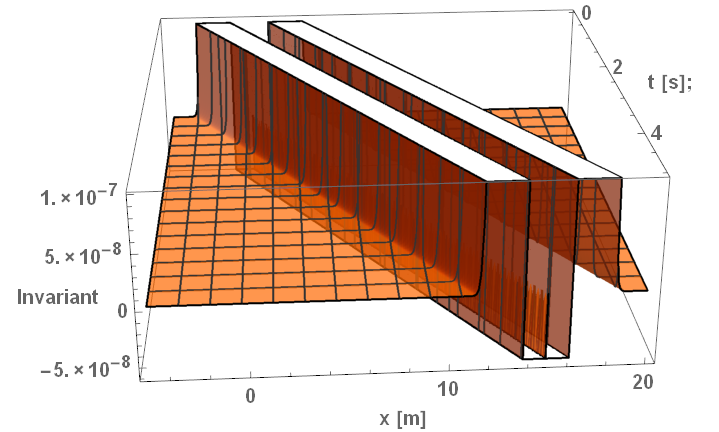}
		\caption{Plot of Alcubierre $w_2$ with $v_s=3 \ c$}
		\label{Aw2s8r1v3}
	\end{subfigure}
	\par \bigskip
	\begin{subfigure}{.48\linewidth}
		\includegraphics[scale=0.27]{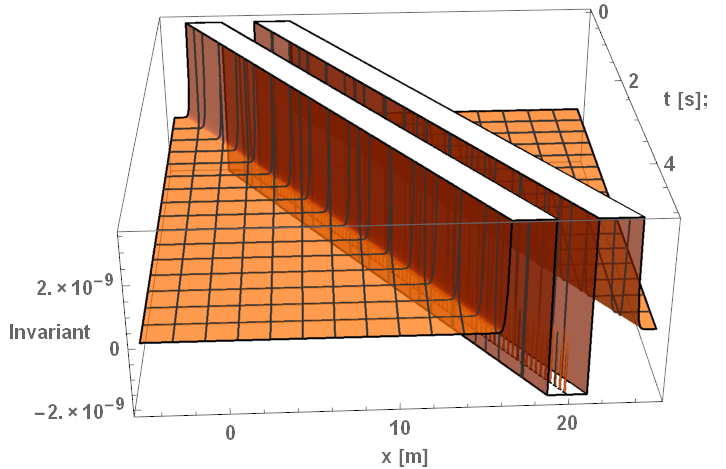}
		\caption{Plot of Alcubierre $w_2$ with $v_s=4 \ c$}
		\label{Aw2s8r1v4}
	\end{subfigure}
	~
	\begin{subfigure}{.48\linewidth}
		\includegraphics[scale=0.28]{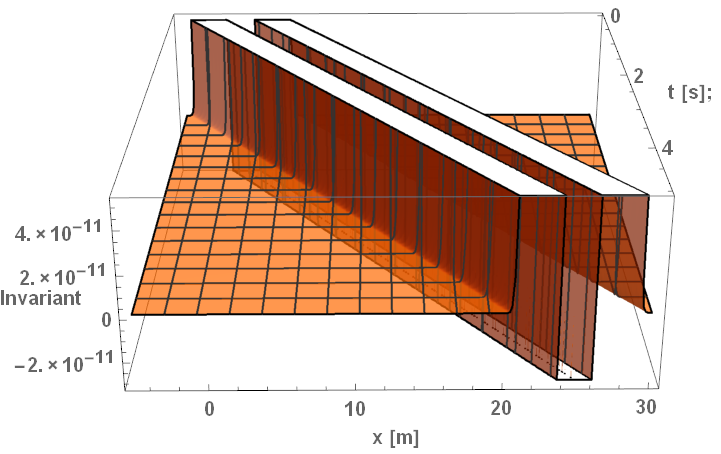}
		\caption{Plot of Alcubierre $w_2$ with $v_s=5 \ c$}
		\label{Aw2s8r1v5}
	\end{subfigure}
	\vspace{6pt}
	\caption{Plots of the $w_2$ invariants for the Alcubierre warp drive while varying velocity. 
	The radius was chosen as $\rho=1$~m to match the parameters Alcubierre originally suggested in his paper \cite{Alcubierre:1994tu}. The skin depth was chosen as $\sigma$ = 2  m$^{-1}$ to keep the plots as machine size numbers. 
	Equation \eqref{14} is in natural units, so the speed of light was normalized out of the equation.
	The factor of $c$ was included in these captions to stress that the plots are of multiples of the speed of light.
	Their units are~ms$^{-1}$} \label{fig:4.4}
    \end{figure}
    
    \FloatBarrier

    \begin{figure}[ht]
	\begin{subfigure}{.5\linewidth}
		\includegraphics[scale=0.33]{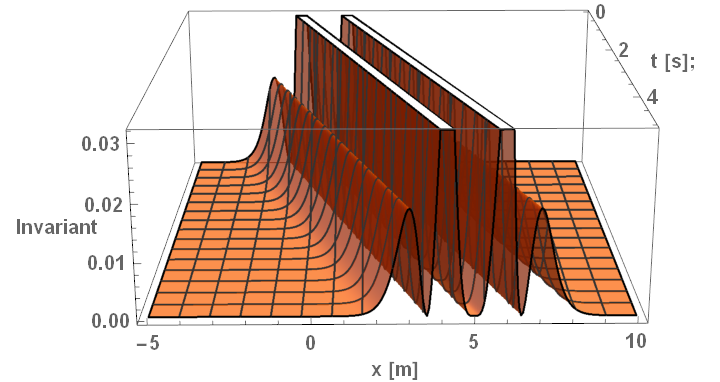}
		\caption{Plot of Alcubierre $r_1$ with $\sigma=1$ m$^{-1}$}
		\label{Ar1s1r1v1 b}
	\end{subfigure}
	~
	\begin{subfigure}{.48\linewidth}
		\includegraphics[scale=0.25]{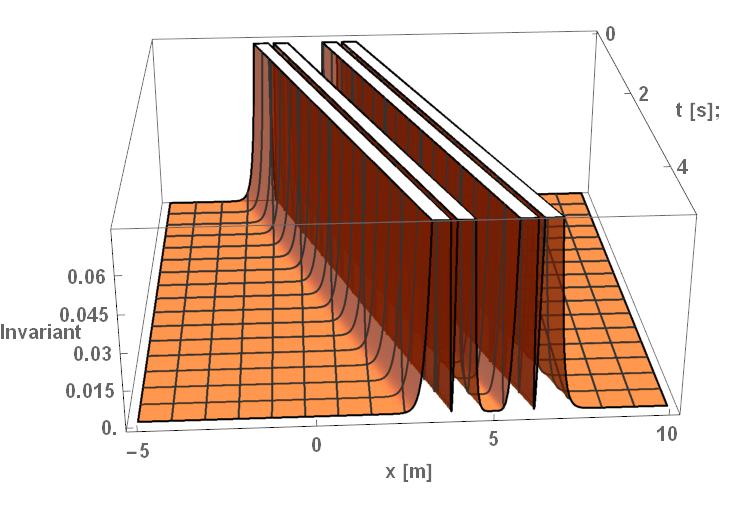}
		\caption{Plot of Alcubierre $r_1$ with $\sigma=2$ m$^{-1}$}
		\label{Ar1s2r1v1}
	\end{subfigure}
	\par \bigskip
	\begin{subfigure}{.5\linewidth}
		\includegraphics[scale=0.33]{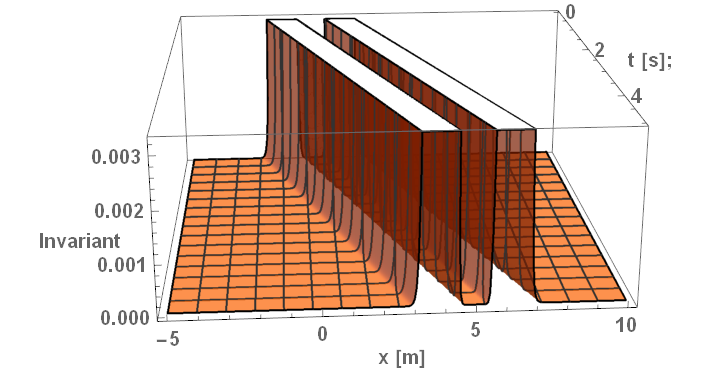}
		\caption{Plot of Alcubierre $r_1$ with $\sigma=4$ m$^{-1}$}
		\label{Ar1s4r1v1}
	\end{subfigure}
	~
	\begin{subfigure}{.48\linewidth}
		\includegraphics[scale=0.25]{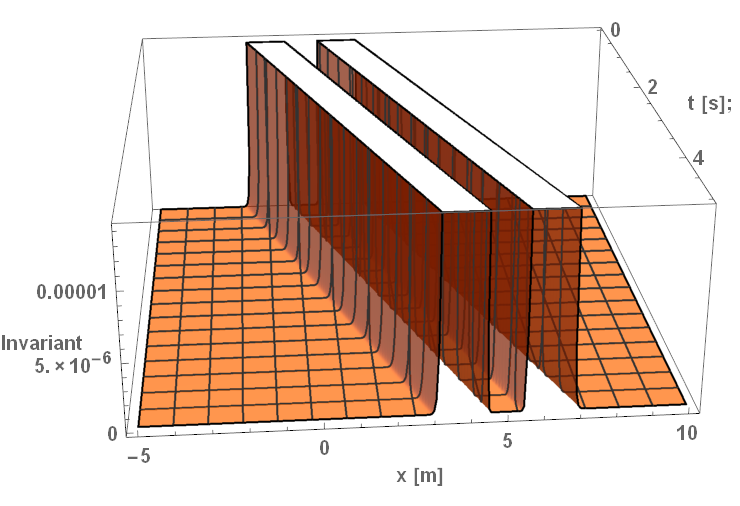}
		\caption{Plot of Alcubierre $r_1$ with $\sigma=6$ m$^{-1}$}
		\label{Ar1s6r1v1}
	\end{subfigure}
	\par \bigskip
	\begin{subfigure}{.48\linewidth}
		\includegraphics[scale=0.28]{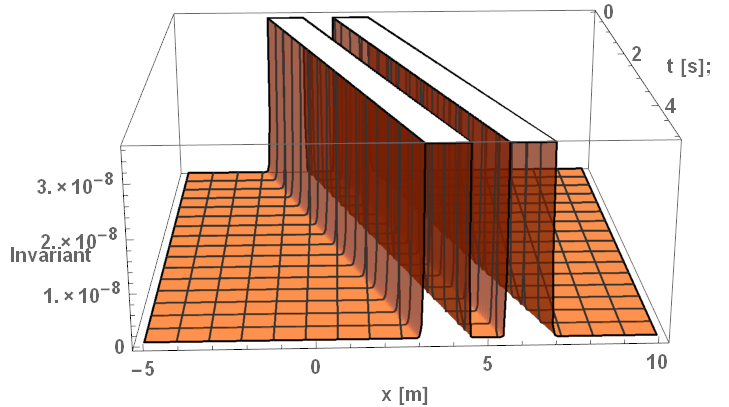}
		\caption{Plot of Alcubierre $r_1$ with $\sigma=8$ m$^{-1}$}
		\label{Ar1s8r1v1}
	\end{subfigure}
	~
	\begin{subfigure}{.48\linewidth}
		\includegraphics[scale=0.28]{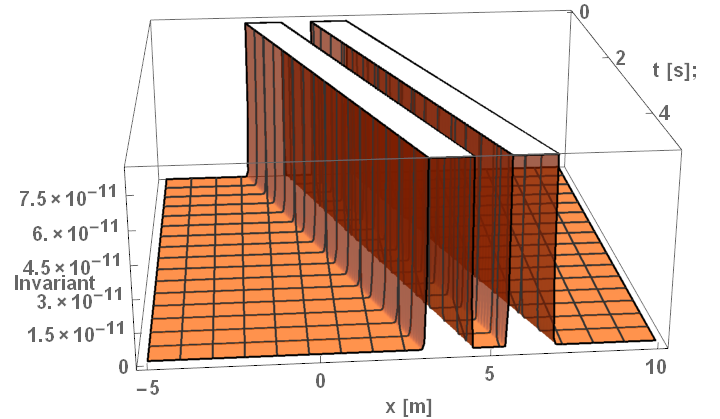}
		\caption{Plot of Alcubierre $r_1$ with $\sigma=10$ m$^{-1}$}
		\label{Ar1s10r1v1}
	\end{subfigure}
	\caption{Plots of the $r_1$ invariants for the Alcubierre warp drive while varying skin depth. 
	The other parameters were chosen as $v_s=1 \ c$ and $\rho=1$~m to match the parameters Alcubierre originally suggested in his paper \cite{Alcubierre:1994tu}.} \label{fig:4.6}
    \end{figure}
    
    \newpage
    
    \begin{figure}[ht]
	\begin{subfigure}{.48\linewidth}
		\includegraphics[scale=0.28]{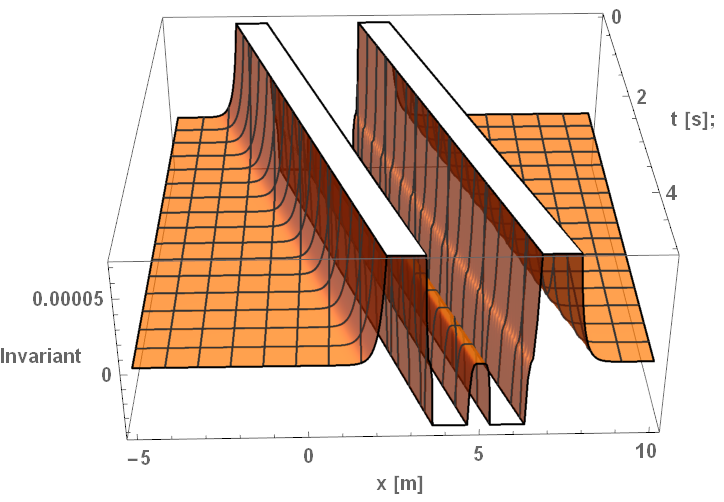}
		\caption{Plot of Alcubierre $w_2$ with $\sigma=1$ m$^{-1}$}
		\label{Aw2s1r1v1}
	\end{subfigure}
	~
	\begin{subfigure}{.48\linewidth}
		\includegraphics[scale=0.28]{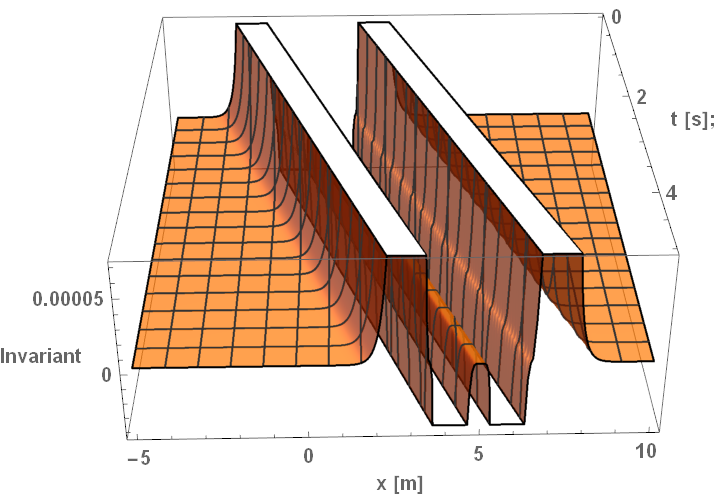}
		\caption{Plot of Alcubierre $w_2$ with  $\sigma=2$ m$^{-1}$}
		\label{Aw2s2r1v1}
	\end{subfigure}
	\par \bigskip
	\begin{subfigure}{.48\linewidth}
		\includegraphics[scale=0.3]{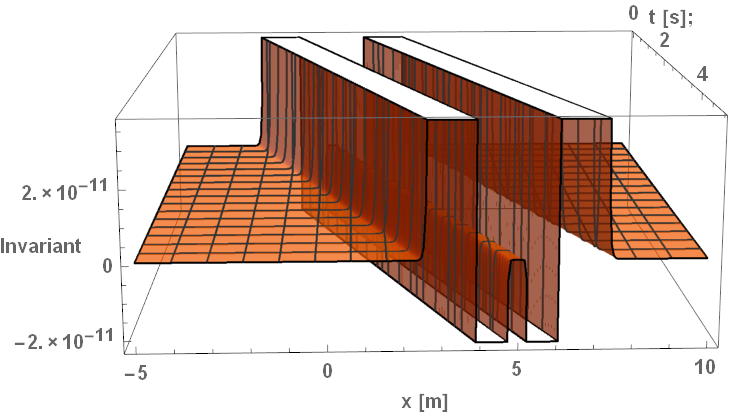}
		\caption{Plot of Alcubierre $w_2$ with $\sigma=4$ m$^{-1}$}
		\label{Aw2s4r1v1}
	\end{subfigure}
	~
	\begin{subfigure}{.48\linewidth}
		\includegraphics[scale=0.28]{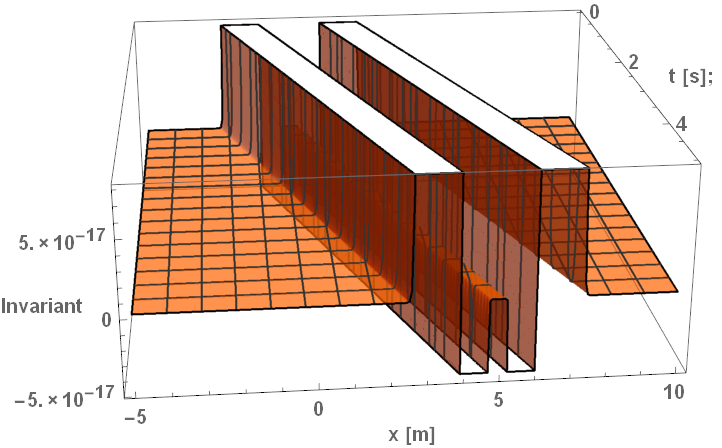}
		\caption{Plot of Alcubierre $w_2$ with $\sigma=6$ m$^{-1}$}
		\label{Aw2s6r1v1}
	\end{subfigure}
	\par \bigskip
	\begin{subfigure}{.5\linewidth}
		\includegraphics[scale=0.33]{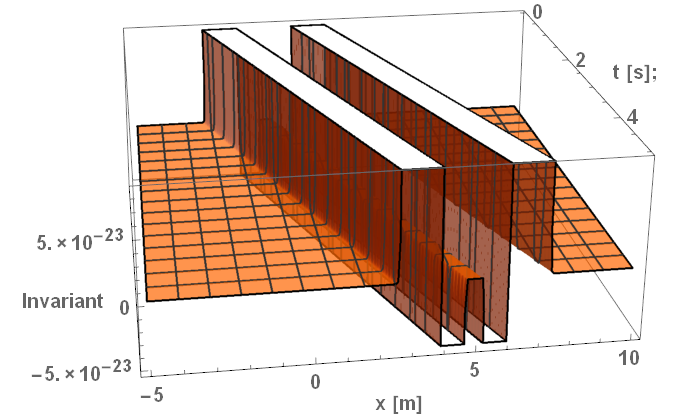}
		\caption{Plot of Alcubierre $w_2$ with $\sigma=8$ m$^{-1}$}
		\label{Aw2s8r1v1}
	\end{subfigure}
	~
	\begin{subfigure}{.48\linewidth}
		\includegraphics[scale=0.27]{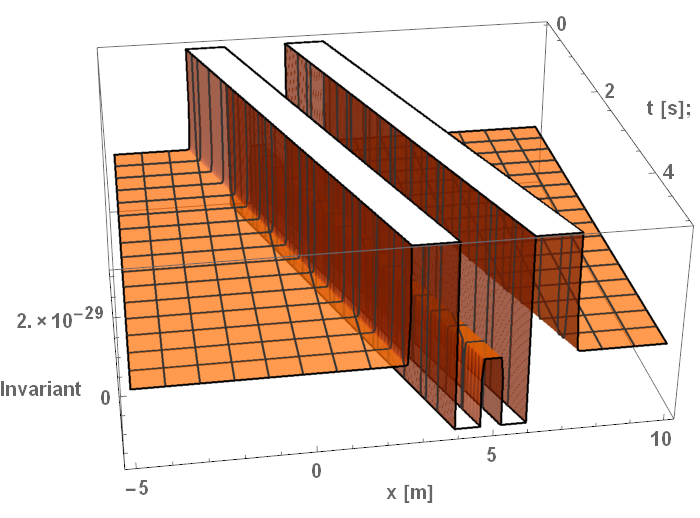}
		\caption{Plot of Alcubierre $w_2$ with $\sigma=10$ m$^{-1}$}
		\label{Aw2s10r1v1}
	\end{subfigure}
	\vspace{6pt}
	\caption{Plots of the $w_2$ invariants for the Alcubierre warp drive while varying skin-depth.
	The radius and velocity were chosen as $\rho=1$~m and $v_s=1 c$ to match the parameters Alcubierre originally suggested in his paper \cite{Alcubierre:1994tu}.} \label{fig:4.7}
    \end{figure}
    
    \FloatBarrier

    \begin{figure}[ht]
	\begin{subfigure}{.48\linewidth}
		\includegraphics[scale=0.27]{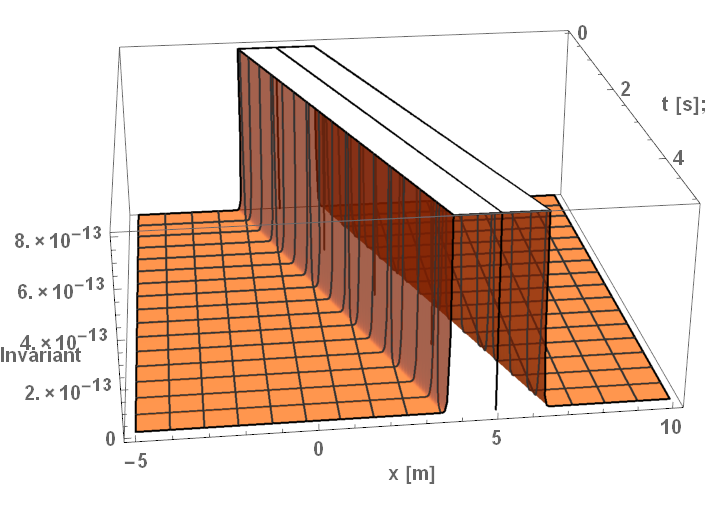}
		\caption{Plot of Alcubierre $r_1$ with $\rho=0.1$  m }
		\label{Ar1s8rp1v1}
	\end{subfigure}
	~
	\begin{subfigure}{.48\linewidth}
		\includegraphics[scale=0.3]{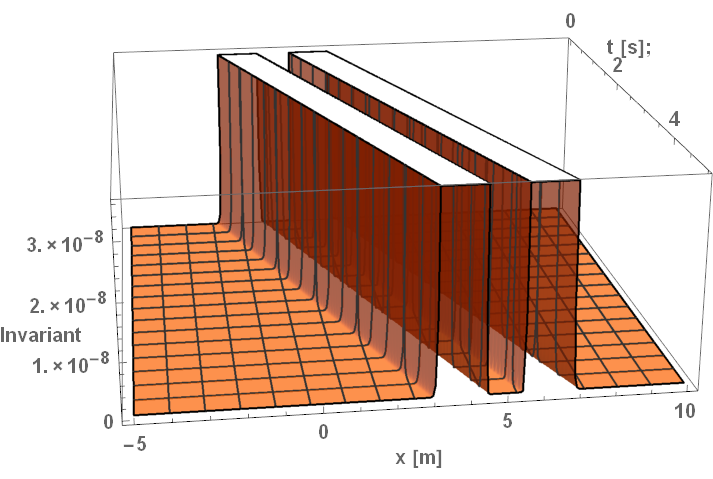}
		\caption{Plot of Alcubierre $r_1$ with $\rho=1$  m }
		\label{Ar1s8r1v1 c}
	\end{subfigure}
	\par \bigskip
	\begin{subfigure}{.5\linewidth}
		\includegraphics[scale=0.32]{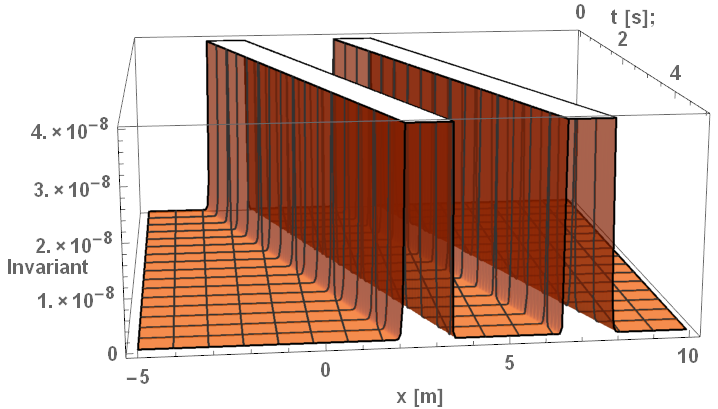}
		\caption{Plot of Alcubierre $r_1$ with $\rho=2$  m }
		\label{Ar1s8r2v1}
	\end{subfigure}
	~
	\begin{subfigure}{.48\linewidth}
		\includegraphics[scale=0.26]{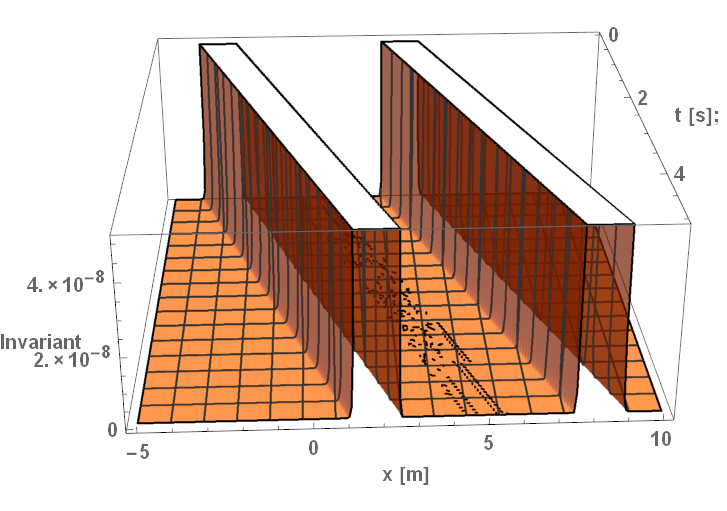}
		\caption{Plot of Alcubierre $r_1$ with $\rho=3$  m }
		\label{Ar1s8r3v1}
	\end{subfigure}
	\par \bigskip
	\begin{subfigure}{.48\linewidth}
		\includegraphics[scale=0.32]{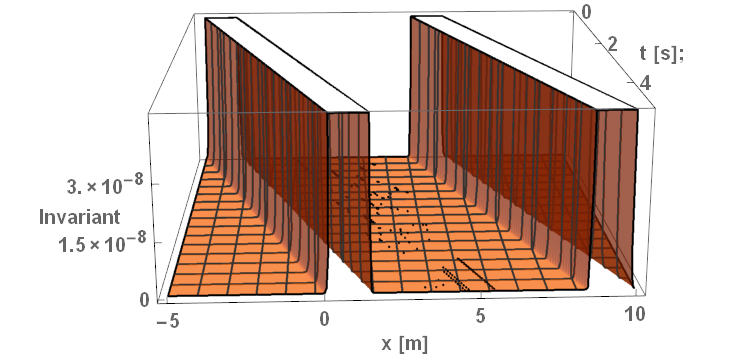}
		\caption{Plot of Alcubierre $r_1$ with $\rho=4$  m }
		\label{Ar1s8r4v1}
	\end{subfigure}
	~
	\begin{subfigure}{.48\linewidth}
		\includegraphics[scale=0.28]{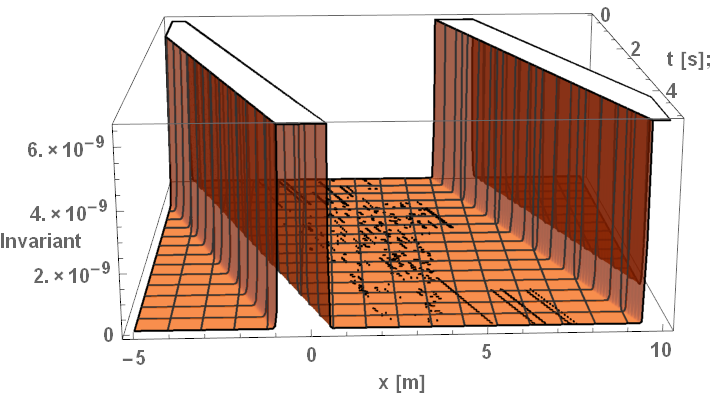}
		\caption{Plot of Alcubierre $r_1$ with $\rho=5$  m }
		\label{Ar1s8r5v1}
	\end{subfigure}
	\vspace{6pt}
	\caption{Plots of the $r_1$ invariants for the Alcubierre warp drive while varying the radius. The other parameters were chosen as $\sigma=8$ m$^{-1}$ and $v_s=1 \ c$ to match the parameters Alcubierre originally suggested in his paper \cite{Alcubierre:1994tu}.} \label{fig:4.9}
    \end{figure}
    
    \newpage
    
    \begin{figure}[ht]
	\begin{subfigure}{.48\linewidth}
		\includegraphics[scale=0.26]{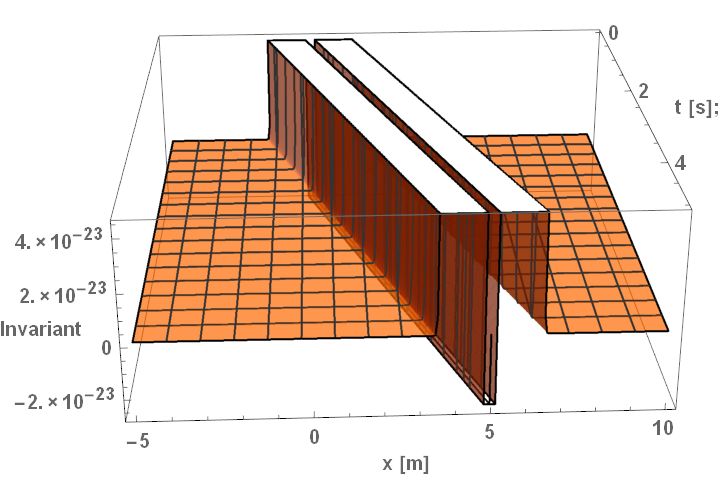}
		\caption{Plot of Alcubierre $w_2$ with $\rho=0.1$  m }
		\label{Aw2s8rp1v1}
	\end{subfigure}
	~
	\begin{subfigure}{.5\linewidth}
		\includegraphics[scale=0.33]{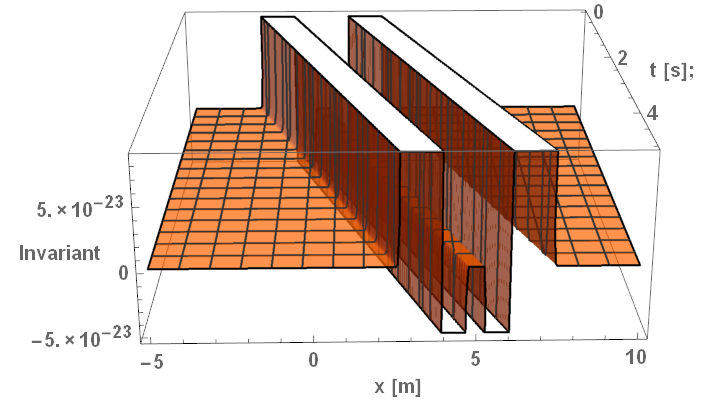}
		\caption{Plot of Alcubierre $w_2$ with $\rho=1$  m }
		\label{Aw2s8r1v1 c}
	\end{subfigure}
	\par \bigskip
	\begin{subfigure}{.5\linewidth}
		\includegraphics[scale=0.32]{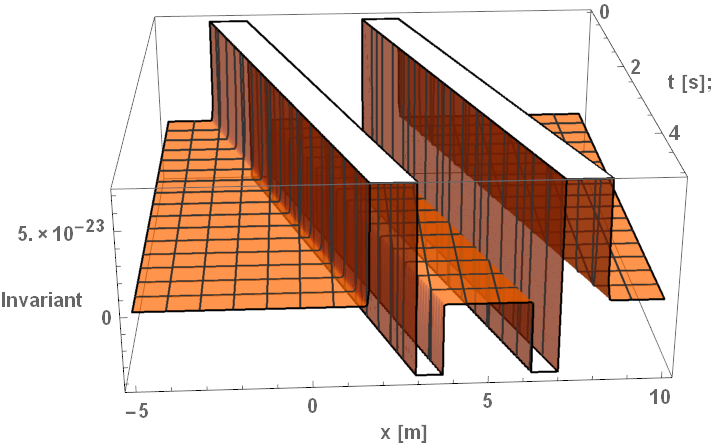}
		\caption{Plot of Alcubierre $w_2$ with $\rho=2$  m }
		\label{Aw2s8r2v1}
	\end{subfigure}
	~
	\begin{subfigure}{.48\linewidth}
		\includegraphics[scale=0.27]{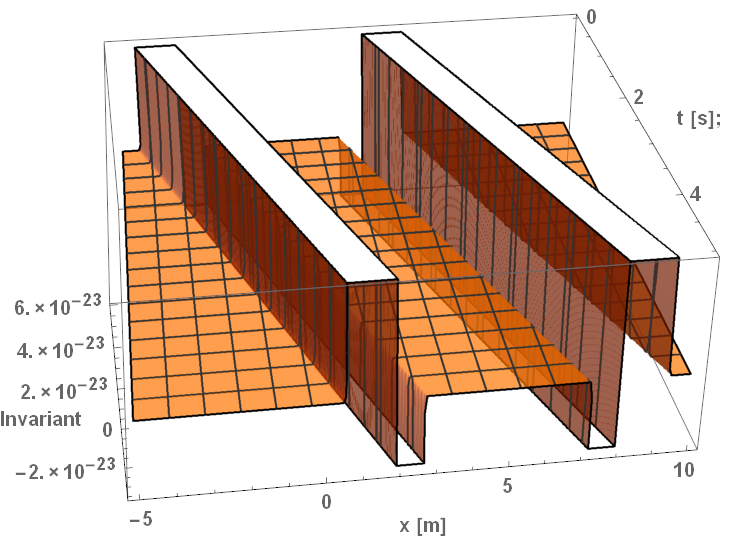}
		\caption{Plot of Alcubierre $w_2$ with $\rho=3$  m }
		\label{Aw2s8r3v1}
	\end{subfigure}
	\par \bigskip
	\begin{subfigure}{.48\linewidth}
		\includegraphics[scale=0.28]{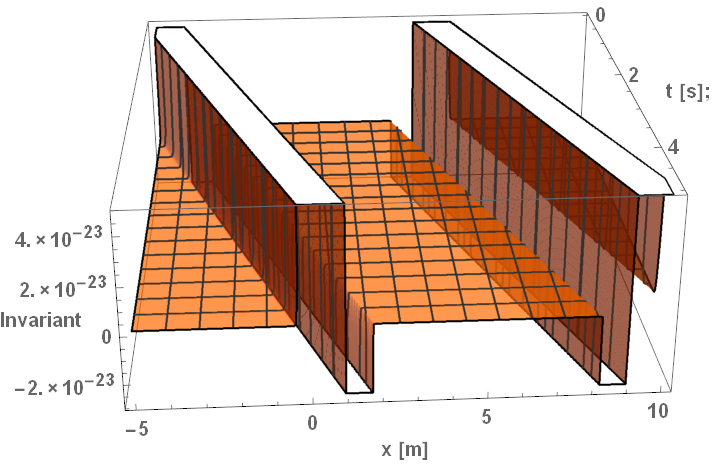}
		\caption{Plot of Alcubierre $w_2$ with $\rho=4$  m }
		\label{Aw2s8r4v1}
	\end{subfigure}
	~
	\begin{subfigure}{.48\linewidth}
		\includegraphics[scale=0.32]{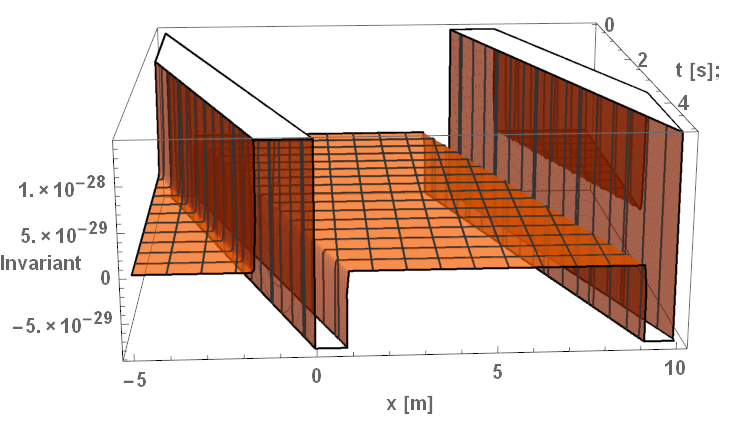}
		\caption{Plot of Alcubierre $w_2$ with $\rho=5$  m }
		\label{Aw2s8r5v1}
	\end{subfigure}
	\caption{Plots of the $w_2$ invariants for the Alcubierre warp drive while varying radius. 
	The other parameters were chosen as $\sigma=8$  m$^{-1}$ and $v_s=1 \ c$ to match the parameters Alcubierre originally suggested in his paper \cite{Alcubierre:1994tu}.} \label{fig:4.10}
    \end{figure}
    
    \FloatBarrier
    \newpage

\section{Invariant Plots for the Nat\'ario Warp Drive} \label{app:NatPlot}

    
    ~
    \begin{figure}[htb]
	\begin{subfigure}{.45\linewidth}
		\includegraphics[scale=0.28]{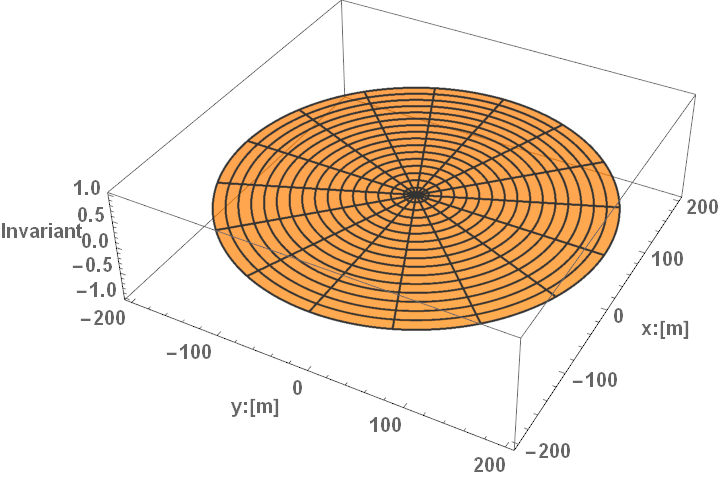}
		\caption{$v=0.0 \ c$}
		\label{fig:4.12a}
	\end{subfigure}
	~
	\begin{subfigure}{.55\linewidth}
		\includegraphics[scale=0.28]{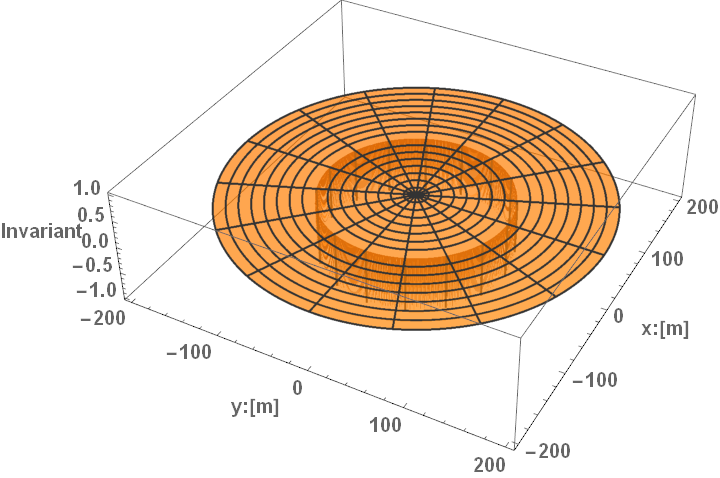}
		\caption{$v=0.01 \ c$}
		\label{fig:4.12b}
	\end{subfigure}
	\par \bigskip
	\begin{subfigure}{.45\linewidth}
		\includegraphics[scale=0.28]{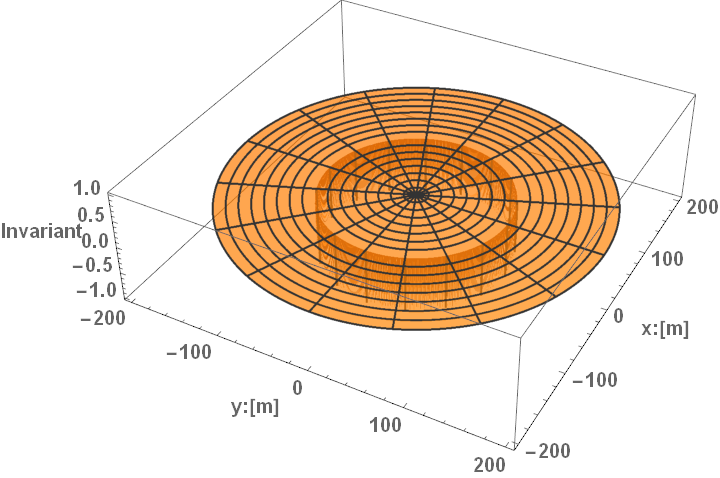}
		\caption{$v=0.1 \ c$}
		\label{fig:4.12c}
	\end{subfigure}
	~
	\begin{subfigure}{.55\linewidth}
		\includegraphics[scale=0.28]{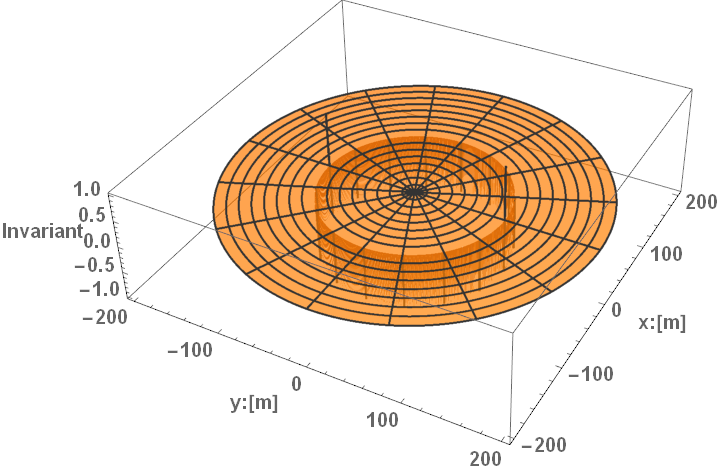}
		\caption{$v=1 \ c$}
		\label{fig:4.12d}
	\end{subfigure}
	\par \bigskip
	\begin{subfigure}{.45\linewidth}
		\includegraphics[scale=0.28]{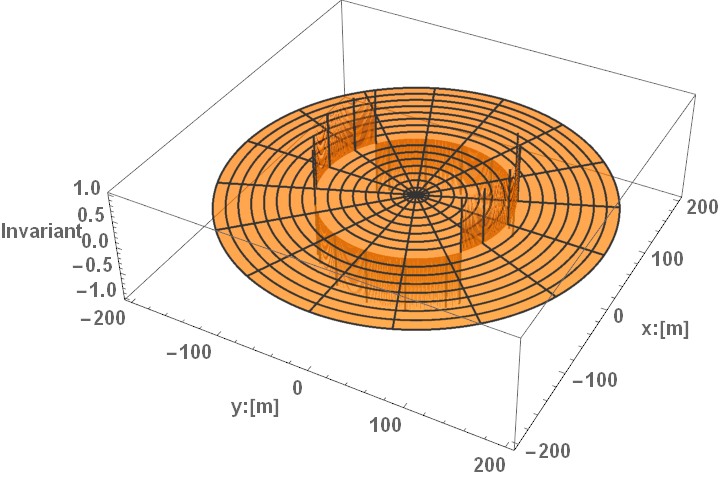}
		   \vspace{3pt}
		\caption{$v=10 \ c$}
		\label{fig:4.12e}
	\end{subfigure}
	~
	\begin{subfigure}{0.55\linewidth}
	    \centering
		\includegraphics[scale=0.28]{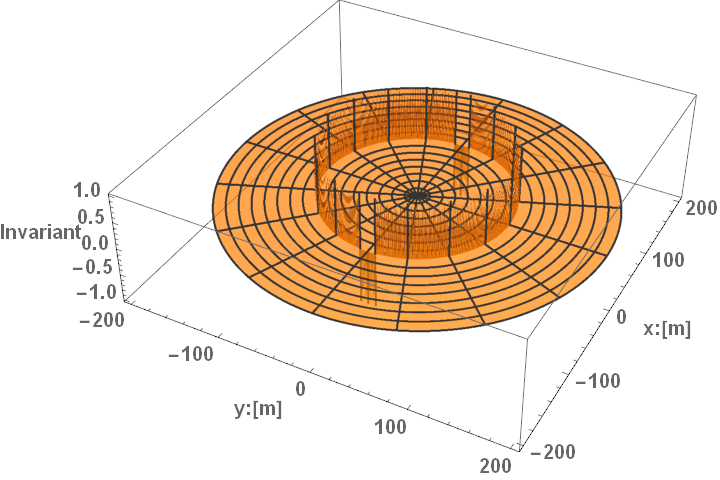}
		\caption{$v=100 \ c$}
		\label{fig:4.12f}
	\end{subfigure}
	\vspace{6pt}
	\caption{The velocity evolution of the $r_1$ invariant for the Nat\'ario warp drive at a constant velocity.
	The other parameters are set to $\sigma$ = 50, 000~$\frac{1}{\mathrm{m}}$ and $\rho$ = 100~m. 
	Equation \eqref{17} is in natural units, so the speed of light was normalized out of the equation.
	The factor of $c$ was included in these captions to stress that the plots are of multiples of the speed of light.
	Their units are ms$^{-1}$.} \label{fig:4.12}
    \end{figure}
    ~
    ~
    \begin{figure}[htb]
	\begin{subfigure}{.45\linewidth}
		\includegraphics[scale=0.28]{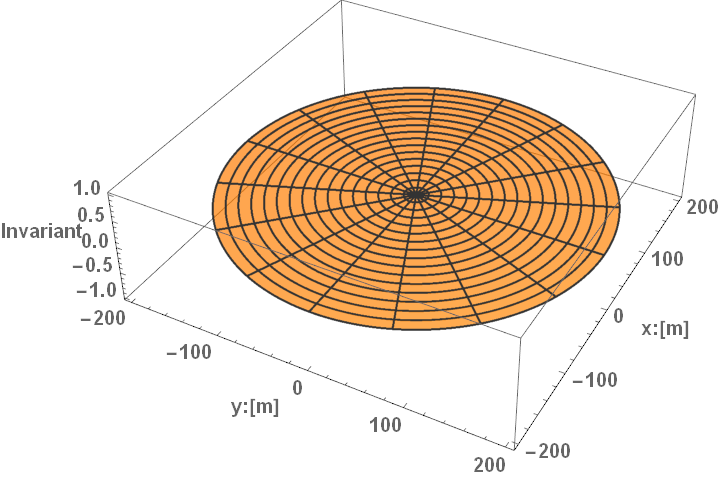}
		\caption{$v=0.0 \ c$}
		\label{fig:4.13a}
	\end{subfigure}
	~
	\begin{subfigure}{.55\linewidth}
		\includegraphics[scale=0.28]{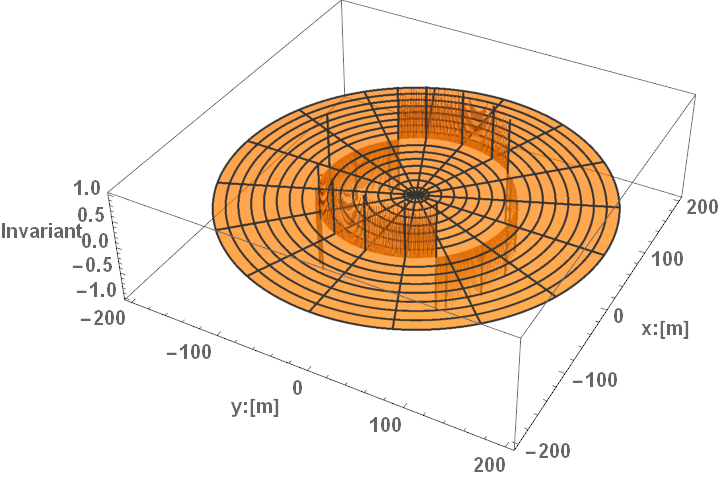}
		\caption{$v=0.01 \ c$}
		\label{fig:4.13b}
	\end{subfigure}
	\par \bigskip
	\begin{subfigure}{.45\linewidth}
		\includegraphics[scale=0.28]{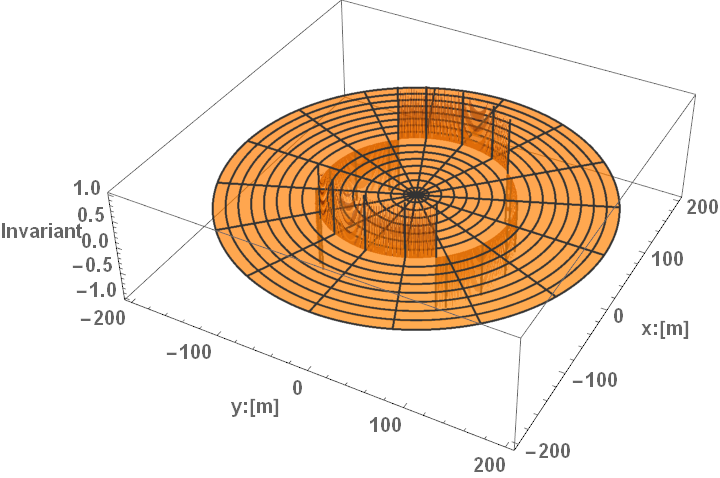}
		\caption{$v=0.1 \ c$}
		\label{fig:4.13c}
	\end{subfigure}
	~
	\begin{subfigure}{.55\linewidth}
		\includegraphics[scale=0.28]{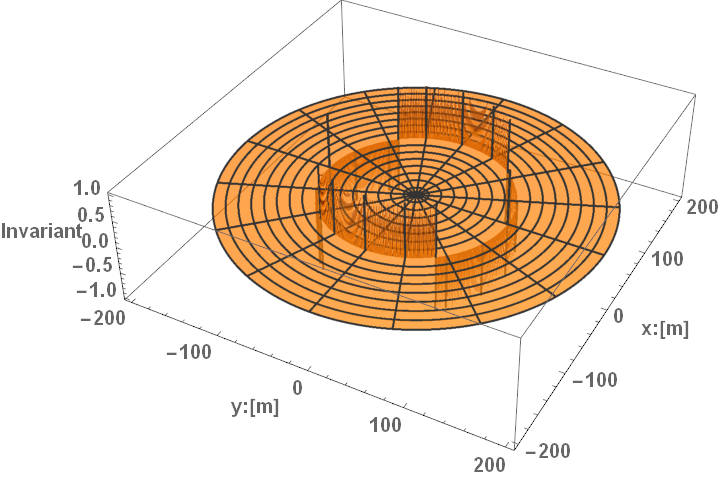}
		\caption{$v=1 \ c$}
		\label{fig:4.13d}
	\end{subfigure}
	\par \bigskip
	\begin{subfigure}{.45\linewidth}
		\includegraphics[scale=0.28]{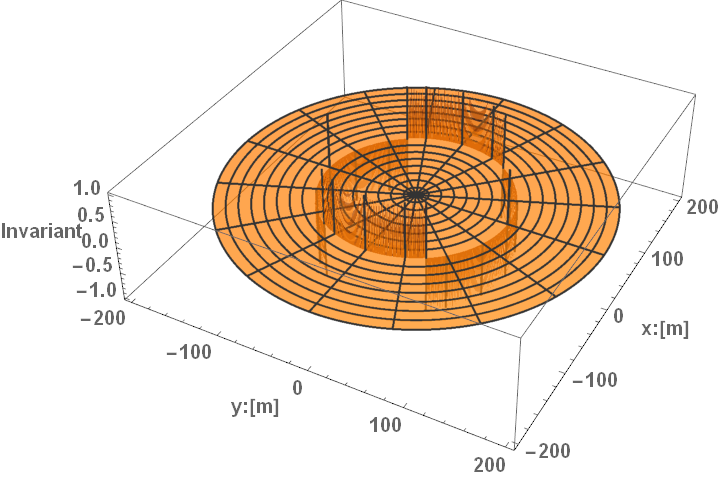}
		\caption{$v=10 \ c$}
		\label{fig:4.13e}
	\end{subfigure}
	~
	\begin{subfigure}{.55\linewidth}
		\includegraphics[scale=0.28]{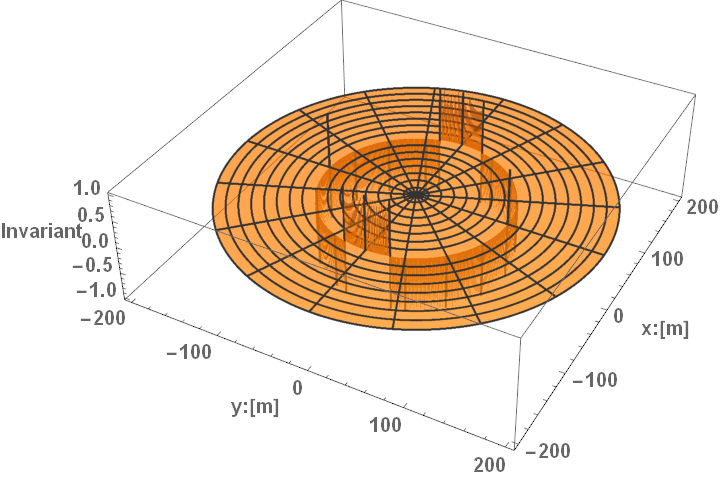}
		\caption{$v=100 \ c$}
		\label{fig:4.13f}
	\end{subfigure}
	\vspace{6pt}
	\caption{The velocity Evolution of the $r_2$ invariant for the Nat\'ario warp drive at a constant velocity.
	The other parameters are set to $\sigma$ = 50, 000~$\frac{1}{\mathrm{m}}$ and $\rho$ = 100~m.
	Equation \eqref{17} is in natural units, so the speed of light was normalized out of the equation.
	The factor of $c$ was included in these captions to stress that the plots are of multiples of the speed of light.
	Their units are ms$^{-1}$.} \label{fig:4.13}
    \end{figure}
    ~
    ~
    \begin{figure}[htb]
	\begin{subfigure}{.45\linewidth}
		\includegraphics[scale=0.28]{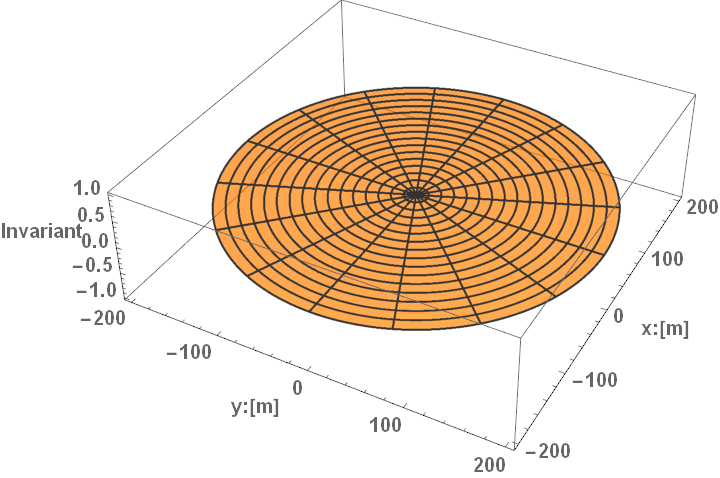}
		\caption{$v=0.0 \ c$}
		\label{fig:4.14a}
	\end{subfigure}
	~
	\begin{subfigure}{.55\linewidth}
		\includegraphics[scale=0.28]{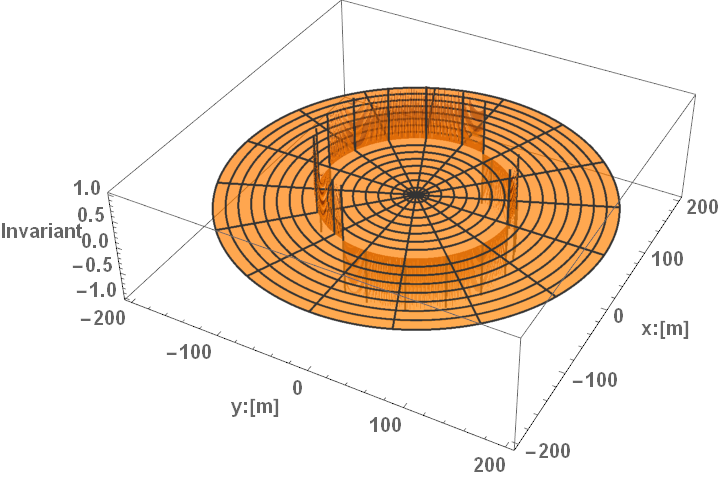}
		\caption{$v=0.01 \ c$}
		\label{fig:4.14b}
	\end{subfigure}
	\par \bigskip
	\begin{subfigure}{.45\linewidth}
		\includegraphics[scale=0.28]{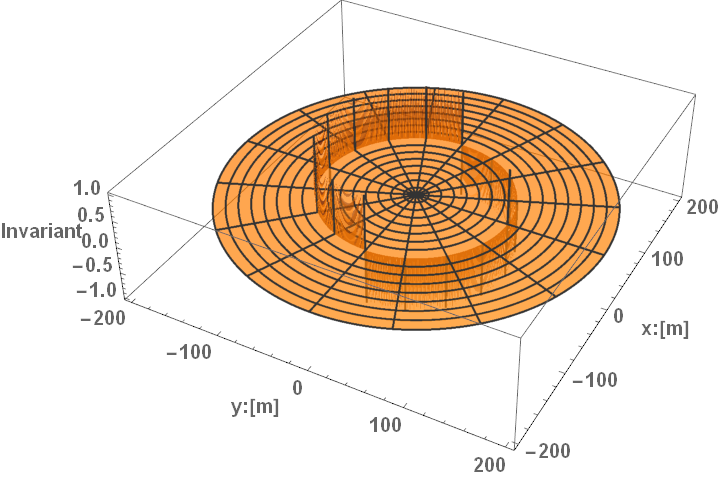}
		\caption{$v=0.1 \ c$}
		\label{fig:4.14c}
	\end{subfigure}
	~
	\begin{subfigure}{.55\linewidth}
		\includegraphics[scale=0.28]{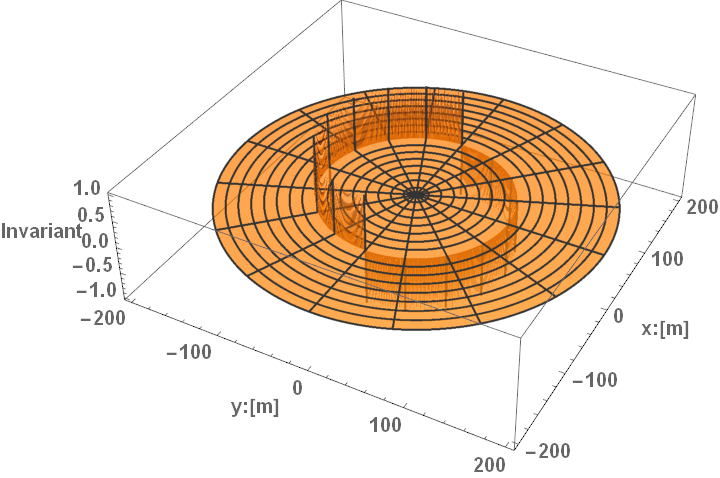}
		\caption{$v=1 \ c$}
		\label{fig:4.14d}
	\end{subfigure}
	\par \bigskip
	\begin{subfigure}{.45\linewidth}
		\includegraphics[scale=0.28]{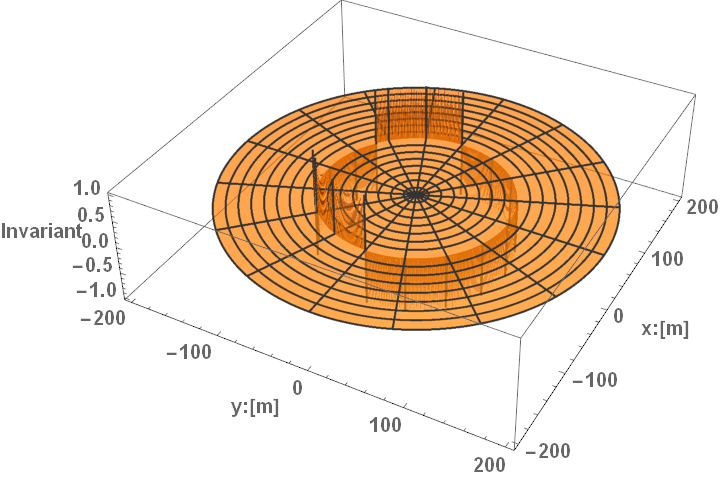}
		\caption{$v=10 \ c$}
		\label{fig:4.14e}
	\end{subfigure}
	~
	\begin{subfigure}{.55\linewidth}
		\includegraphics[scale=0.28]{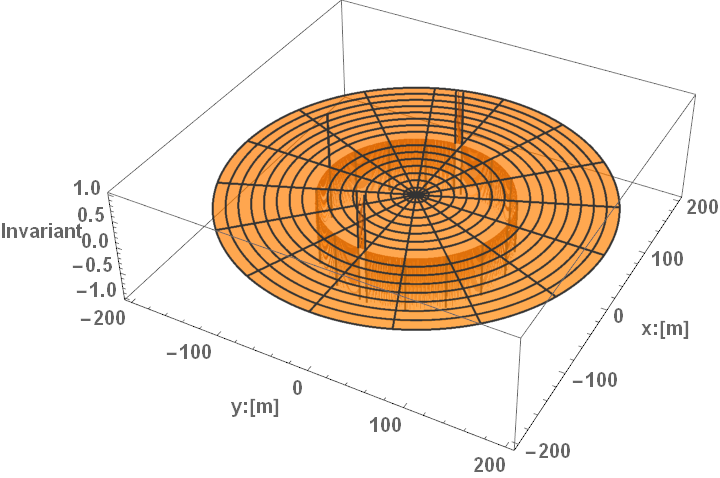}
		\caption{$v=100 \ c$}
		\label{fig:4.14f}
	\end{subfigure}
	\vspace{6pt}
	\caption{The velocity evolution of the $w_2$ invariant for the Nat\'ario warp drive at a constant velocity.
	The other parameters are set to $\sigma$ = 50, 000~$\frac{1}{\mathrm{m}}$ and $\rho$ = 100~m.
	Equation \eqref{17} is in natural units, so the speed of light was normalized out of the equation.
	The factor of $c$ was included in these captions to stress that the plots are of multiples of the speed of light.
	Their units are ms$^{-1}$.} \label{fig:4.14}
    \end{figure}
    
    \FloatBarrier
    
    \begin{figure}[hb] \label{fig:4.16p1}
    	\begin{subfigure}{.45\linewidth}
    		\includegraphics[scale=0.25]{Images/Chapter4/Natario/NcV-Rs50000p100v1.png}
    		\caption{The invariant $R$ with $\sigma$ = 50, 000 $\frac{1}{\mathrm{m}}$}
    		\label{fig:4.15a}
    	\end{subfigure}
	~
    	\begin{subfigure}{.55\linewidth}
    		\includegraphics[scale=0.25]{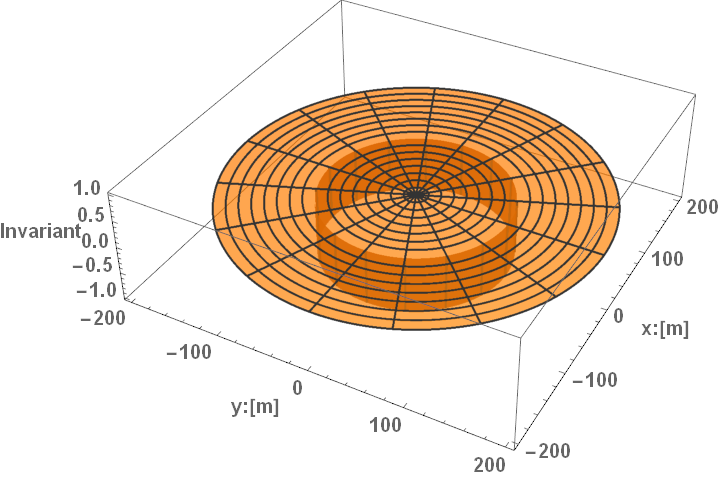}
    		\caption{The invariant $R$ with $\sigma$ = 100, 000 $\frac{1}{\mathrm{m}}$}
    		\label{fig:4.15b}
    	\end{subfigure}
	\par \bigskip
    	\begin{subfigure}{.45\linewidth}
    		\includegraphics[scale=0.25]{Images/Chapter4/Natario/NcV-r1s50000p100v1.png}
    		\caption{The invariant $r_1$ with $\sigma$ = 50, 000 $\frac{1}{\mathrm{m}}$}
    		\label{fig:4.15c}
    	\end{subfigure}
	~
    	\begin{subfigure}{.55\linewidth}
    		\includegraphics[scale=0.25]{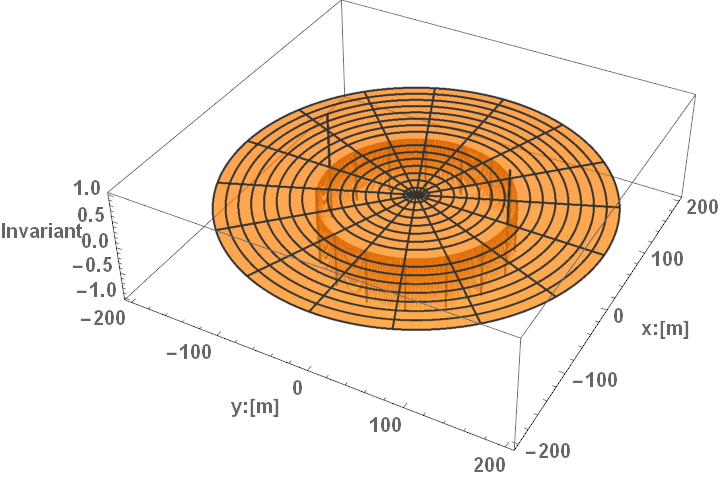}
    		\caption{The invariant $r_1$ with $\sigma$ = 100, 000 $\frac{1}{\mathrm{m}}$}
    		\label{fig:4.15d}
    	\end{subfigure}
    \par \bigskip
    	\begin{subfigure}{.45\linewidth}
    		\includegraphics[scale=0.25]{Images/Chapter4/Natario/NcV-r2s50000p100v1.png}
    		\caption{The invariant $r_2$ with $\sigma$ = 50, 000 $\frac{1}{\mathrm{m}}$}
    		\label{fig:4.16a}
    	\end{subfigure}
	~
    	\begin{subfigure}{.55\linewidth}
    		\includegraphics[scale=0.25]{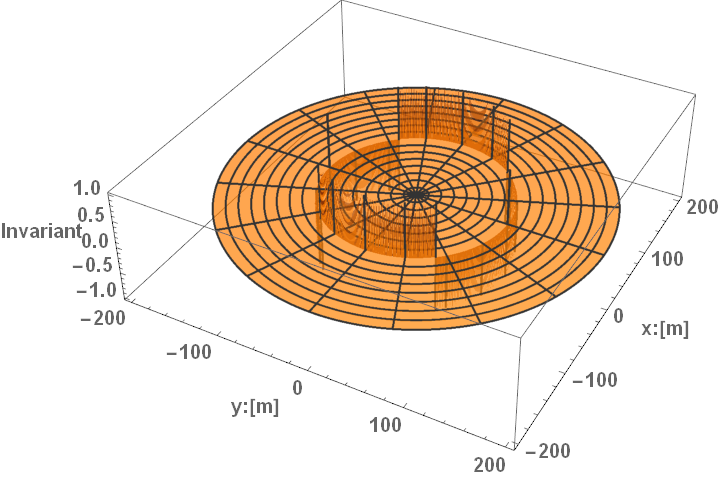}
    		\caption{The invariant $r_2$ with $\sigma$ = 100, 000 $\frac{1}{\mathrm{m}}$}
    		\label{fig:4.16b}
    	\end{subfigure}
	\par \bigskip
    	\begin{subfigure}{.45\linewidth}
    		\includegraphics[scale=0.25]{Images/Chapter4/Natario/NcV-w2s50000p100v1.png}
    		\caption{The invariant $w_2$ with $\sigma$ = 50, 000 $\frac{1}{\mathrm{m}}$}
    		\label{fig:4.16c}
    	\end{subfigure}
	~
    	\begin{subfigure}{0.55\linewidth}
    		\includegraphics[scale=0.25]{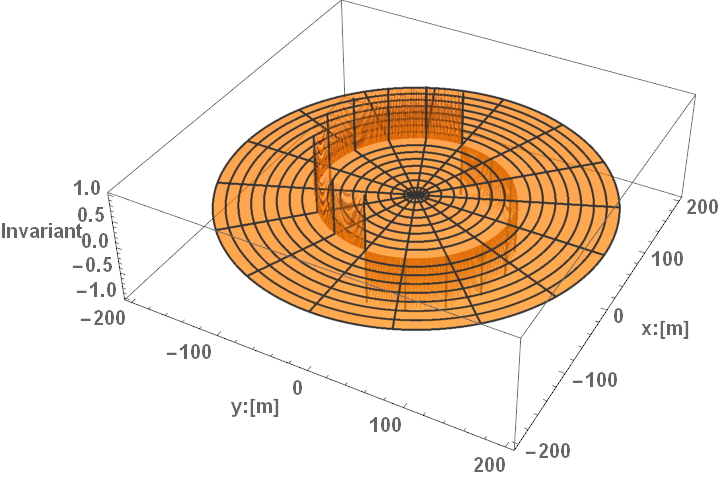}
    		\caption{The invariant $w_2$ with $\sigma$ = 100, 000 $\frac{1}{\mathrm{m}}$}
    		\label{fig:4.16d}
    	\end{subfigure}
    	\vspace{6pt}
    	\caption{The warp bubble skin depth for $R$, $r_1$, $r_2$ and $w_2$ for the Nat\'ario warp drive at a constant velocity.
    	The other parameters were chosen to be $v=1 \ c$, and $\rho$ = 100~m in natural units.} \label{fig:4.16}
    \end{figure}
    
    \FloatBarrier
    
    \newpage
    
    \begin{figure}[h]
    	\begin{subfigure}{.45\linewidth}
    		\includegraphics[scale=0.26]{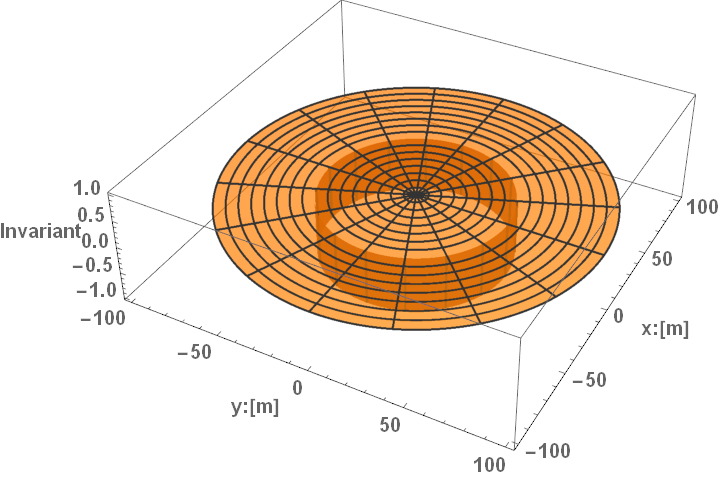}
    		\caption{The invariant $R$ with $\rho$ = 50~m}
    		\label{fig:4.17.a}
    	\end{subfigure}
	~
    	\begin{subfigure}{.55\linewidth}
    		\includegraphics[scale=0.26]{Images/Chapter4/Natario/NcV-Rs50000p100v1.png}
    		\caption{The invariant $R$ with $\rho$ = 100~m}
    		\label{fig:4.17.b}
    	\end{subfigure}
	\par \bigskip
    	\begin{subfigure}{.45\linewidth}
    		\includegraphics[scale=0.26]{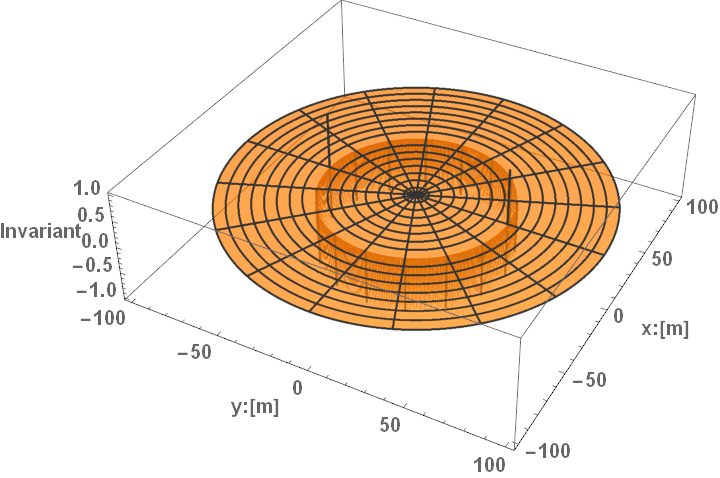}
    		\caption{The invariant $r_1$ with $\rho$ = 50~m}
    		\label{fig:4.17.c}
    	\end{subfigure}
	~
    	\begin{subfigure}{.55\linewidth}
    		\includegraphics[scale=0.26]{Images/Chapter4/Natario/NcV-r1s50000p100v1.png}
    		\caption{The invariant $r_1$ with $\rho$ = 100~m}
    		\label{fig:4.17.d}
    	\end{subfigure}
    \par \bigskip
    	\begin{subfigure}{.45\linewidth}
    		\includegraphics[scale=0.26]{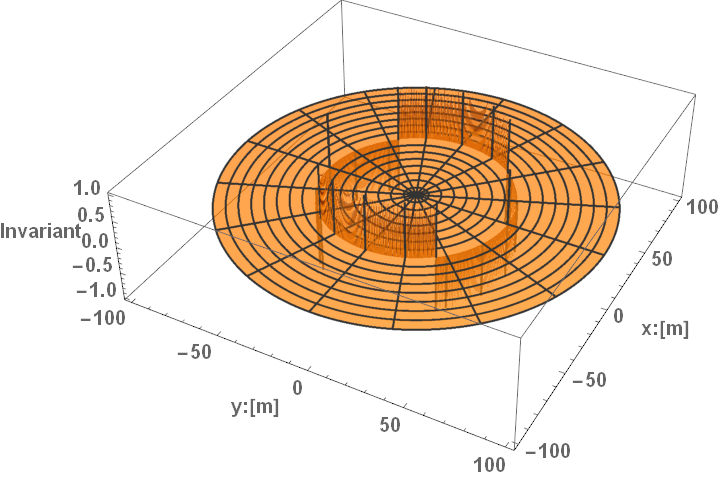}
    		\caption{The invariant $r_2$ with $\rho$ = 50~m}
    		\label{fig:4.18.a}
    	\end{subfigure}
	~
    	\begin{subfigure}{.55\linewidth}
    		\includegraphics[scale=0.26]{Images/Chapter4/Natario/NcV-r2s50000p100v1.png}
    		\caption{The invariant $r_2$ with $\rho$ = 100~m}
    		\label{fig:4.18.b}
    	\end{subfigure}
	\par \bigskip
    	\begin{subfigure}{.45\linewidth}
    		\includegraphics[scale=0.26]{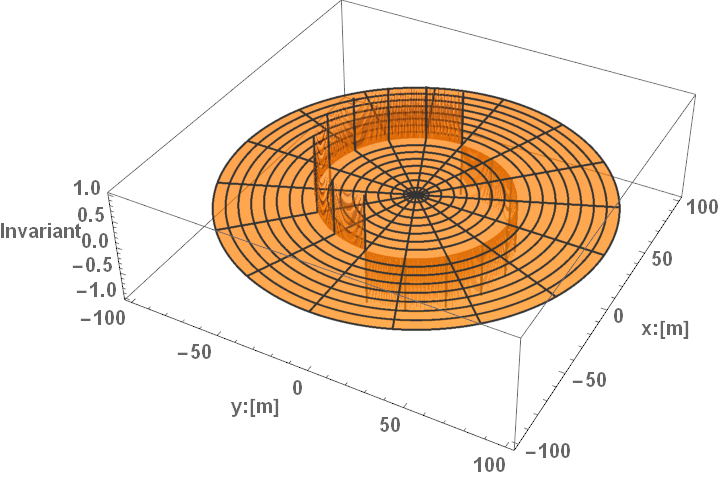}
    		\caption{The invariant $w_2$ with $\rho$ = 50~m}
    		\label{fig:4.18.c}
    	\end{subfigure}
	~
    	\begin{subfigure}{0.55\linewidth}
    		\includegraphics[scale=0.26]{Images/Chapter4/Natario/NcV-w2s50000p100v1.png}
    		\caption{The invariant $w_2$ with $\rho$ = 100~m}
    		\label{fig:4.18.d}
    	\end{subfigure}
    	\vspace{6pt}
    	\caption{The warp bubble radius for the $R$, $r_1$, $r_2$, and $w_2$ for the Nat\'ario warp drive at a constant velocity.
    	The other parameters were chosen to be $v=1 \ c$, and $\sigma $ = 50, 000  m$^{-1}$ in natural units.} \label{fig:4.18}
    \end{figure}
    
    \FloatBarrier



\end{document}